\documentclass{aa}
\usepackage[utf8]{inputenc}
\usepackage{graphicx}

\begin{document}

\title{Multi-wavelength observations and modelling of a quiescent cloud LDN1512}

\author{Mika Saajasto\inst{1}
  \and Mika Juvela\inst{1}
  \and Charlène Lefèvre\inst{2}
  \and Laurent Pagani\inst{3}
  \and Nathalie Ysard\inst{4}
  }
%\offprints{M. Saajasto, \email{mika.saajasto@helsinki.fi}}

\institute{Department of Physics, P.O.Box 64, FI-00014, University of Helsinki, Finland
\and Institut de Radioastronomie Millimétrique (IRAM), 300 rue de la Piscine, 38406 Saint-Martin d’Hères, France
\and LERMA \& UMR 8112 du CNRS, Observatoire de Paris, PSL Research University, CNRS, Sorbonne Universités, UPMC Univ. Paris 06, 75014 Paris, France
\and Institut d’Astrophysique Spatiale, CNRS, Univ. Paris-Sud, Université Paris-Saclay, B$\hat{a}$t. 121, 91405 Orsay cedex, France
}

\date{Received day month year / Accepted day month year}

\abstract
{Light scattering at near-infrared wavelengths has been used to study the optical properties of the interstellar dust grains, but these studies are limited by the assumptions on the strength of the radiation field. On the other hand, thermal dust emission can be used to constrain the properties of the radiation field, although this is hampered by uncertainty about the dust emissivity.}
{Combining light scattering and emission studies allows us to probe the properties of the dust grains in detail. We wish to study if current dust models allow us to model a molecular cloud simultaneously in the near infrared (NIR) and far infrared (FIR) wavelengths and compare the results with observations. Our aim is to place constraints on the properties of the dust grains and the strength of the radiation field.}
{We present computations of dust emission and scattered light of a quiescent molecular cloud LDN1512. We use NIR observations covering the J, H, and K$\rm _S$ bands, and FIR observations between 250 $\mu$m and 500 $\mu$m from Herschel space telescope. We construct radiative transfer models for LDN1512 that include an anisotropic radiation field and a three-dimensional cloud model.}
{We are able to reproduce the observed FIR observations, with a radiation field derived from the DIRBE observations, with all of the tested dust models. However, with the same density distribution and the assumed radiation field, the models fail to reproduce the observed NIR scattering in all cases except for models that take into account dust evolution via coagulation and mantle formation. The intensity from the diffuse interstellar medium (ISM) like, dust models can be increased to match the observed one by reducing the derived density, increasing the intensity of the background sky and the strength of the radiation field between factors from 2 to 3. We find that the column densities derived from our radiative transfer modelling can differ by a factor of up to two, compared to the column densities derived from the observations with modified blackbody fits. The discrepancy in the column densities is likely caused because of temperature difference between a modified blackbody fit and the real spectra. The difference between the fitted temperature and the true temperature could be as high as $\Delta T = +1.5$ K.}
{We show that the observed dust emission can be reproduced with several different assumptions about the properties of the dust grains. However, in order to reproduce the observed scattered surface brightness dust evolution must be taken into account.}

\keywords{Interstellar medium (ISM): Clouds -- Physical processes: Emission -- Physical processes: Scattering -- Methods: Radiative Transfer -- Stars: Formation}

\maketitle

\section{Introduction}

Understanding how stars are formed is one of the crucial questions in astronomy. The Herschel space observatory has provided us with detailed observations of nearby molecular clouds and shown that star forming regions have vastly diverse morphologies, from dynamically active filamentary fields to more quiescent clouds with simple geometries \citep{Molinari2010, Menshchikov2010, Juvela2012}. These far infrared (FIR) observations can be used to derive column density estimates and to study possible variations in dust properties. However, these studies are limited by our understanding of the emission properties of the grains, in particular the dust opacity and to a lesser degree the dust opacity spectral index.  

The light scattered by dust grains at near-infrared (NIR) and mid-infrared (MIR) wavelengths can be studied and analysed without explicit assumptions of the FIR thermal emission properties of the grains, thus the scattering observations can be used to place additional constraints on dust properties and the density distribution. \citet{Lehtinen1996} were the first to study the extended surface brightness of dense interstellar cores at NIR wavelengths and later \citet{Foster2006} showed that it is possible to make large-scale maps of this extended surface brightness, naming it 'Cloudshine', and attributed it to scattered light. \citet{Padoan2006} used radiative transfer modelling to show that the observed Cloudshine was consistent with the hypothesis of light scattering and could be used for studies of cloud structure. Star-forming clouds, but not the cores, are usually only moderately optically thick in the NIR and therefore there is a near-linear dependence between the surface brightness and the column density, assuming that the dust properties do not change with column density.

At MIR wavelengths, especially in the Spitzer 3.6 and 4.5 $\mu$m bands, a surprisingly high surface brightness was detected towards several cloud cores \citep{Steinacker2010, Pagani2010, Juvela2012}. Explaining the higher-than-expected surface brightnesses with the classical grain size distributions \citep{Mathis1977} proved difficult, implying the presence of larger grains with sizes of the order of $\sim 1 \, \mu$m \citep{Steinacker2010, Pagani2010, Andersen2013, Lefevre2014, Steinacker2015}. The high surface brightness towards cloud centres was named 'Coreshine' and is considered to be a direct evidence of dust growth in dense clouds.

The comparison of thermal dust emission and light scattering provides crucial information on the dust properties. On the other hand, because of limited resolution, or issues caused by anisotropic illumination and line-of-sight confusion, careful radiative transfer modelling is often needed to place constraints on theoretical models of dust.

We have chosen the cloud LDN1512 (hereafter L1512) to study and model NIR light scattering and thermal dust emission at FIR wavelengths. The cloud has a simple cometary morphology, is nearby \citep[140 $\pm \, 20$ pc,][]{Launhardt2013}, and based on the reported small line width of $\rm N_2H^+$ \citep{Caselli1995, Caselli2002, Lin2020}, appears to be quiescent. 

The cloud has been previously mapped with the \textit{Herschel} space observatory using both the photodetecting array camera and spectrometer (PACS) \citep{Poglitsch2010} and the spectral and photometric imaging receiver (SPIRE) \citep{Griffin2010} instruments. As discussed by \citet{Launhardt2013} and \citet{Lippok2013} the \textit{Herschel} observations show a single starless core surrounded by a diffuse envelope. Based on the fitting of the $\rm ^{13} CO$ line, \citet{Lippok2013} showed that the envelope of the core has a higher gas temperature compared to the central regions and their stability analysis shows that the core is thermally supercritical. The $\rm N_2H^+$ observations of \citet{Lin2020} show a low kinetic temperature of $\sim8 \, \pm \, 1$ K within the innermost $\sim 0.017$ pc of the core, and based on their chemical modelling, the cloud is sufficiently evolved that the $\rm N_2$ chemistry has reached a steady state. Their results suggest that L1512 is likely older than 1.4 Myr and that ambipolar diffusion has led to the formation of the core.

\begin{figure*}
\sidecaption
\includegraphics[width=11.5cm]{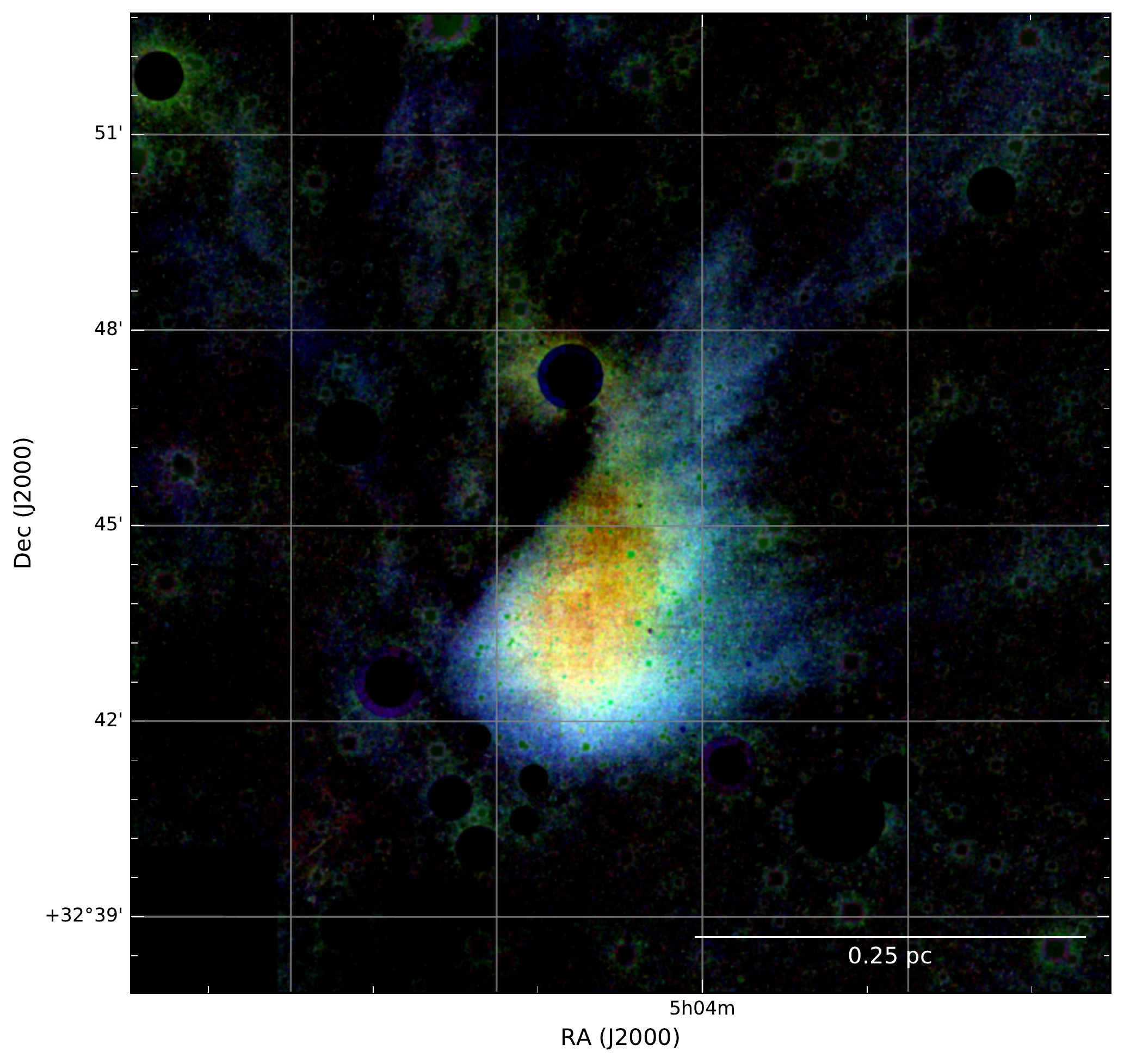}
\caption{Three colour image of the field L1512. The colours correspond to the NIR surface brightness at J (blue colour), H (green), and K$\rm _S$ bands (red). All point sources have been masked.}
\label{fig:RGB}
\end{figure*}

We use J, H, and K$\rm _S$ bands images, shown as a three-colour image in Fig. \ref{fig:RGB}, from the Wide InfraRed CAMera (WIRCAM) on the Canada-France-Hawaii Telescope (CFHT). The observations are presented in \citet{Lin2020}, where only the stellar content of the images has been exploited. in this paper we concentrate on the extended emission. After background subtraction, all three bands show a clear extended surface brightness component.

In this study, we will simultaneously model the cloud at NIR and FIR wavelengths using an anisotropic radiation field and a cloud model derived from the \textit{Herschel} observations together with dust models for the diffuse ISM dust, based on the model described by \citet{Compiegne2011} and three models that take into account the evolution of dust grains in shape and size. The aim of our study is to find a solution, or place constraints, for the cloud, the radiation field and the properties of the dust grains, that would give a consistent explanation for the observed extinction, NIR scattering, and dust emission.

This paper is organised as follows. In section 2, we give an overview of the archival observations used in the paper and describe our NIR observations. In section 3, we describe our radiative transfer methodology and explain our radiation field and cloud models. In section 4 we present our results and discuss our findings in section 5. Finally, in section 6, we summarise our findings and provide our conclusions.

\section{Observations} \label{sect:obs}

The \textit{Herschel} observations were downloaded from the Herschel science archive and are a part of the \textit{Herschel} key project The Earliest Phases of Star Formation (EPoS,  PI: O. Krause). The EPoS project used both the PACS and the SPIRE instruments to cover a total of 12 different fields between wavelengths from 100 $\mu$m to 500 $\mu$m. In the projects source list, the source CB 27 corresponds to the cloud L1512. The nominal scanning speeds for the PACS and SPIRE instruments were set to 20$\arcsec$ s$^{-1}$ and 30$\arcsec$ s$^{-1}$, respectively. The general noise level in the EPoS projects maps is $\sim 6$ mJy/beam, but for the L1512 the noise levels are slightly higher $\sim 11$ mJy/beam. A more detailed description on the data reduction is given by \citet{Launhardt2013}.

The NIR observations cover the wide filters J (1.25 $\mu$m), H (1.6 $\mu$m), and K$\rm _S$ (2.15 $\mu$m) (see Fig. \ref{fig:NIR_obs}) and were obtained in 2013 using the WIRCAM instrument on the CFHT. The observations were carried out using the Sky-Target-Sky (STS) dithering mode, to subtract the atmospheric IR-emission and to preserve any extended scattered light from the source. The seeing conditions during the observations were good, with typical values less than $1 \arcsec$. Data reduction was performed at the TERAPIX center using a specific reduction method to recover the extended emission. 

\section{Methods}

Our aim is to simultaneously model the cloud in NIR and FIR and to compare our modelling results with observations. In this section we describe our model of the density distribution within the cloud and our radiation field model.

\subsection{Cloud model}

We use a three-dimensional model cloud discretised onto a grid of 225$^3$ cells with a Gaussian density distribution along the line-of-sight (LOS). The angular cell size of our model cloud is 6$\arcsec$ which corresponds to the pixel size of the \textit{Herschel} SPIRE 250 $\mu$m map. Thus, the total angular extent of our model is $22.5\arcmin \times 22.5\arcmin$, which at a distance of 140 pc corresponds to a physical size of $\sim 0.4 \times 0.4$ pc. The density distribution is based on the \textit{Herschel} SPIRE 350 $\mu$m surface brightness (determining the column density distribution on the plane of the sky) and scaled along the LOS (z-coordinate) by a Gaussian distribution

\begin{equation}
\rho(z) = \frac{1}{\sigma \sqrt{2 \pi}} \exp \, \left( - \frac{(z - \mu)^2 }{2 \sigma^2} \right),
\end{equation}
where $\mu = 0.0$ pc and with either $\sigma=0.07$ pc, labeled Narrow, or with $\sigma=0.11$ pc, labeled Wide. The column density distribution is optimised during the model fitting, taking into account the temperature structure of the model cloud. 

%\begin{figure}
%\sidecaption
%\includegraphics[width=8.5cm]{LOS_dists.pdf}
%\caption{Gaussian LOS density distributions used to create the cloud model. The distributions have been normalised.}
%\label{fig:LOS}
%\end{figure}

\subsection{Radiation field}

\begin{figure}
%\sidecaption
\includegraphics[width=8.5cm]{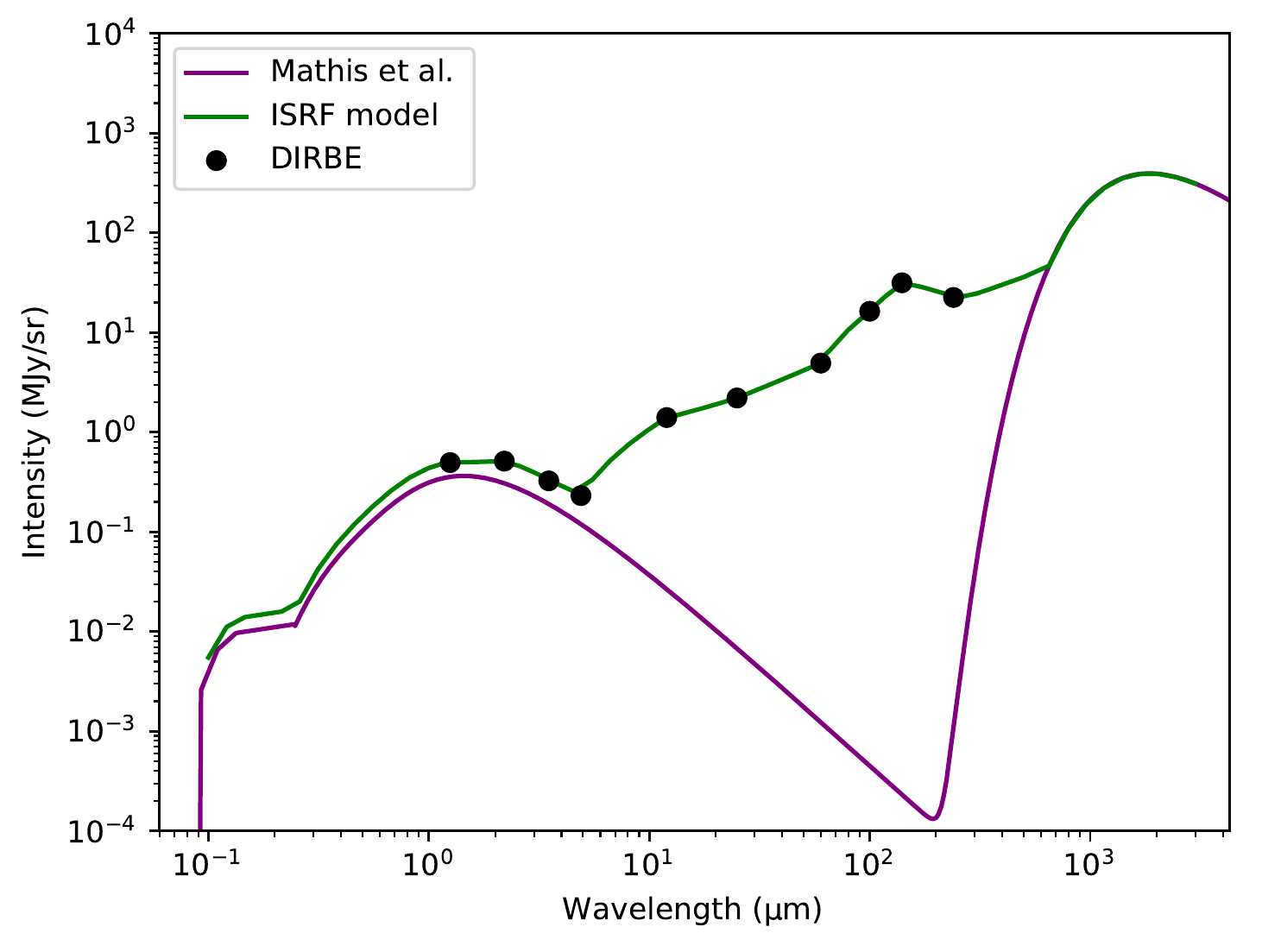}
\caption{Intensity of the radiation field, green line, as a function of wavelength. The purple line shows the \citet{Mathis1983} model. The black dots show the intensity of the DIRBE observations averaged over the HEALPIX map.}
\label{fig:ISRF} 
\end{figure}

\begin{figure}
%\sidecaption
\includegraphics[width=8.5cm]{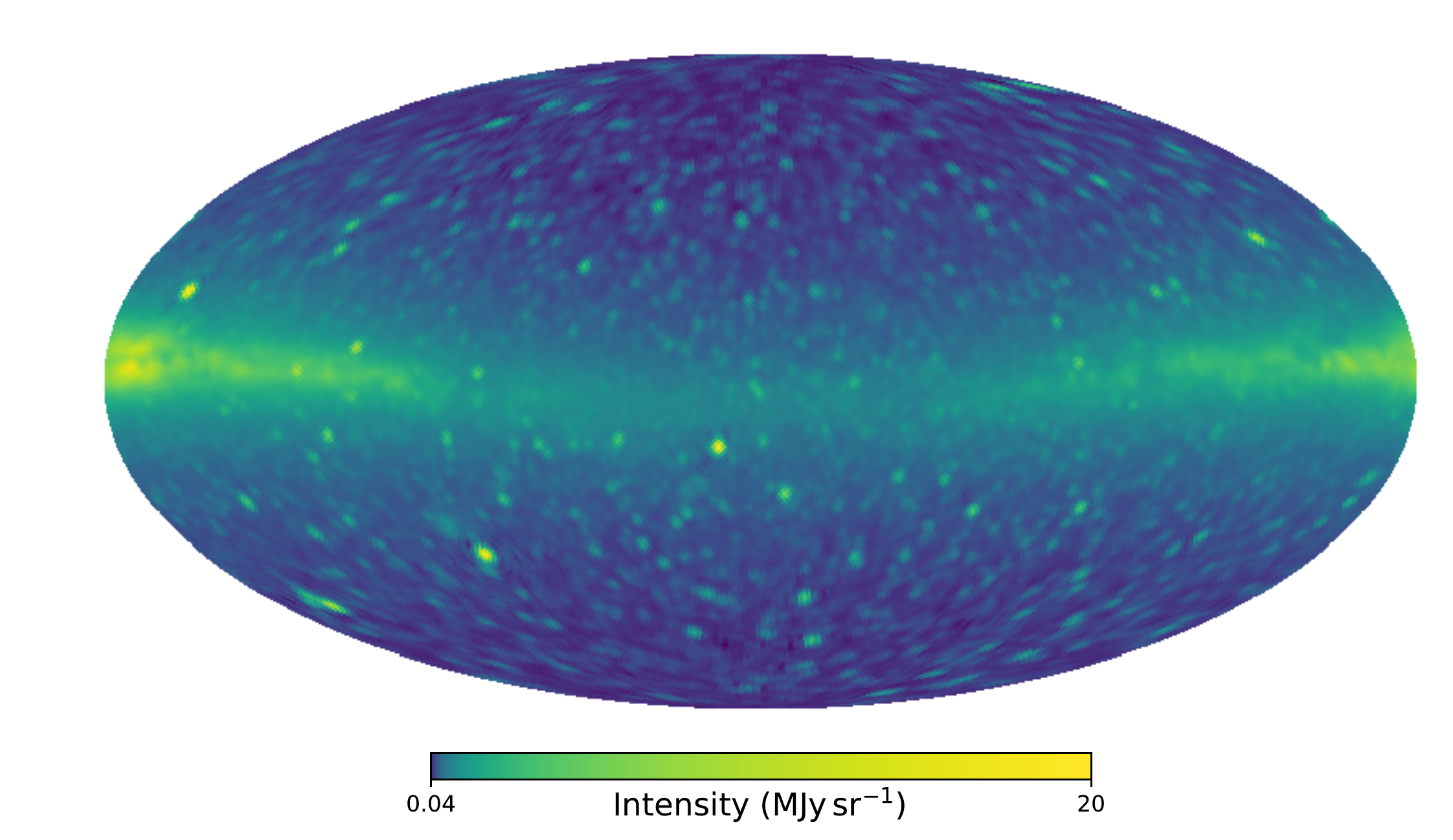}
\caption{Model for the all-sky sky brightness distribution in the J band. The map has been rotated so that the direction towards L1512 is in the centre. The intensity of the model has been truncated at 20 MJy sr$^{-1}$ for plotting and the image has been smoothed by a Gaussian with $\rm FWHM = 2^{\circ}$.}
\label{fig:J_ISRF}
\end{figure}

The radiation field used in our modelling is based on the Diffuse Infrared Background Experiment (DIRBE\footnote{\url{http://lambda.gsfc.nasa.gov/product/cobe/dirbe_exsup.cfm}}) Zodi-Subtracted Mission Average (ZSMA) maps \citep{Hauser_DIRBE} at 1.2 - 240 $\mu$m. Outside the DIRBE wavelengths we follow the \citet{Mathis1983} model, scaling the values below 1 $\mu$m by 1.4 to match the level of the DIRBE data. Similarly, we use linear interpolation in the range from 240 $\mu$m to 650 $\mu$m, to avoid a discontinuity in our model (Fig. \ref{fig:ISRF}). The intensity is distributed over the map pixels following the surface brightness distribution of the DIRBE maps. The spatial distribution at wavelengths shorter than 1 $\mu$m follows the distribution of the DIRBE J band (Fig. \ref{fig:J_ISRF}). At wavelengths longer than 240 $\mu$m, we assume the spatial distribution of the DIRBE 240 $\mu$m band. Compared to the \citet{Mathis1983} model, our radiation field has significantly higher intensity in the range [10, 240] $\mu$m, but the contribution from these wavelengths to the heating of the dust particles is not significant. However, the increased radiation field strength at NIR and MIR wavelengths will affect our light scattering modelling.

\subsection{Radiative transfer}\label{ssect:RT}

To solve the dust emission and the surface brightness of the scattered radiation, we use the Monte Carlo radiative transfer program SOC \citep{Juvela_SOC}. A schematic overview of our modelling is shown in Fig \ref{fig:work_flow}. 

\begin{figure}
%\sidecaption
\includegraphics[width=8.5cm]{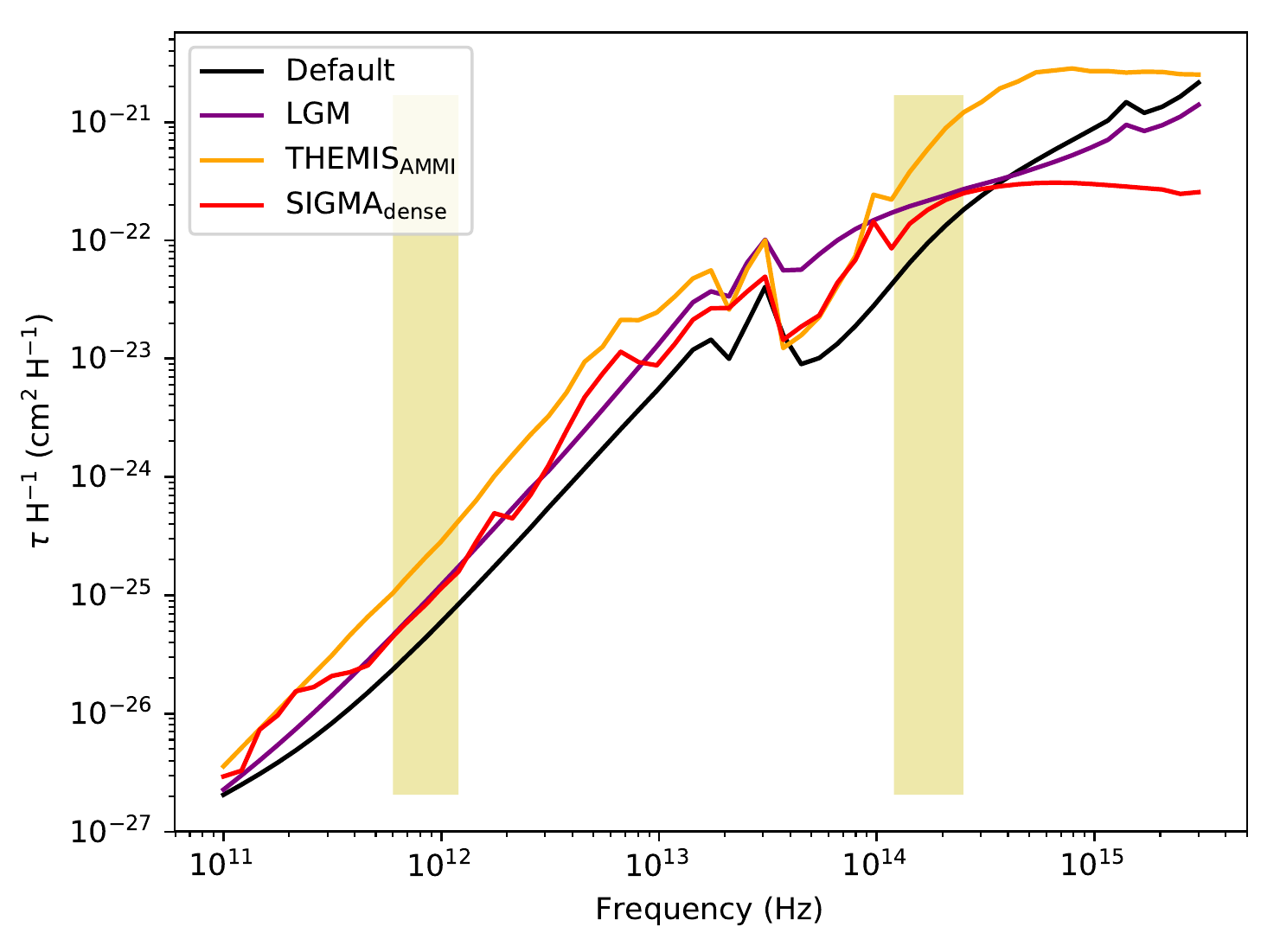}
\caption{Comparison between the optical depth per H as a function of frequency for the Default, LGM, SIGMA, and THEMIS models. The yellow shaded regions show the frequency range from 250 to 500 $\mu$m and from 1.2 to 2.5 $\mu$m.}
\label{fig:Qabs_comp}
\end{figure}

The dust grain properties are defined by the absorption and scattering efficiencies, $\rm Q_{\rm abs}$ and $\rm Q_{\rm sca}$, and a scattering function described using the asymmetry parameters of the Henyey–Greenstein scattering functions, $g$, summed over the individual dust components. The default parameters are taken from the \citet{Compiegne2011} model for the diffuse ISM. Variations in these parameters, as well as three models including dust evolution, are also considered. A more detailed description of the dust models is provided in Appendix \ref{sec:SDM}. The absorption and scattering properties of the dust grains vary significantly between the models, an example of the optical properties of selected models is shown in Fig. \ref{fig:Qabs_comp}.

As a first step, the dust properties, the radiation field, and cloud model are used to compute the dust emission at the \textit{Herschel} SPIRE wavelengths of 250, 350, and 500 $\mu$m. We fit our model to the observed dust emission by an iterative process where the density of the model cloud and the strength of the radiation field are fitted to the observations. For each iteration step, we compare the difference between the observed and simulated surface brightness at 350 $\mu$m, $I_{\rm obs}(350) - I_{\rm sim}(350)$ and, for each map pixel, scale the density of each model cloud cell along the corresponding LOS by 

\begin{equation}
K_\rho = \frac{I_{\rm obs}(350)}{I_{\rm sim}(350)}.
\end{equation}
Similarly, we scale the radiation field by an average scalar factor

\begin{equation}
K_{ISRF}= \biggl< \frac{I_{\rm obs}(250 \, \mu \rm{m})/I_{\rm obs}(500 \, \mu \rm{m})}{I_{\rm sim}(250 \, \mu \rm{m})/I_{\rm sim}(500 \, \mu \rm{m})} \biggr>.
\end{equation}
In the $K_{\rm ISRF}$ computation, we use only the pixels in the central part of the map, within the red circle shown in Fig. \ref{fig:heating}. The scaled radiation field and cloud model are then used to compute a new estimate for the dust emission.

Once the emission fit has converged, we use the final scaled cloud model and radiation field  in a separate radiative transfer calculation to predict the scattered light at NIR wavelengths. 

The observed surface brightness excess is relative to the background sky

\begin{equation} \label{eq:RT}
I_{\nu}^{\Delta} = I_{\rm sca} + I_{\rm bg} \times (e^{-\tau}-1),
\end{equation}
where $I_{\rm sca}$ is the intensity of the scattered light, $I_{\rm bg}$ is the absolute value of the background radiation seen towards the cloud and $\tau$ is the optical depth of the cloud. To compare the observations and simulations a background needs to be subtracted from the simulated maps. For the J and K$_{\rm S}$ bands, the background is estimated from the DIRBE ZSMA maps as an average over the four closest pixels to L1512 and for the H band we estimate the value from the J and K$\rm _S$ band values following the \cite{Mathis1983} model. However, the DIRBE observations include the contribution from point sources which for Eq. \ref{eq:RT} needs to be removed. We estimate this contribution from the 2MASS point source catalogue by integrating the intensity of all point sources that are within the DIRBE pixels (see \citet{Lefevre2014} for details). The integrated intensity values are then subtracted from the average DIRBE values resulting in background levels of 0.059, 0.061, and 0.040 $\rm MJy/sr$ for the J, H, and K$_{\rm S}$ bands, respectively. The pixel-to-pixel variations in the DIRBE observations are at most $\sim 15 \, \%$ for the J and K$_{\rm S}$ bands. Because of the relative proximity of the cloud, $\rm D = 140$ pc, we assume that all of the background radiation is emitted from behind the cloud (i.e. there is no foreground emission).

\section{Results}

In this section we report the main results of our study. In Section \ref{sect:OBS}, we analyse the NIR and \textit{Herschel} observations and describe the results of our modelling in Section \ref{sect:MOD}.

\subsection{Observations} \label{sect:OBS}

\begin{figure*}
\includegraphics[width=17.8cm]{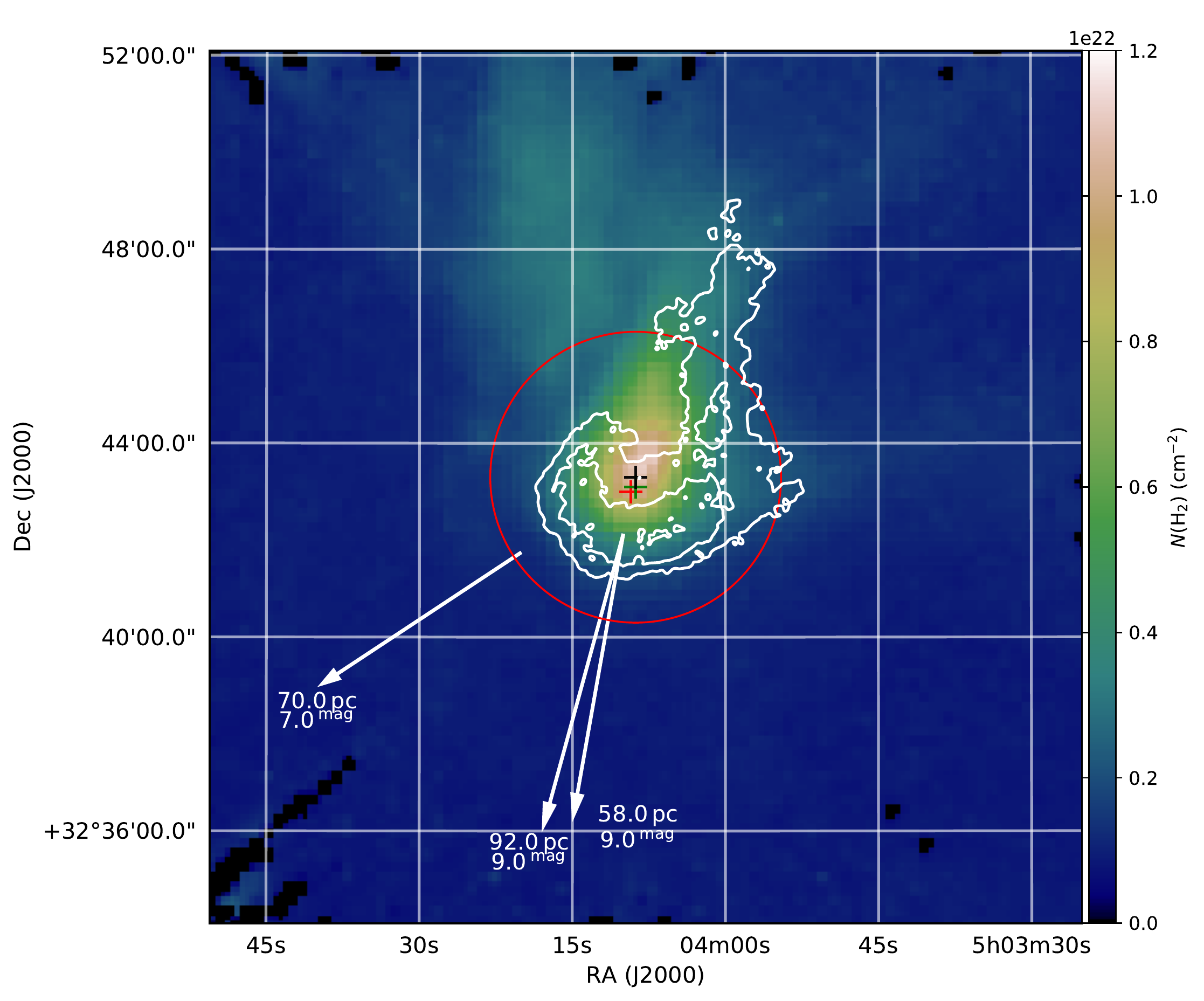}
\caption{Contours of J-band surface brightness on $Herschel$ column density map. The plus signs show the locations of the 250 $\mu$m, 350 $\mu$m, and 500 $\mu$m emission maxima (red, green, and black, respectively). The white arrows indicate the projected direction of the three brightest nearby stars and the numbers next to the arrows indicate the 3D distance to the star and the Gaia G band average magnitude. The red circle shows the area that is used to compute the $K_{\rm ISRF}$ scaling factor.}
\label{fig:heating}
\end{figure*}

Figure \ref{fig:heating} shows the column density map obtained with modified blackbody (MBB) fits of the $Herschel$ SPIRE data. We assume a spectral index $\beta=1.8$, which corresponds to the average in nearby molecular clouds \citep{PlanckXXV2011}. The SPIRE observations were convolved to a common resolution of 40$\arcsec$ and colour corrected using the factors of \citet{Sadavoy2013}. The J band surface brightness has a local minimum at the location of the column density maximum, showing that the J-band scattering has saturated because of high optical depth. The dust emission peaks south of the column density peak in all the SPIRE channels, indicating stronger heating from that direction.

The Gaia data release 2 \citep[DR2,][]{GAIADR2} provides the parallaxes and magnitudes of over 1.6 billion sources, a catalogue that can be used to estimate distances within the Galaxy. To locate stars that could produce additional illumination, we select all stars that are within two degrees of the cloud centre and brighter than 15$^{\rm \, mag}$ in the Gaia G band. We convert the parallaxes to distance estimates assuming $r = 1 / \bar{\omega}$, where $\bar{\omega}$ is given in seconds of arc (see discussion by \citet{Bailer-Jones2015} and \citet{Luri2018} for the accuracy of this method). We find nine stars that are likely to be closer than 100 pc to the cloud and have Gaia G band magnitude brighter than 10 magnitudes. Only three of these stars are on the southern side of the cloud (Fig. \ref{fig:heating}). However, their estimated G band illumination compared to the observed NIR surface brightness is $I_{\rm stars} / I_{\rm SB} = 0.001$, with $I_{\rm stars} = 5.6 \times 10^{-5}$ MJy/sr and assuming an average surface brightness of 0.07 MJy/sr. Thus, compared to the illumination from the ISRF, the contribution from these stars to the illumination of the cloud is minimal.

We derive the J band optical depth from the WIRCam observations using the Near-Infrared Color Excess Revisited  \citep[\textit{NICER;}][]{Lombardi2001} method and assume a standard extinction curve \citep{Cardelli1989}. The submillimetre optical depth was obtained by fitting the spectral energy distribution (SED) of the \textit{Herschel} SPIRE observations with MBB curves, assuming an opacity spectral index $\beta = 1.8$ and a dust opacity of $\kappa_{\nu} = 0.1 (\nu / 1000 \rm GHz)^{\beta}$ cm$^{2} \,$g$^{-1}$ \citep{Beckwith1990}, thus

\begin{equation}
\tau_{250} = \frac{I_{250}}{B_{250}(T)},
\end{equation}
where $I_{250}$ is the fitted 250 $\mu$m intensity, $B_{250}$ is the Planck function at 250 $\mu$m, and $T$ is the colour temperature from the SED fit.
%Our $\tau_{250}$ estimates are likely systematically underestimating the true values because of line-of-sight temperature variations \citep{Shetty2009, Malinen2011, Pagani2015}. 

%We expect that our J band extinction estimates are insensitive to the ratio between total to selective extinction $R_{\rm V}$, since we only use the NIR bands to estimate the optical depth.

\begin{figure*}
%\sidecaption
\includegraphics[width=17.8cm]{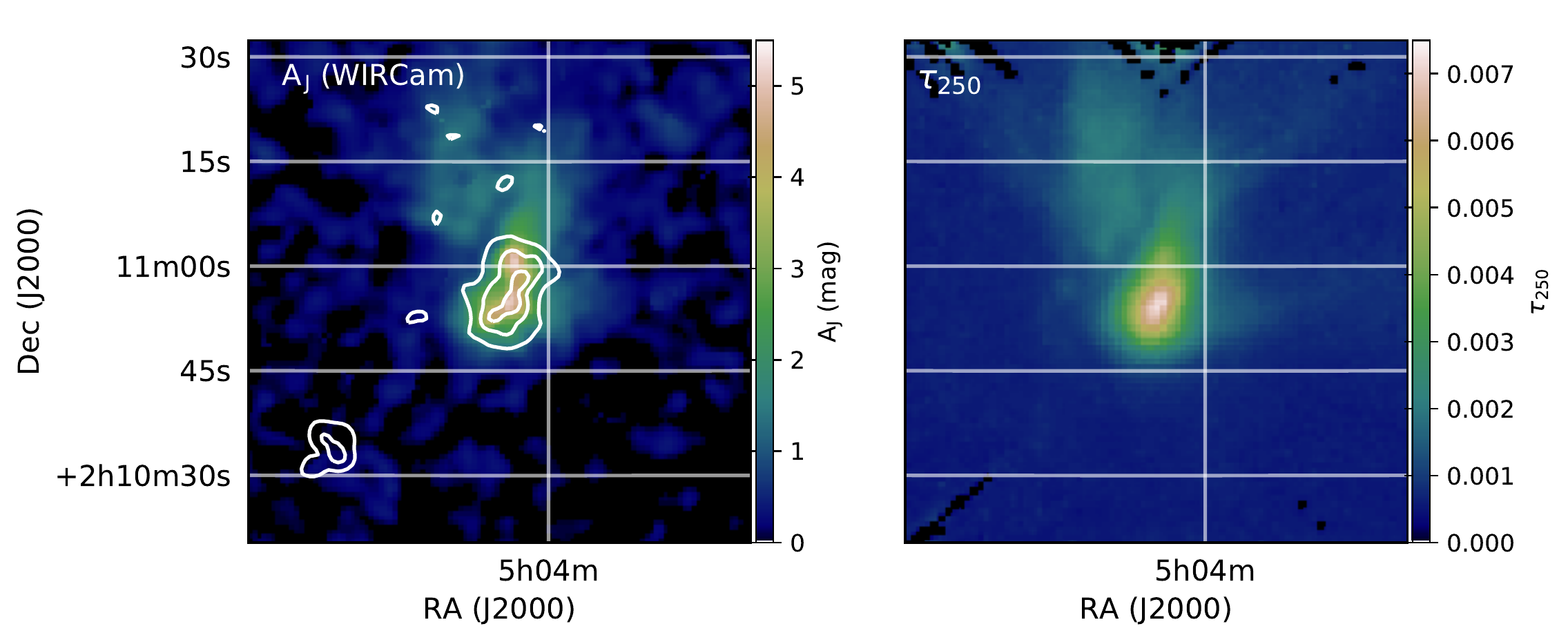}
\caption{J band extinction map and the optical depth at 250 $\mu$m. The white contours in the J band extinction map show the estimated uncertainties of $0.21, 0.3, 0.41$ mag, from lowest to highest contour level}
\label{fig:opacities}
\end{figure*}

\begin{figure}
%\sidecaption
\includegraphics[width=8.5cm]{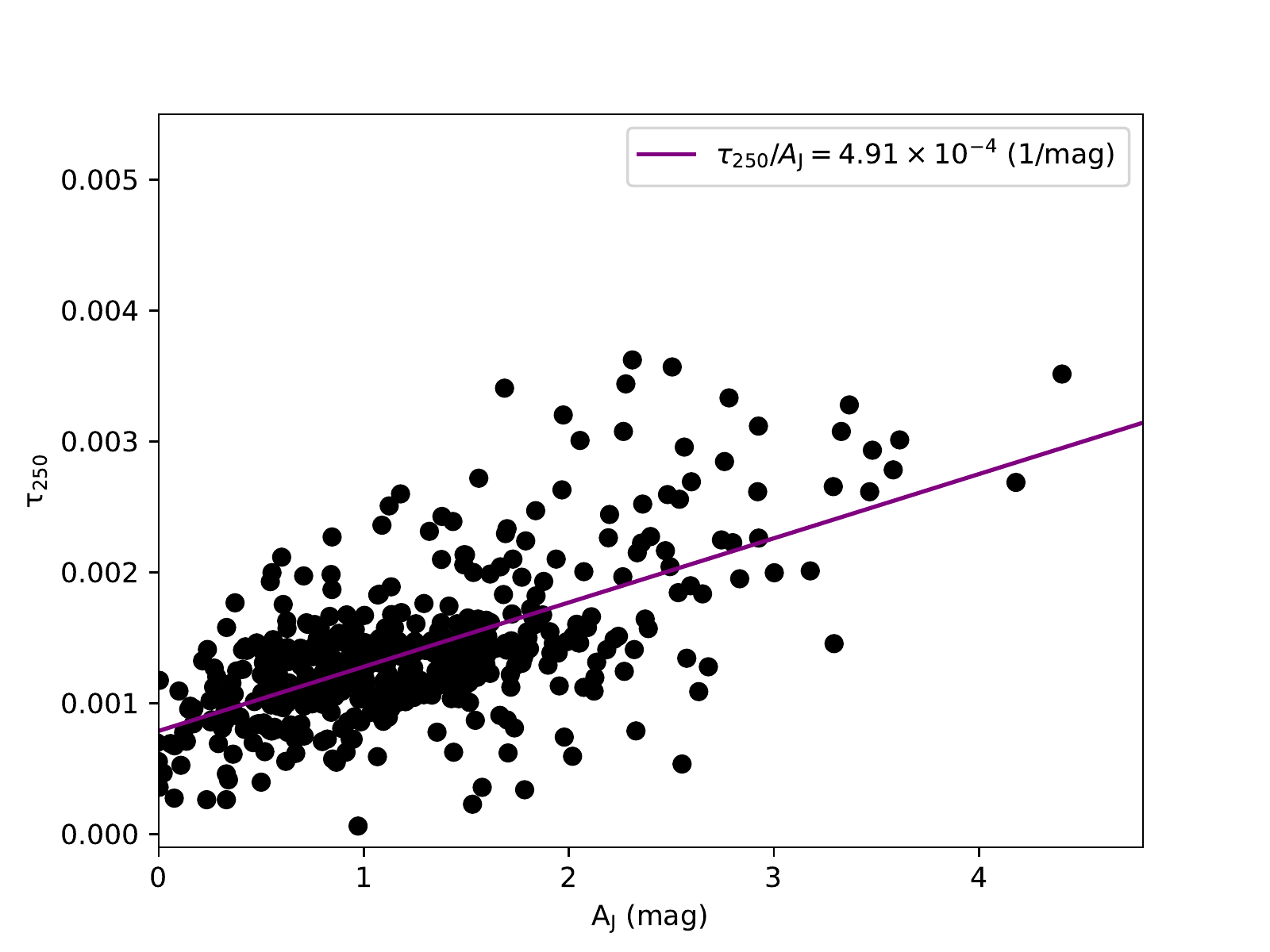}
\caption{Correlation between the optical depth at 250 $\mu$m and the J band extinction. The purple line is a least squares fit to the data and the Pearson correlation coefficient is 0.65 (p $<<$ 0.01).}
\label{fig:ext_corr}
\end{figure}

The maps for J-band extinction and $\tau_{\rm 250}$ are shown in Fig. \ref{fig:opacities} and the correlation between the two quantities in Fig. \ref{fig:ext_corr}. In Fig. \ref{fig:ext_corr}, we have excluded the central region where the extinction estimates are uncertain due to the low number of background stars (the region within the contours in Fig \ref{fig:opacities}). A linear least squares fit gives $\tau_{250} / A_{\rm J} = (4.9 \pm 0.3)  \times 10^{-4}$ $\rm mag^{-1}$. Masking the high values above $\tau_{250} > 0.002$, reduces the ratio to $\tau_{250} / A_{\rm J} = (2.3 \pm 0.3) \times 10^{-4}$ $\rm mag^{-1}$. The ratio of $\tau_{250} / A_{\rm J} = 4.9 \times 10^{-4}$ $\rm mag^{-1}$ is slightly higher than what one would expect for diffuse clouds \citep{Bohlin1978,PlanckXI2014}, but is lower than the average value in dense clumps found in \citet{Juvela2015}.

\subsection{Radiative transfer models} \label{sect:MOD}

In this section we describe the results of our emission and scattering modelling. We use the \citet{Compiegne2011} dust model as the default model, but will also test variations, for example the effects of different size distributions and grain optical properties. We will refer to these as 'COM models'.

Table \ref{tab:models} lists the dust models used, and a more detailed explanations are provided in Appendix \ref{sec:SDM}. We computed for each model the column density, background-subtracted intensity in the J, H, and K$_{\rm S}$ bands, the J band and 250 $\mu$m optical depths ($\tau_{\rm J, em}$ and $\tau_{250}$), and the scaling factor of the radiation field $K_{\rm ISRF}$. The values in Table \ref{tab:models} are averages over $5 \times 5$ map pixels centred on point 1. The temperature of the core is computed as an average value over $5^3$ cells, centred at the core. The values derived from observations are all averages over 5$\times$5 pixels centred on point 1. The core temperature is based on the $\rm N_2H^+$ observations by \citet{Lin2020}. In addition to the models in Table 1, we tested three further modifications to the dust properties. The differences to the Default model were minor, and these results are presented only in Appendix \ref{sec:ADDmodels}.

Our radiative transfer computations do not include stochastically heated grains (SHG), as the emission from these grains is minimal in the SPIRE bands and including the stochastic heating would increase the computations times significantly. We computed stochastic heating only for two test cases, for models that were previously fitted assuming dust at an equilibrium temperature. The results are shown in Appendix \ref{STOKA} (Figs. \ref{fig:stoka_def} and \ref{fig:stoka_the}). Including the stochastic heating decreases the emission in the SPIRE bands by $\sim 10 - 15 \, \%$, because some of the energy is now emitted in the MIR wavelengths. The emission is at 100 $\mu$m 30-45 $\%$ and at 160 $\mu$m ~15$\%$ above the observed values. For the Default model, the morphology of the simulated 100 and 160 $\mu$m maps agrees with the observed maps. For the THEMIS model, the bright rim in both maps is up to 40 MJy$\,$sr$^{-1}$ brighter than observed. For these two cases, we also computed the predicted SEDs for the full wavelength range 3.6-500 $\mu$m (Fig. \ref{fig:SED}).

%The background seen around the cloud is 0.059, 0.061, and 0.040 $\rm MJy/sr$ for the J, H, and K$_{\rm S}$ bands, respectively, and is described in Sect. \ref{ssect:RT}.

%\begin{table*}
\begin{sidewaystable*}
\caption{Summary of the radiative transfer models, including the column densities, NIR intensities, J band optical depth, 250 $\mu$m optical depth, radiation field scaling, and the core temperature.}
\centering
\begin{tabular}{c c c c c c c c c c}
\hline
\hline
& & & & & & & & &\\
Model name & Description  & $N(\rm H_2)$ \tablefootmark{\rm (1)} & J\tablefootmark{\rm (1)} & H\tablefootmark{\rm (1)} & K$\rm _S$\tablefootmark{\rm (1)} & $\tau_{\rm J, em}$\tablefootmark{\rm (1)} & $\tau_{250}$\tablefootmark{\rm (1)} & $K_{\rm ISRF}$ & $\rm T_{core}$\tablefootmark{\rm (2)} \\
 &   & $(\rm cm^{-2})$ & $(\rm MJy / sr)$ & $(\rm MJy / sr)$ & $(\rm MJy / sr)$ & & ($ \times 10^{-3}$) & & $\rm (K)$ \\
\hline
& & & & & & & & &\\
OBS& Values derived from observations & $4.51 \times 10^{21}$  & 0.08 & 0.15 & 0.10 & 1.50 & 2.56 & - & 7.5 \\

& & & & & & & & &\\
\hline
& & & & & & & & &\\
& COM models & & & & & & & &\\
& & & & & & & & &\\

Default & \citet{Compiegne2011}  & $1.59 \times 10^{22}$ & 0.055 & 0.047 & 0.054 & 5.80 & 2.67 & 0.447 & 9.34\\

Scaled2 & Emissivity of $\lambda > 60$ $\mu$m scaled by 2  & $8.31 \times 10^{21}$ & 0.074 & 0.057 & 0.053 & 3.03 & 2.80 & 0.515 & 10.31\\

Scaled4 & Emissivity of $\lambda > 60$ $\mu$m scaled by 4  & $4.26 \times 10^{21}$ & 0.082 & 0.051 & 0.040 & 1.55 & 2.87 & 0.663 & 10.66\\

LG & Included grains up to a size of 5 $\mu$m  & $1.62 \times 10^{22}$ & 0.112 & 0.111 & 0.128 & 4.50 & 2.93 & 0.462 & 10.36\\

LGM & Mass of large grains $\times$2, PAH mass $\times$0.5  & $9.21 \times 10^{21}$ & 0.111 & 0.106 & 0.131 & 5.00 & 3.21 & 0.440 & 10.24\\

& & & & & & & & &\\
\hline
& & & & & & & & &\\
& Models with dust evolution & & & & & & & &\\
& & & & & & & & &\\

SIGMA & Two components, diffuse and dense components\tablefootmark{\rm (3)} & $9.58 \times 10^{21}$ & 0.187 & 0.212 & 0.220 & 3.42 & 3.73 & 0.574 & 10.58\\

THEMIS & Dust model using the THEMIS framework\tablefootmark{\rm (4)} & $1.44 \times 10^{22}$ & 0.082 & 0.127 & 0.173 & 1.05 & 2.74 & 0.471 & 7.61\\

DDust & Two dust components, Default and LG & $1.82 \times 10^{22}$ & 0.079 & 0.085 & 0.108 & 14.38 & 3.53 & 0.745 & 9.31\\
& & & & & & & & &\\
\hline
& & & & & & & & &\\
\end{tabular}
\tablefoot{(1) The column density, background subtracted intensity, and the optical depths of J band and 250 $\mu$m band have been computed as average values over $5 \times 5$ map pixels centred on point 1 (see left panel of Fig. \ref{fig:scat_loc_spec}). \\
(2) The value derived from observations based on the $\rm N_2H^+$ line observations by \citet{Lin2020}. The modelled values are computed as averages over $10^3$ cells centred at the core.\\
(3) The diffuse component is the Default model and the dense component is built with SIGMA (Lefèvre et al, in prep.). \\
(4) For details see \citet{Kohler2015, Ysard2016}}
\label{tab:models}
%\end{table*}
\end{sidewaystable*}

\subsubsection{COM models: dust emission}

\begin{figure*}
%\sidecaption
\includegraphics[width=17.8cm]{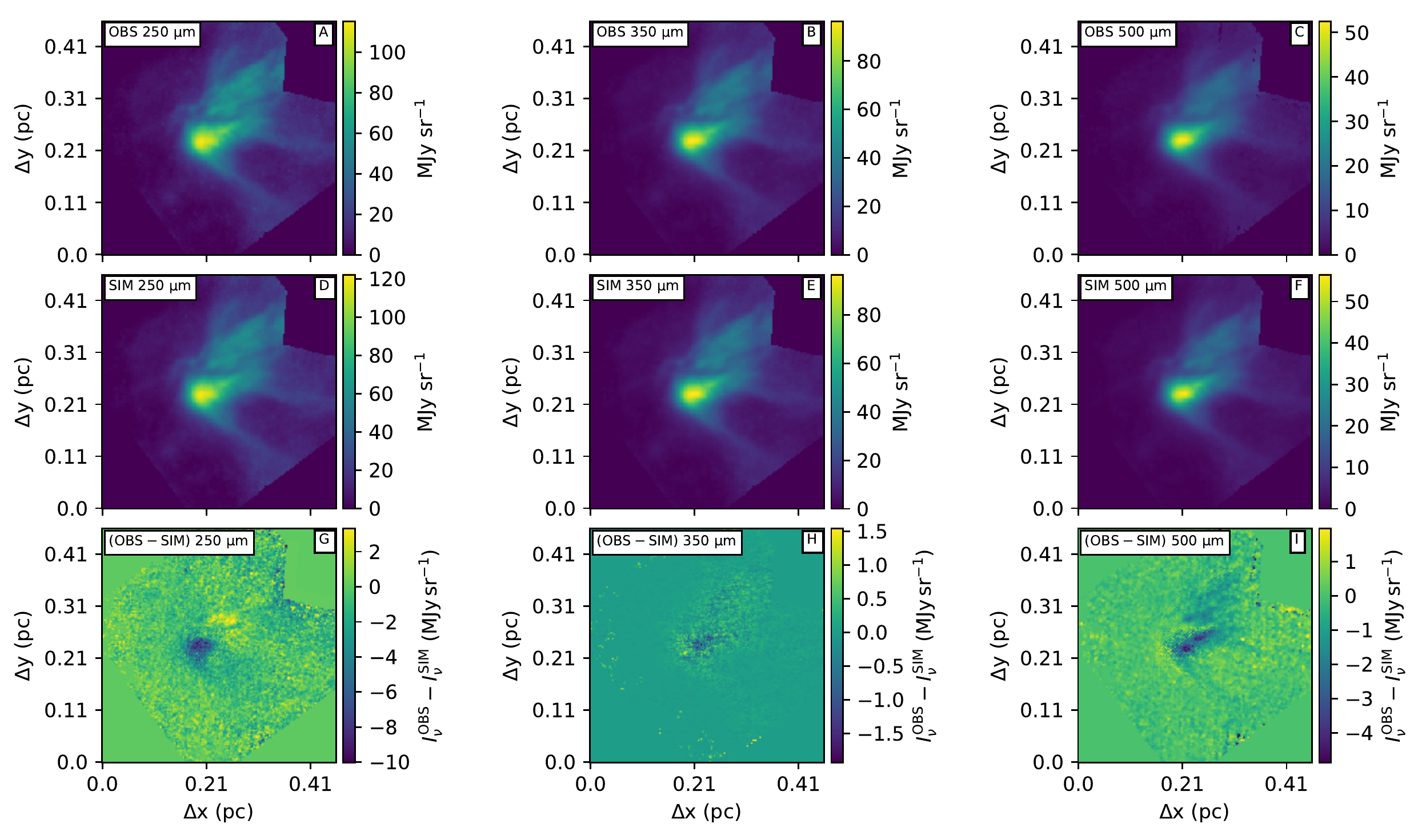}
\caption{Observed (first row) and simulated (second row) emission maps using the Default model at 250, 350, and 500 $\mu$m. The third row shows the difference between the observations and simulations.}
\label{fig:emission_default}
\end{figure*}

Figure \ref{fig:emission_default} compares the dust emission in the Default model with the observations. The fit residuals (observed value minus the model prediction) of the simulated 350 $\mu$m emission are $\pm 1.5  \, \%$. For the 250 and 500 $\mu$m bands, the residuals are on average $\pm 3  \, \%$, but increase up to  $ -10 \, \% $ in the densest region.

The fitted radiation field strength is lower than our reference model for all test cases, but on average, the observations can be fitted with a scaling factor of $K_{\rm ISRF} = 0.5$. The lowest value is $K_{\rm ISRF} \sim 0.44$ for the model LGM, while the highest value is $K_{\rm ISRF} \sim 0.66$ for the Scaled4 model. Thus, increasing the emissivity of the grains, models Scaled2 and Scaled4, increases the required $K_{\rm ISRF}$ to compensate for the grains being colder for a given radiation field. The $\tau_{250}$ values of the models are within higher but within $15 \, \%$ of the values derived from the observations with SED fits. In models with larger grain sizes, such as LG and LGM, the difference increases to $\sim 30 \, \%$.

The column density of the simulated Default case has a maximum of $\sim 2.6 \times 10^{22}$ cm$^{-2}$, whereas the maximum value derived from the \textit{Herschel} observations via MBB analysis is $1.2 \times 10^{22}$ cm$^{-2}$. Changes in the assumed emissivity of the dust grains naturally affect the derived column density, see also the discussion by \citet{Malinen2011} and \citet{Ysard2012} on the uncertainties of the MBB analysis. 

%For example, in the Scaled2 case the resulting column density is much smaller $8.3 \times 10^{21}$ cm$^{-2}$. Increasing the scaling factor of the emissivity will further reduce the estimated column density and for case Scaled4 the column density is a factor of $\sim 7$ lower $\sim 4.2 \times 10^{21}$ cm$^{-2}$ compared to the Default case.

A low kinetic temperature of $\sim8 \, \pm \, 1$ K within the innermost $\sim 0.017$ pc of the core was derived by \citet{Lin2020} using $\rm N_2H^+$ line observations. We computed for each model an average temperature $T_{\rm core}$ over a cube of $5^3$ cells centred on the core. For the Default model $\rm T_{\rm core} = 9.34$ K, which is $\sim 1$ K higher than the $\rm N_2H^+$ estimate. The other models also show temperatures 1-2.5 K above the N$_2$H$^+$ estimate. The models with higher emissivity, Scaled2 and Scaled4, and with larger grains, LG and LGM, are $0.7 - 1.5$ K warmer than the Default model. This is caused by the higher submillimetre emissivity leading to lower column density and lower cloud optical depth at the short wavelengths responsible for dust heating. This, in turn, results in higher dust temperatures in the core. On the other hand, the increased LOS illumination, model Wide, only increases the core temperature marginally to 9.7 K.

\subsubsection{COM models: light scattering} \label{ssect:LS}

We calculated the scattered surface brightness for the J, H, and K$\rm _S$ bands using the density distribution and the radiation field obtained from the fits to the dust emission and by adding net effect of the background $I_{\rm BG}\times (e^{-\tau} -1)$, resulting in simulated surface brightness maps.

\begin{figure*}
%\sidecaption
\includegraphics[width=17.8cm]{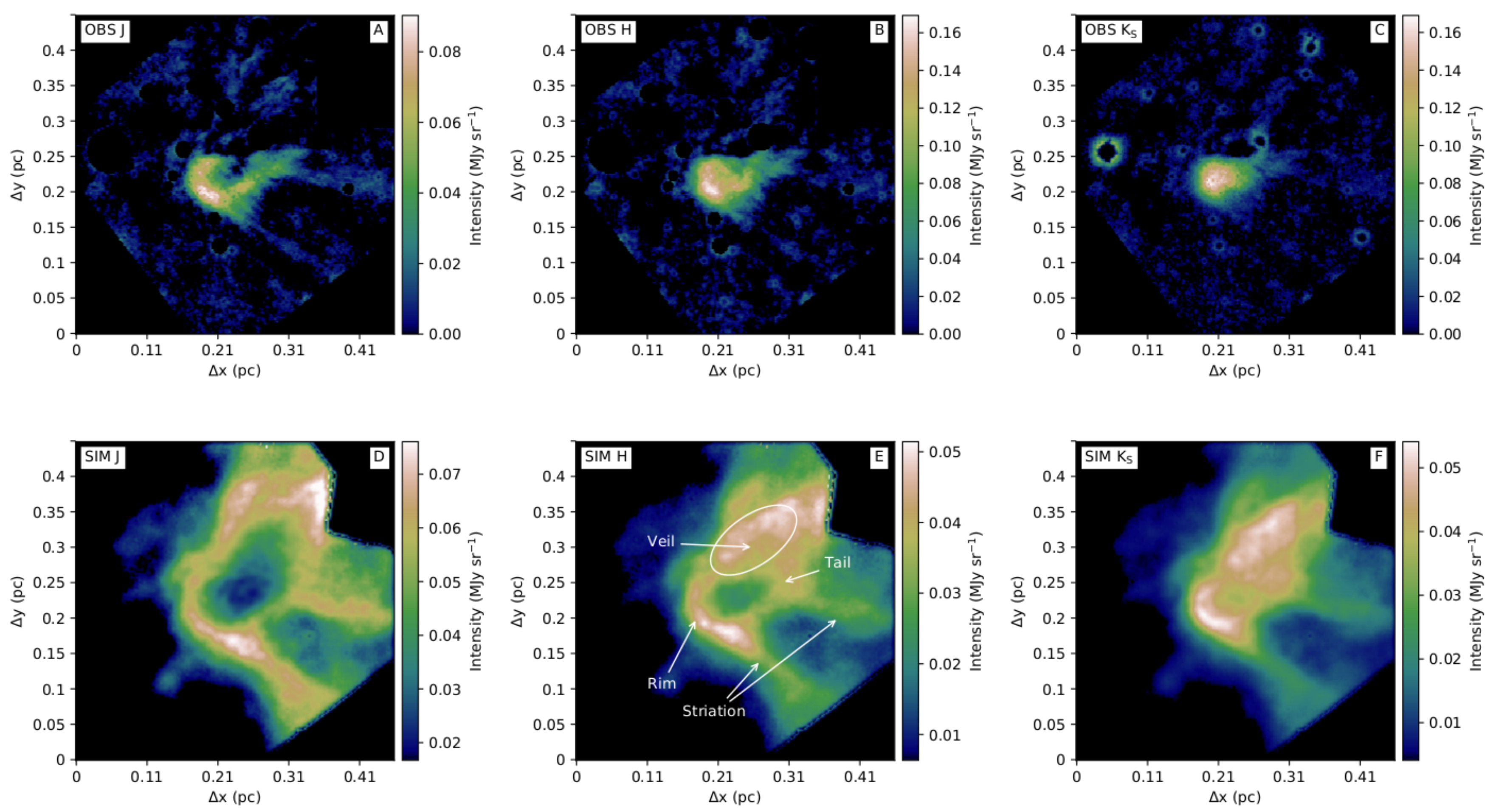}
\caption{Observed (first row) and estimated surface brightness maps from the Default model at the J, H, and K$_{\rm S}$ bands (second row). All maps have been background substracted.}
\label{fig:scat_def_surf}
\end{figure*}

The computed surface brightness maps are shown in Fig. \ref{fig:scat_def_surf} and the individual components of the signal in Fig. \ref{fig:NIR_comparison}. The modelled surface brightnesses are up to a factor of four lower compared to the observed surface brightness. Furthermore, the general morphology of the simulated maps does not match the observed maps. The faint striations (see Fig. \ref{fig:scat_def_surf} E) are more pronounced and the morphology of the bright rim is narrower in the simulated maps. The central part of the cloud has a low intensity, $\sim 0.01$ $\rm MJy \, sr^{-1}$, but the low-surface-brightness area is more extended. In the simulated H and K$_{\rm S}$ band maps, the rim and the striations are clearly visible, unlike in the observations. As in the case of the J band, the morphology of the core does not match the observations. In all simulated maps, the Veil is considerably more prominent than in the observations. Since the Veil cannot be seen in the WIRCam observations but is clearly seen in the \textit{Herschel} observations, it is possibly not connected to the L1512 but is further away on the same line-of-sight.

% On the other hand, the  observed J, H, and K$_{\rm S}$ show less surface brightness on the northern side, thus the Veil might be a more extended diffuse component close to the cloud.

The increase in emissivity, and thus lower column density, of the Scaled2 and Scaled4 models has increased the surface brightness by up to $40 \, \%$ in the J band. In the H and K$\rm _S$ bands the intensity has decreased by $10 \, \%$ in the diffuse regions and increased by up to $30 \, \%$ in the dense regions. This produces a more compact core and the Veil is less prominent. However, the intensities are still a factor of 2-3 below the observed values. For the models that include larger grains, LG and LGM, the surface brightness values are only $\sim 30 \, \% $ lower in the H and K$_{\rm s}$ bands and $\sim 20 \, \%$ higher in the J band than the observed values. However, the morphology of the maps does not match the observations: the core is less compact and the Striations and the Veil more prominent.

The values of $\tau_{\rm J, em}$ are on average three times higher than the NICER estimates (see Table 1). The model Scaled4 is an exception, the optical depth $\tau_{\rm J, em} = 1.55$ being within $5 \, \%$ of the value derived from observations.

\subsubsection{Models with dust evolution}

The Default model is designed for diffuse lines of sight, but in dense cores dust grains are expected to grow due to coagulation and mantle formation. Thus, we also test dust models that take into account the evolution of dust grains. A comparison between the optical properties of the models is shown in Fig. \ref{fig:Qabs_comp}.

We test three cases where the dust evolution is modelled by changing the properties of the dust grains with increasing density. A more detailed explanation of the models is provided in the Appendix \ref{sec:SDM}. As a summary: the model DDust is a combination of models Default and LG, where the relative abundance of the LG component increases with increasing density. The SIGMA model uses the Default model for diffuse gas and a combination of aggregate silicate and carbon grains with ice mantles in the dense regions. Finally, the THEMIS model uses a diffuse component with core-mantle (CM) grains, an intermediate component with core-mantle-mantle grains (CMM), and a dense dust component with amorphous core-mantle-mantle aggregates that have ice mantles (AMMI) \citep{Kohler2015,Ysard2016}. The evolution of the dust grains, with increasing density, will affect all dust parameters. Thus, one cannot identify a single parameter or parameters that change, as not only the physical parameters, emissivity absorption and scattering properties, and size distribution, but also chemical properties evolve due to formation of aggregate grains and formation of (ice)mantles. Furthermore, the changes in the chemical composition are also expected to effect the optical properties, thus the affect of dust evolution is intricate.

In the fits of the dust emission, the 250 and 500 $\mu$m residuals in the central region are smaller for SIGMA (Fig. \ref{fig:EM_3}) than for the default dust model (e.g. Fig. \ref{fig:emission_default}). The residuals are slightly larger in the case of the THEMIS model (Fig. \ref{fig:EM_3}) and in the case of the DDust (Fig. \ref{fig:EM_3}) model are similar to the Default model. 

Compared to the Default case, the SIGMA and DDust models have higher $K_{\rm ISRF}$ factor, 0.67 and 0.75 respectively, while the $K_{\rm ISRF}$ factor of the THEMIS model is 0.47. As with the models that include larger grains, LG and LGM, $\tau_{250}$ of the SIGMA, THEMIS, and DDust models are higher by $\sim 30 \, \%$ compared to the Default model. The NIR optical depths are $\tau_{\rm J}$=3.4, 1.05, and 14.4 for the SIGMA, THEMIS, and DDust models, respectively. The corresponding core temperatures read from the 3D models are 10.6 K, 7.6 K, and 9.3 K.  Thus, in spite of the lowest $\tau_{\rm J}$ value (even below the NICER measurement), THEMIS results in the lowest dust temperatures, some 0.5 K below the estimate derived from N$_2$H$^+$ line observations.

%The J band optical depth of the SIGMA model is similar to the model Scaled2 with $\tau_{\rm J}= 3.4$ and the optical depth of the THEMIS model is close to the value derived using the \citet{Cardelli1989} extinction curve with $\tau_{\rm J}= 1.05$. The optical depth of the DDust model is considerably higher, $\tau_{\rm J}= 14.4$, but the core temperatures of the model is similar to the Default model, 9.31 K. The temperature of the SIGMA model is $\sim 1$ K higher than the Default model with $\rm T_{\rm core} = 10.58$, and the THEMIS model is colder, 7.61 K, which is $\sim 0.5$ K colder than the value derived from $\rm N_2H^+$ line observations.

\begin{figure*}
%\sidecaption
\includegraphics[width=17.8cm]{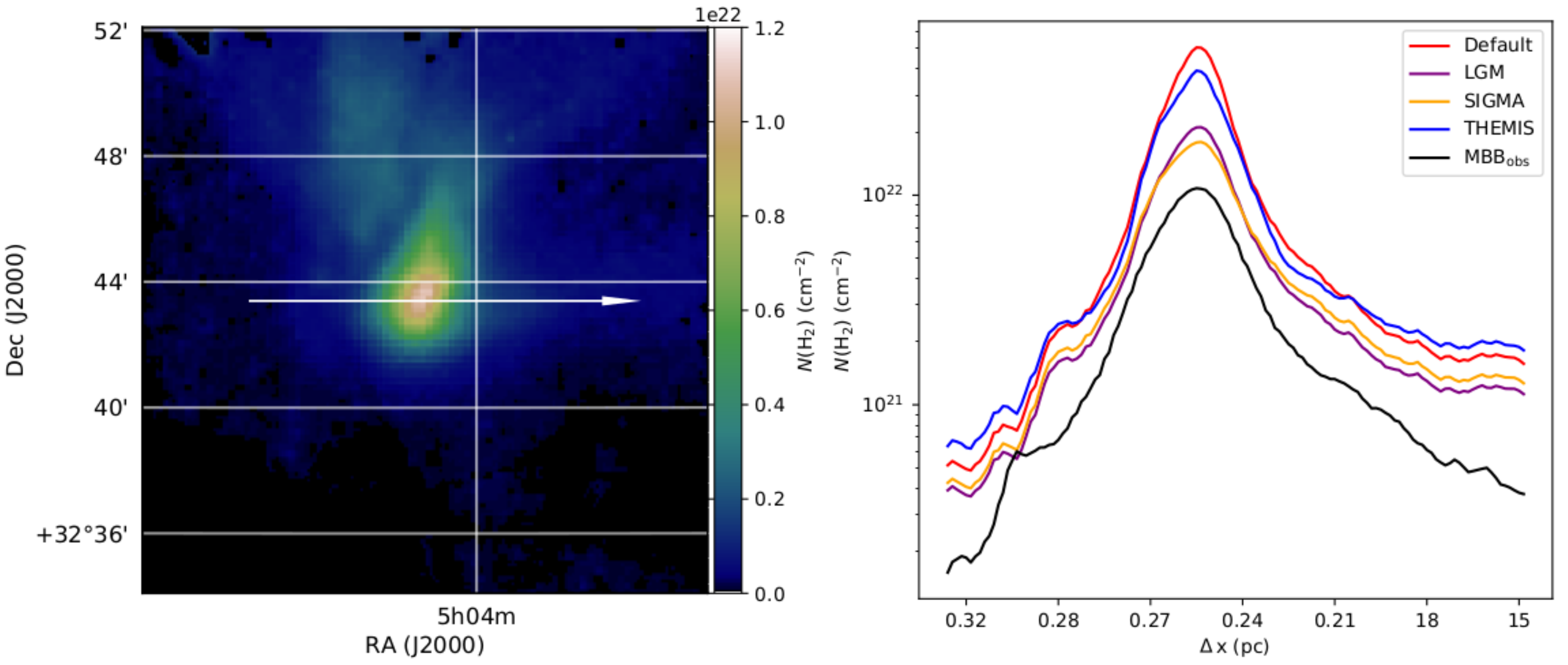}
\caption{Column density of H$_2$ derived from \textit{Herschel} observations and the comparison between column density profiles for selected models. The profiles have been computed as average values over $3 \times 3$ pixels along the arrow in the left panel.}
\label{fig:col_cut}
\end{figure*}

The THEMIS, and DDust models produce column densities close to the Default model with $N(\rm H_2) = 1.4 \times 10^{22}$ cm$^{-2}$ and $1.8 \times 10^{22}$ cm$^{-2}$, respectively. The column density of the SIGMA model is lower, with $N(\rm H_2) = 9.61 \times 10^{21}$ cm$^{-2}$. The model column density profiles (Fig. \ref{fig:col_cut}) agree with the profile derived from the \textit{Herschel} observations via MBB fits, but are narrower than the FWHM=2.0$\arcmin$ estimated from the \textit{Herschel} estimate. The profile of the Default model, FWHM $=$ $1.4 \, \arcmin$,  is close to the modelled N$_2$H$^+$ profile with FWHM $=$ $1.32 \, \arcmin$ \citep{Lin2020}.

THEMIS produces two to three times higher NIR surface brightness than the Default model (Fig. \ref{fig:NIR_sim_3}), but the SED shape and the morphology still do not match the observations. Although the central region of the cloud in H and K$_{\rm S}$ bands has surface brightness values within 0.02 $\rm MJy \, sr^{-1}$ of the observed value, the more diffuse regions are brighter than observed regions. The striations are clearly visible in the THEMIS maps (H and K$_{\rm S}$ bands), whereas in observations no extended surface brightness is seen. The model SIGMA has the highest surface brightness of all our test cases, the intensities in the diffuse regions are in excess of 0.1 $\rm MJy \, sr^{-1}$ for the H and K$_{\rm S}$ bands and even for the J band the values are in range [0.04,0.12] MJy sr$^{-1}$. The morphology is close to the observed morphology, with a compact core in the H and K$_{\rm S}$ bands and the surface brightness of the Veil is lower. The model DDust shows a clear dip in the core, even in the K$_{\rm S}$ band map, a factor of 2 to 3 lower surface brightness compared to the observations and the model clearly overestimates the surface brightness of the diffuse regions.

\subsection{NIR Spectra}

\begin{figure*}
%\sidecaption
\includegraphics[width=17.8cm]{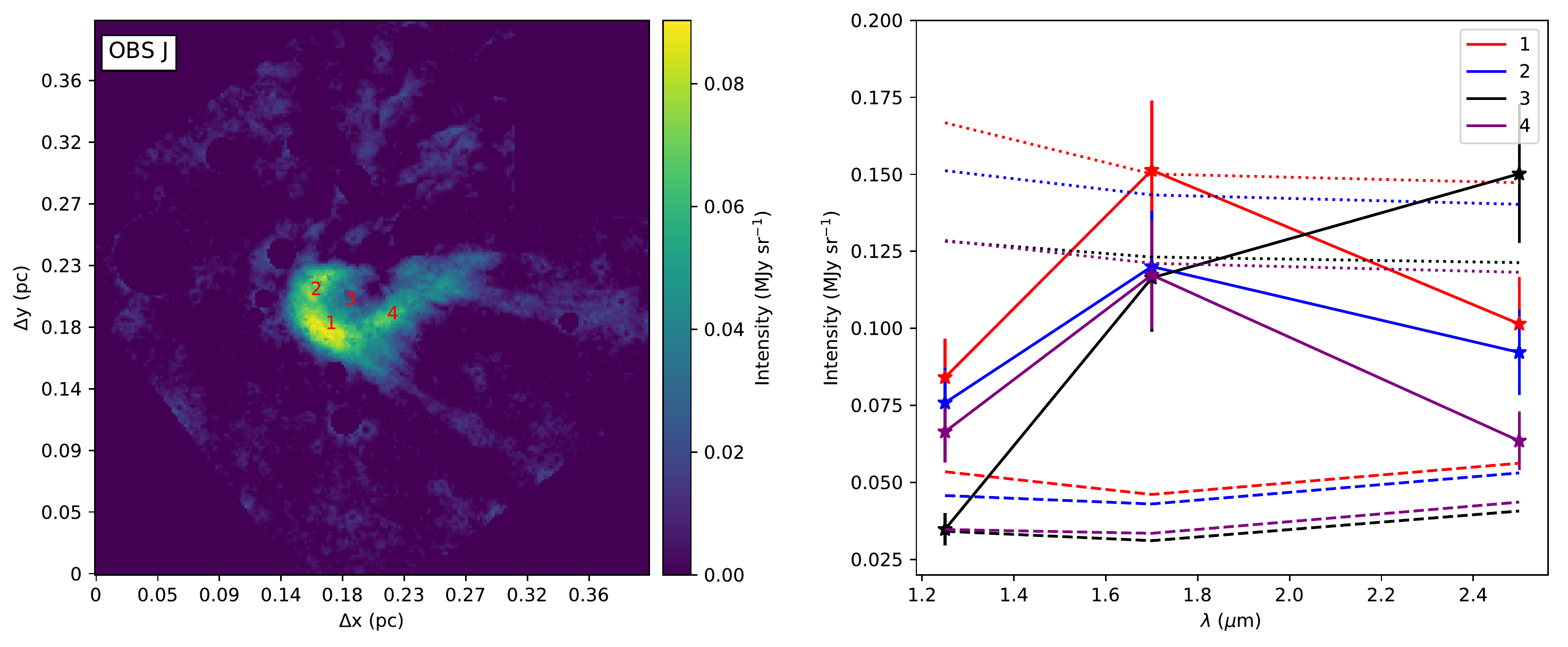}
\caption{Observed map of the J band scattered light (left panel), the red numbers from one to four indicate the locations from where we have extracted NIR spectra, shown in the panel on the right. The solid lines show the observed spectra and the dashed lines show the spectra derived from the Default model. The dotted lines show the simulated spectra derived from the Default model, but the intensity of the scattered light has been multiplied by a factor of three before background subtraction.}
\label{fig:scat_loc_spec}
\end{figure*}

Figure \ref{fig:scat_loc_spec} shows NIR spectra for observations and the Default model, for the four positions indicated in the figure. Point 1 corresponds to the brightest part of the J band map and the three other points trace density variations across the densest part of the cloud. In addition to the low intensities, the general shapes of the simulated spectra do not match the observations, the relative brightness of the K$_{\rm S}$ and J bands being too high. A factor of three increase in the scattered light would increase the net surface brightness by a larger factor but would not improve the match with the SED shape. However, the observed H-band value has somewhat higher uncertainty because of the lack of a direct background sky brightness measurements at that wavelength.

\begin{figure*}
\sidecaption
\includegraphics[width=12.1cm]{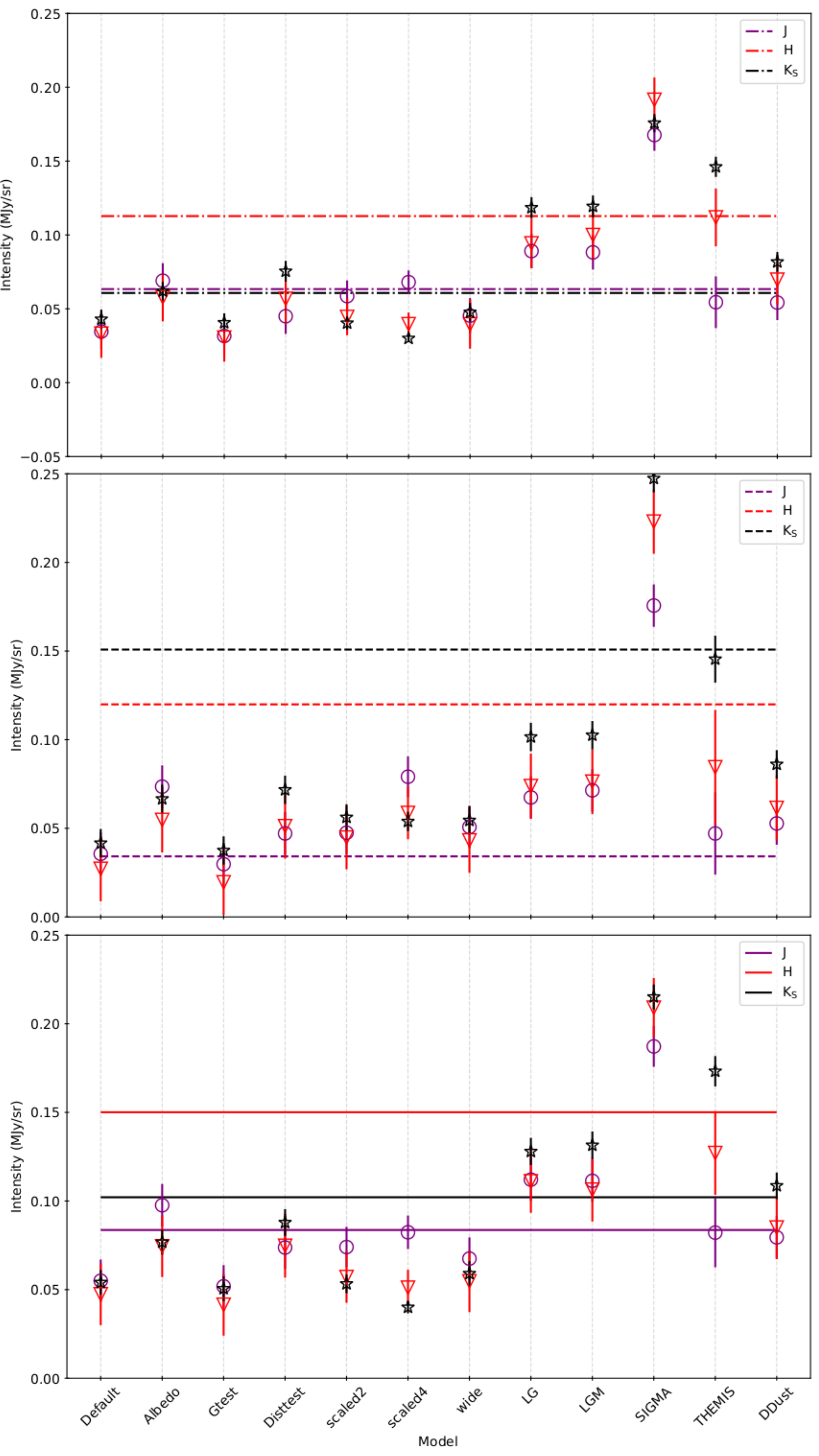}
\caption{J, H, and K$_{\rm S}$ band intensities from observations (horizontal lines) and different models (symbols) for the map point 1 (lower panel), point 3 (middle panel), and point 4 (upper panel). The colours correspond to the J (purple), H (red), and K$_{\rm S}$ (black) bands. All intensity values have been background subtracted. We assume a $20 \, \%$ uncertainty in background sky estimates for the J and K$_{\rm S}$ bands and an uncertainty of $30 \, \%$ for the H band.}
\label{fig:all_spec_com}
\end{figure*}

Figure \ref{fig:all_spec_com} shows the NIR spectra for all test cases and points 1, 3, and 4. Point 2 is similar to point 4 and its spectra are shown in the appendix (Fig. \ref{fig:all_spec_P2}). 

The error bars assume an uncertainty in the background sky brightness that is $30 \, \%$ in the H and $20 \, \%$ in the other bands. The Default-case simulated intensities are a factor two to three lower than the observed values. The models LG (Default model with grains up to 5 $\mu$m in size) and LGM (As LG but relative amount of large grains increased) produce higher intensities, but the intensity of the H band is $\sim 30 \, \%$ lower than the observed value while the J and K$\rm _S$ bands are 30 to  $40 \, \%$ brighter. The SIGMA model (two dust components, Default for diffuse and a component derived with SIGMA for dense LOS) is clearly overestimating the intensity of the scattered light in all three channels. In point 1, the THEMIS model (three dust component based on the THEMIS framework) is closest to the observations, but the intensity of the K$\rm _S$ band is 0.05 $\rm MJy \, sr^{-1}$ too bright. However, in the point 3, the intensity and the shape of the spectra of the THEMIS model is within $10 \, \%$ of the observed intensity. In general the spectra from point 3 is more easily reproduced and the models LG, LGM, and DDust all produce approximately the correct SED shape but do not match the level of intensity. Like in the point 1, the SIGMA model is overestimating the intensity, but the shape of the spectra is correct.

For point 4, the COM models tend to underestimate the NIR intensities, with the exception of the LG and LGM models that overestimate the J and K$_{\rm S}$ bands, but underestimate the H band. However, models Scaled2 and Albedo give very similar results.

%Furthermore, the shape of the spectra do not match the observed spectra. The DDust model is similarly underestimating the H band intensity and the shape of the spectra does not match the observed, but the intensity of the J band spectra is well reproduced. The SIGMA model is still clearly overestimating the intensity of all three bands, but the shape of the spectra is close to the observed spectra with the H band being the strongest. The model predictions of the THEMIS model for the J and H band are in agreement with the observed values, but the intensity of the K$_{\rm S}$ band is over estimated by a factor of $\sim 2$.

%,height=23cm,keepaspectratio

%\begin{figure*}
%\sidecaption
%\includegraphics[width=17.8cm]{all_spectra_P1_error.pdf}
%\caption{J, H, and K$_{\rm S}$ band intensities from observations (horizontal lines) and different models (symbols) for the map position 1. The colours correspond to the J (purple), H (red), and K$_{\rm S}$ (black) bands. All intensity values have been background subtracted. We assume a $20 \, \%$ uncertainty in background sky estimates for the J and K$_{\rm S}$ bands and an uncertainty of $30 \, \%$ for the H band.}
%\label{fig:all_spec_P1}
%\end{figure*}

The increased FIR/submillimetre emissivity in the cases Scaled2 (emissivity for $\lambda > 60$ scaled by a factor of 2) and Scaled4 (as Scaled 2 but scaled by a factor of 4) decreases the column density, thus, the higher intensity of the H band compared to the K$_{\rm S}$ band can be understood as a saturation effect. However, increasing the emissivity further does not improve the match as the column density becomes too low and the intensity of the scattered light is reduced. The increased emissivity of the grains decreases the J band optical depth, which for model Scaled4 is close to the NICER estimate, but the morphology of the surface brightness maps and the shape of the NIR spectra are further away from the observations. Thus, emissivity alone can not reconcile observations and simulations.

%\begin{figure*}
%\sidecaption
%\includegraphics[width=17.8cm]{all_spectra_P3_error.pdf}
%\caption{As Fig. \ref{fig:all_spec_P1}, but for the map position 3.}
%\label{fig:all_spec_P3}
%\end{figure*}

%\begin{figure*}
%\sidecaption
%\includegraphics[width=17.8cm]{all_spectra_P4_error.pdf}
%\caption{As Fig. \ref{fig:all_spec_P1} but for point 4.}
%\label{fig:all_spec_P4}
%\end{figure*}

\section{Discussion}

Based on our modelling of the LDN1512 observations, its thermal dust emission can be fitted with many different assumptions on dust properties. However, we can not simultaneously fit both the dust emission and NIR scattered light with the COM models. The Default model was designed for high-latitude diffuse lines of sight, while the L1512 cloud has a dense central core. The disparity between the observations and our radiative transfer models based on the Default model is a clear indication for the need of evolution of the dust grains and we thus tested additional models, DDust, SIGMA, and THEMIS. We are able to reproduce the observed intensity of dust emission and scattered light at NIR wavelengths with the THEMIS model. However, the morphology of the surface brightness maps shows considerable deviation from the observations. In this section we discuss these results in more detail.

\subsection{Uncertainties of the radiation field}

The ISRF used in our simulations is based on the DIRBE observations and the shape of the \citet{Mathis1977} model. We did not test changes in the anisotropy of the external field. However, based on Gaia point sources, the contribution from the nearby stars is minimal.

%\begin{figure*}
%\sidecaption
%\includegraphics[width=17.8cm]{spectra_local_BG.png}
%\caption{Comparisson of observed NIR spectra (solid lines) with the simulated Default case but assuming a higher (dashed lines) or a lower (dotted lines) local background. The local background values have been scaled by $\pm 25\%, 20\%$, and $5\%$ for the J, H, and K$\rm _S$ bands, respectively. The simulated intensities have been multiplied by a factor of three.}
%\label{fig:local_BG}
%\end{figure*}

Based on the intensity variation between the DIRBE pixels, the background uncertainty is of the order of $\sim 20 \, \%$, but for the H band, for which there are no direct DIRBE measurements, we have assumed a value of $30 \, \%$. With the exception of the model SIGMA (model with two dust components, Default dust for diffuse regions and dust created with SIGMA for dense LOS), the spectra derived from both points 1 and 3 are systematically underestimating the intensity of the H band. For the THEMIS model (model with three dust components based on the THEMIS framework), the intensity of the J and K$_{\rm S}$ bands from point 3 are comparable with the observed values. However, the H band intensity from both points is still significantly below the observed values. For point 4, the J and H band intensities of the THEMIS model are comparable with the observed values, but the K$_{\rm S}$ band is over-estimated by a factor of 2. Furthermore, the shape of the spectra of the model SIGMA agree with the observed spectra, although the intensity is clearly overestimated. Thus, the discrepancy between the observed and simulated values can be explained to a certain degree with uncertainties in the background sky estimate, but since the shape of the simulated spectra also depends on the optical depth of the cloud, uncertainties in the optical depth of the model or variations in the relative abundances of the dust components will also affect the estimated intensity.

\subsection{NIR extinction}

In the models fitted to dust emission observations, the NIR extinction is typically three times higher than the direct NICER estimate of $\tau_{\rm J}$. In addition to NICER estimates using the \citet{Cardelli1989} extinction curve, we also calculated estimates for the extinction curves of the respective dust models. The values are computed as averages over a $5 \times 5$ pixel region around point 1 (the resolution of the optical depth maps was set to 40$\arcsec$). For the Default model $\tau_{\rm J, ext} = 1.32$, which is lower compared to the value derived using the \citet{Cardelli1989} extinction curve, $\tau_{\rm J, Card.} = 1.50$. For the models LG (model with grains up to 5 $\mu$m in size), LGM (as the LG model but the relative amount of large grains increased), and SIGMA, the optical depths are $\tau_{\rm J, ext} = 2.19 - 3.0$. For the model THEMIS the $\tau_{\rm J, ext} = 1.42$, which is within $\sim 10 \, \%$ of the \citet{Cardelli1989} estimate.

\begin{table}
\caption{Comparison of $\tau_{\rm J}$ values derived from the dust-emission models and from the observations of background stars. The values are computed as averages over a $5 \times 5$ region centred on point 1.}
\centering
\begin{tabular}{c c c c}
\hline
\hline
& & & \\
Model & $\tau_{\rm J,em}$ & $\tau_{\rm J,ext}$ & $\tau_{\rm J,em} / \tau_{\rm J,ext}$ \\
\hline
& & & \\
Default  &  5.80 & 1.32 & 4.39\\
LG       &  4.50 & 2.90 & 1.55\\
LGM      &  5.00 & 3.00 & 1.66 \\
SIGMA    &  3.42 & 2.19 & 1.56 \\
THEMIS   &  1.05 & 1.42 & 0.74 \\
\hline
& & & \\
\end{tabular}
\label{tab:taus}
\end{table}

The ratio for the Default model is $\tau_{\rm J,em} / \tau_{\rm J,ext}  \sim 4.39$, while it is lower, $\tau_{\rm J,em} / \tau_{\rm J,ext} = 1.55 - 1.66$, for the models LG and LGM (Table \ref{tab:taus}). For the models with dust evolution, the ratios are smaller, 0.74 and 1.56 for THEMIS and SIGMA, respectively. Thus the models with larger grains, or some form of dust evolution, are more consistent between the sub-millimetre emission and NIR extinction.

\subsection{Uncertainties of cloud models derived from FIR fits}

We derived the cloud density distribution and strength of the external radiation field by fitting the observations of FIR dust emission. The estimated column density is sensitive to the dust temperature which in turn is sensitive to the strength of the radiation field. In this section, we quantify the dependences between the radiation field, the dust temperatures, and the optical depth of the cloud.

\begin{figure}
%\sidecaption
\includegraphics[width=8.5cm]{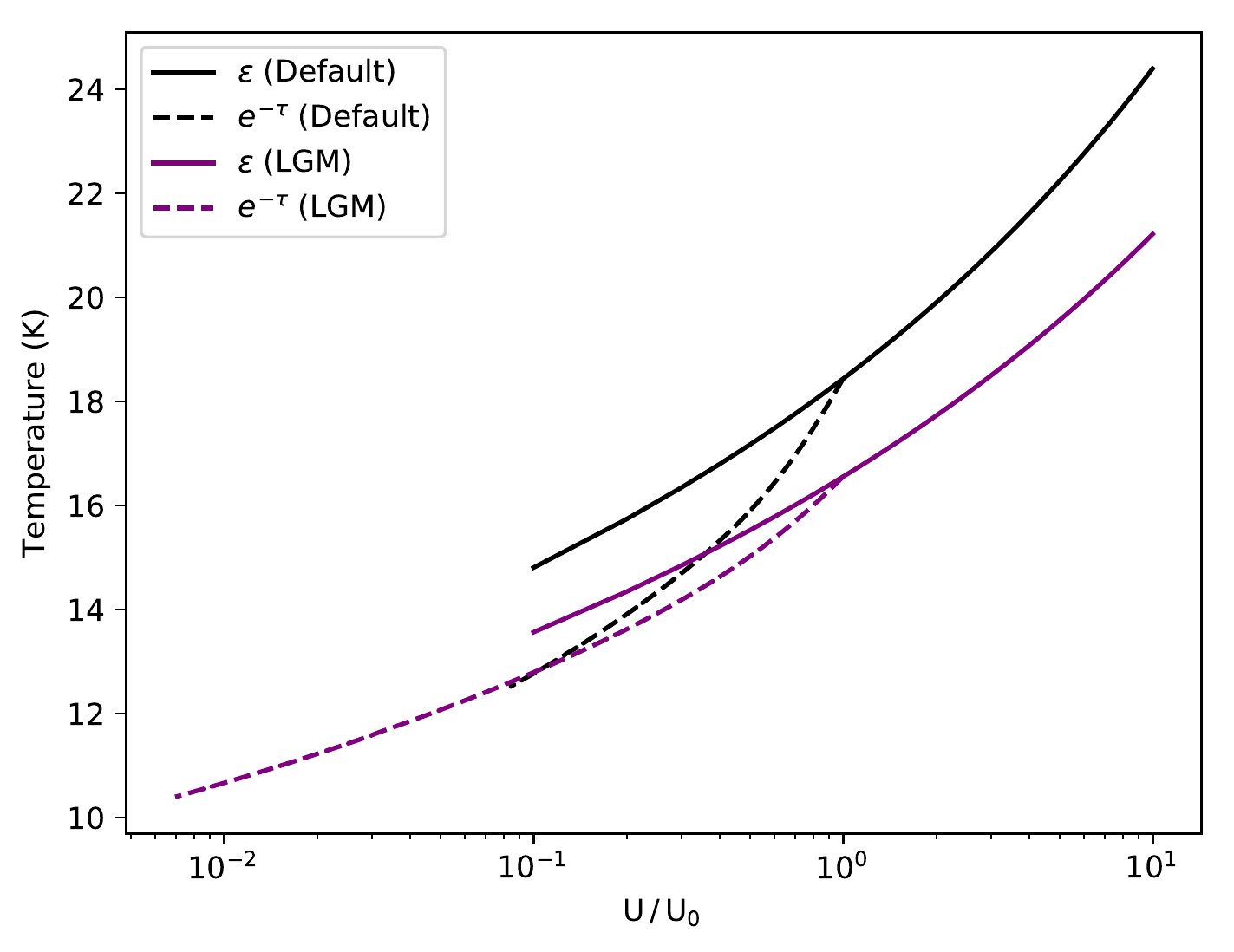}
\caption{Comparison between the dust temperature and the energy density of the radiation field for the Default and LGM models. The strength of the radiation field has been normalised by the \citet{Mathis1983} model.}
\label{fig:T_ISRF}
\end{figure}

The effects of scaling and attenuation of the radiation field on the dust temperature can be solved from the equilibrium equation

\begin{equation}
\int_{0}^{\infty} Q_{\rm abs}(\nu) \times I_{\rm ISRF} \, d\nu = \int_{0}^{\infty} Q_{\rm abs}(\nu) \times B_{\nu}(T) \, d\nu,
\end{equation}
where Q($\nu$) are the dust absorption efficiencies, $I_{\rm ISRF}$ is the intensity of the radiation field, and $B_{\nu}(T)$ is a black-body function at temperature $T$. Figure \ref{fig:T_ISRF} shows the dependence of the temperature of the Default and LGM dust grains on the energy density of the radiation field. Results are calculated for grains at an equilibrium temperature and using the \citet{Mathis1983} radiation field with a linear scaling factor $\epsilon$ and an attenuation by by $e^{-\tau}$. We use values in the range of [0.1,10] for both $\epsilon$ and $\tau / \tau_{\rm J}$.

Compared to the Default model, the LGM model has more large grains and to reach similar temperature, a higher radiation field is required. Because of the higher emissivity, larger grains will have a lower temperature for a given radiation field and a lower column density is required to reach similar temperature as with the Default model.

%Thus, at NIR wavelengths and assuming scattering is not saturated, the lower column density will decrease the intensity of the scattered light. However, the models can have different relative abundances of dust components, for example carbon and silicate grains, which will complicate the above picture. 

\begin{figure*}
\sidecaption
\includegraphics[width=12.1cm]{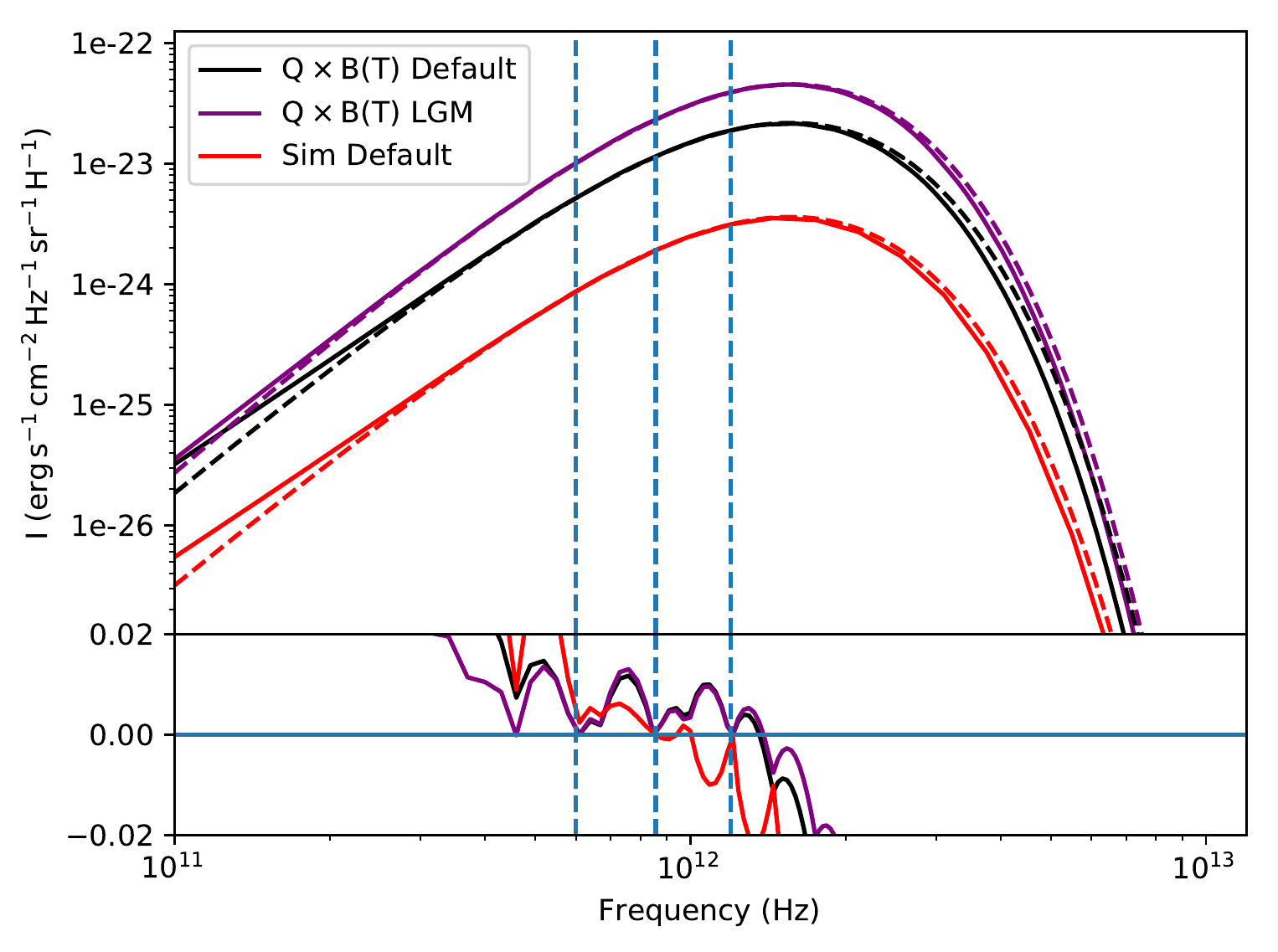}
\caption{Comparison of dust emission spectra and their MBB fits. The black and purple lines show the emission spectra for Default and LGM models with grains at 15 K temperature. The dashed lines show MBB spectra fitted to the 250, 350, and 500 $\mu$m points. The resulting fit parameters are $T_{\rm c} = 16.48$, $\beta=1.62$ and $T_{\rm c} = 15.85$, $\beta=1.79$, for the Default and LGM models, respectively. The red curve (scaled by a factor of $1.0 \times 10^{-30}$ to fit the figure) is from a simulated Default model map pixel where the fitted temperature is similar to that of the black curve. The blue vertical lines show the 250, 350, and 500 $\mu$m frequencies. The relative differences between the black-bodies and the MBB fits near the 250, 350, and 500 $\mu$m points are shown in the plot below the emission spectra curves.}
\label{fig:KAPPA}
\end{figure*}

Dust emission spectra are typically analysed as modified black-body radiation. However, the spectra will always deviate from this model, because of temperature variations in the sources and because the spectral index of dust opacity is not constant over the examined wavelength range. Figure \ref{fig:KAPPA} shows emission spectra obtained by multiplying the absorption cross sections of the Default and LGM models with a $\rm T=15$ K black-body function. The figure also shows the MBB fits using three points at 250, 350, and 500 $\mu$m and keeping both $T$ and $\beta$ as free parameters. The fitted temperatures $T_{\rm c} = 16.48$ K and $T_{\rm c} = 15.85$ K for the Default model and LGM model, respectively, are higher than the true temperature of 15 K. The fitted $\beta$ values are $\beta = 1.62$ and $\beta = 1.80$, for the Default and LGM models respectively, lower than the $\beta$ values of the dust models in the $250-350$ and $350-500$ $\mu$m intervals, $\beta \sim 1.88$ and $\beta \sim 1.94$ for the Default and LGM models, respectively. Thus, an observer relying on MBB fits would underestimate the column density (see also Fig. \ref{fig:col_cut}). Similar results of dust models containing only bare astrosilicates \citep{Draine1984, Draine2001, Draine2007} failing to reproduce the SED of the observed dust emission, have been discussed by \citet{Fanciullo2015, PlanckXXII2015, PlanckXXIX2016}.

The errors in column density and radiation field will also affect the predictions of NIR scattering. However, the low intensity of our modelled NIR surface brightness is not caused by the uncertainties in the column density, but is related to the scattering efficiency of the dust models, or in the case of the evolved dust models (e.g. SIGMA and THEMIS) the chosen relative abundances of the different grain populations. Furthermore, MBB results are sensitive to temperature variations that can lead to a severe underestimation of dust column densities \citep{Shetty2009, Juvela2012TB}. For example, \citet{Pagani2015}, who studied the cloud LDN 183, showed that Herschel observations can not be used to set strong constraints on the amount of very cold dust. Additional methods may be needed, such as observations of molecular lines and NIR/MIR extinction.

%On the other hand, in our simulations we are also estimating the strength of the radiation field, thus, if we are overestimating the temperature we will underestimate the strength of the radiation field as we do not need as much heating to reach sufficient dust temperature. Underestimating the radiation field will then cause us to overestimate the column density. 

\begin{figure}
%\sidecaption
\includegraphics[width=8.5cm]{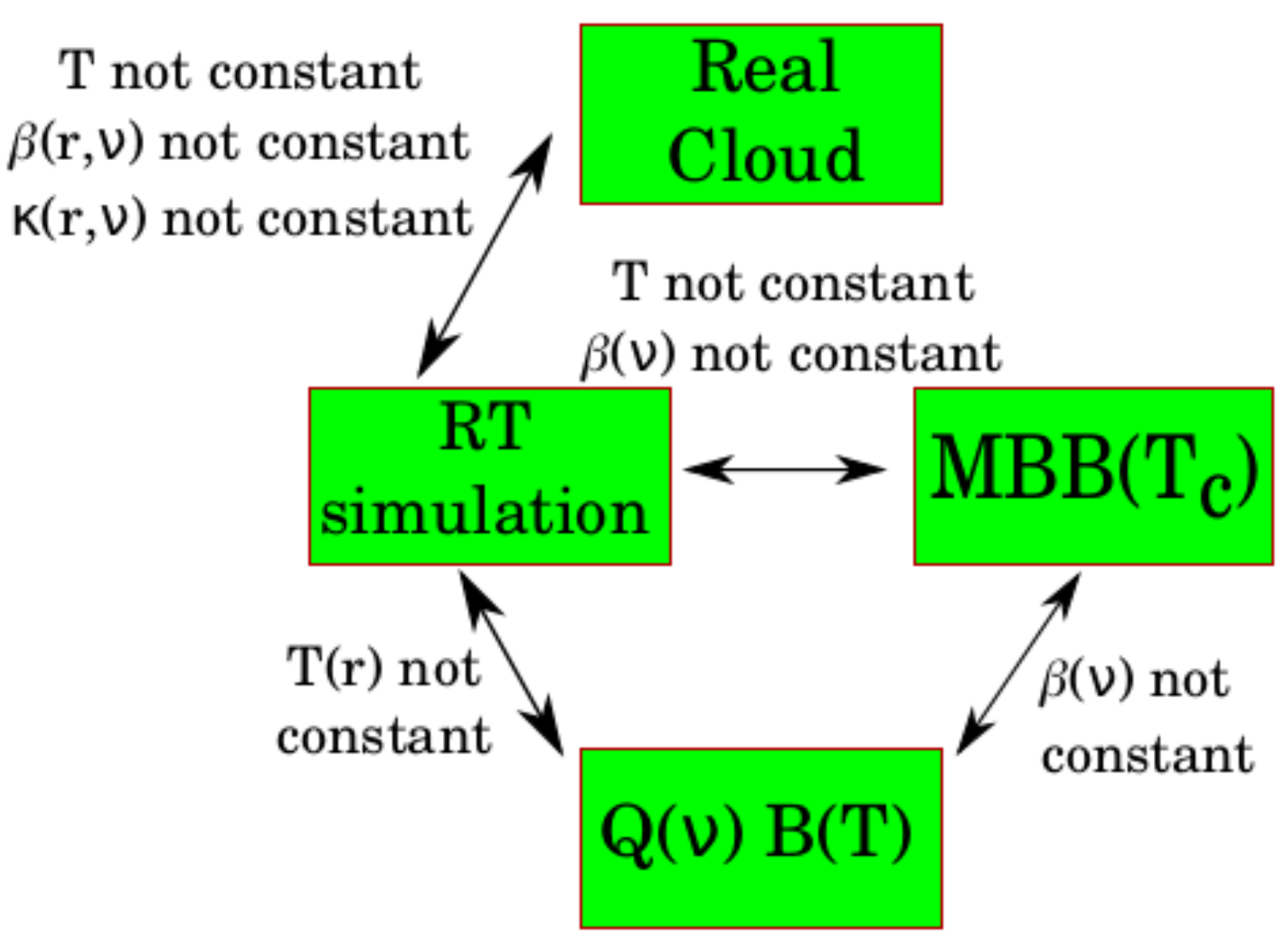}
\caption{Schematic overview between the assumptions that are made during a MBB fit and their comparison to the simulated spectra and a real cloud.}
\label{fig:MBB_problem}
\end{figure}

The column densities based on the Default model vary within a factor of 2, while the variation in the radiation field strengths is within a factor of $\sim 1.5$. The column density differences are not much larger between the SIGMA, THEMIS, and DDust models, but the the differences in the radiation field strength reach a factor of two. However, the relationships between the true $Q(\nu)B(T_{\rm d})$ of the dust, the emission predicted by RT models, and the results of MBB fits are complex (Fig. \ref{fig:MBB_problem}). The emission from the dust grains might not follow any MBB curve, $T_{\rm d}$ and $T_{\rm c}$ differ and the fitted spectral index $\beta$ differs from the opacity spectral index of the dust grains. These problems are acerbated when temperatures also vary within the beam.

\begin{figure*}
%\sidecaption
\includegraphics[width=17.8cm]{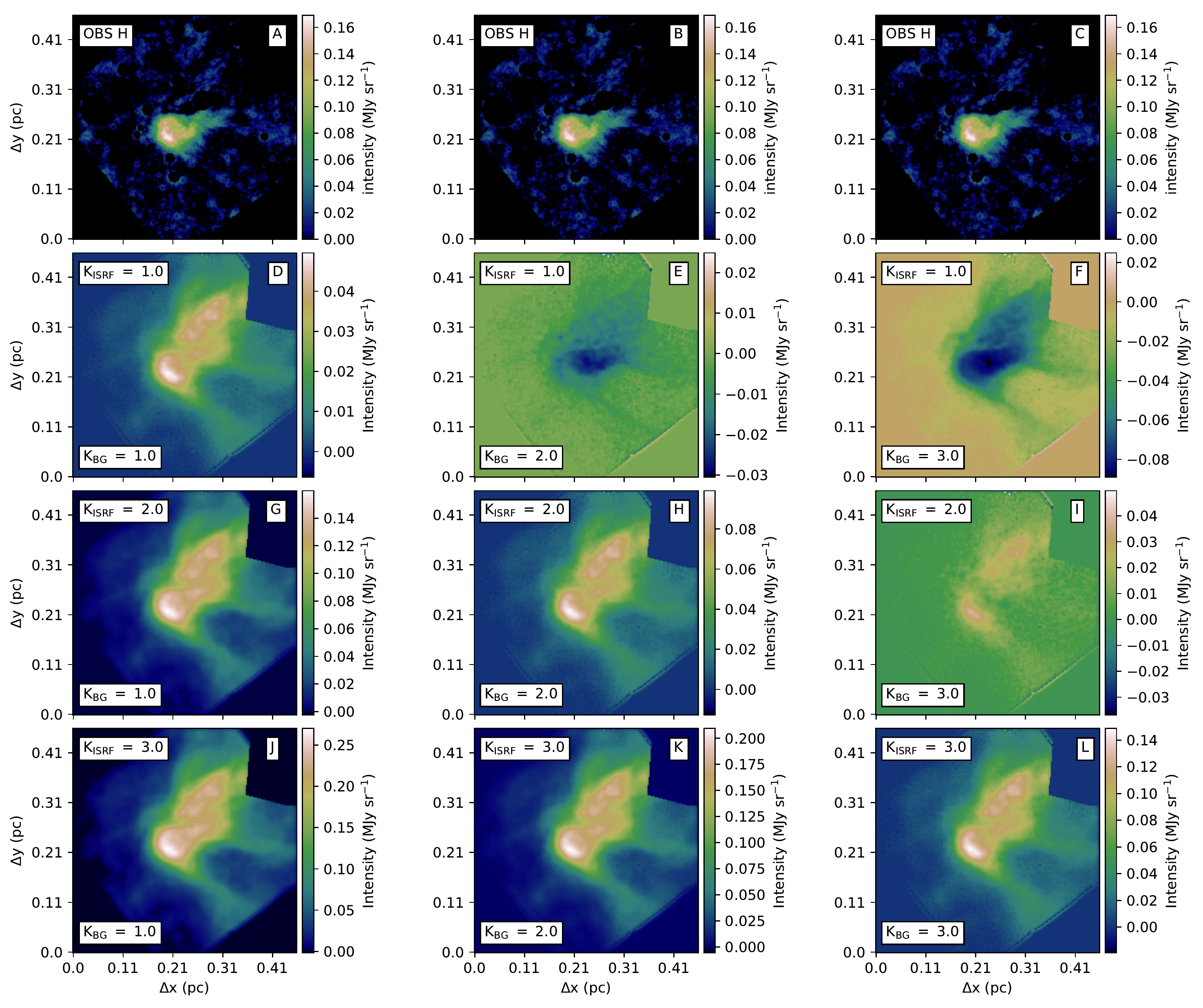}
\caption{Simulated H band surface brightness maps for the Default model with density scaled by a factor of 0.6 during the scattering computations and with different assumptions on the strength of the radiation field and on the sky brightness behind the cloud. Shown on the first row is the observed H band surface brightness map. Shown on the rows 2 to 4 are the simulated H band maps, for which the strength of the radiation field has been scaled with a factor between 1 and 3, and the intensity of the background has been scaled between factors 1 and 3.0.}
\label{fig:ISRF_test2}
\end{figure*}

We further quantify the discussion above by computing H band surface brightness maps using the Default model, with the assumption that we have underestimated the strength of the radiation field and overestimated the column density. Our aim is to see if by modifying these parameters, the Default model can be made to agree with the observations. We use a radiation field that is 1, 2, or 3 times and a column density that is 1, 0.6, or 0.3 times the value previously obtained for the Default model. Another component affecting the shape and strength of the NIR scattered light is the sky background behind the cloud, we assume that this is 1.0, 2.0, or 3.0 times the value derived from the DIRBE observations. The resulting H band surface brightness maps are shown in Fig. \ref{fig:ISRF_test2} and in Figs. \ref{fig:ISRF_test1}, \ref{fig:ISRF_test3}. Increasing the strength of the radiation field will increase the surface brightness excess, but increasing the background sky brightness will decrease the surface brightness. The two values can be used to fix the level of the modelled surface brightness, but they are not enough to reproduce the observed surface brightness morphology (Fig. \ref{fig:ISRF_test1}). For that, also the cloud column density needs to be decreased.

Increasing the strength of the radiation field by a factor of 2 and simultaneously decreasing the column density by $\sim 30  \, \%$, (e.g. Fig. \ref{fig:ISRF_test2} panel G), the intensity of the simulated H band surface brightness is within $15 \, \%$ of the observed one. This can also be achieved by increasing the strength of the radiation field and the assumed background sky brightness by a factor of 3 and decreasing the column density by $\sim 30  \, \%$ (e.g. Fig. \ref{fig:ISRF_test2} panel L)). In both cases, the morphology of the surface brightness map in the central part is comparable with the observations, with a bright rim towards the south and decreasing surface brightness towards the core. However, the Veil (Fig. \ref{fig:scat_def_surf}) is still prominent. Decreasing the density of the cloud further (Fig. \ref{fig:ISRF_test3}) the central region of the cloud becomes excessively compact, although the correct intensity can be reached by scaling the radiation field or the assumed intensity of the background sky brightness (e.g. Fig. \ref{fig:ISRF_test2} panels G, K, and L).

Although suitable parameters can be found to correct the H band intensity of the Default model, the required changes are substantial. Thus, it is evident that dust evolution needs to be taken into account, since the models SIGMA and THEMIS produce results closer to the observations without the need to fine tune other parameters.

\subsection{Effects of grain properties} \label{EGP}

%In the test case where we increased the maximum grain size of the Default model to 5 $\mu$m, case LG, the resulting column density was lower by $25 \, \%$ and the surface brightness of the scattered light was higher in the central region of the cloud (see Fig. \ref{fig:NIR_sim_2} third row). However, the morphology of the scattered light did not match the observed morphology. Similarly, the $\tau_{\rm J,ext}$ values, derived using the extinction curve of the models are higher by a factor of 2 ($\tau_{\rm J,ext} \sim 3.1$) compared to the value derived using a model without large grains ($\tau_{\rm J,ext} = 1.40$).

\citet{Launhardt2013} showed that L1512 is a cold ($\rm T \approx 12$ K) core with high column density ($N(\rm H_2) \approx \sim 2.0 \times 10^{22} cm^{-2}$) and in the study by \citet{Lippok2013}, the envelope of the core was better fitted by assuming a higher temperature compared to the core, which is consistent with the absence of internal heating. However, to explain the observed molecular abundances, \citet{Lippok2013} had to increase the hydrogen density derived by \citet{Launhardt2013} and the abundances of all modelled chemical species between a factor of 2 to 3. In their chemical models \citet{Lippok2013} used a slightly coagulated grain model for the whole cloud with a coagulation time of $10^5$ yr at a gas density of $10^5$ cm$^{-3}$ \citep{OH1994}. \citet{Lippok2013} discuss that by using a model of non-coagulated grains, they would have increased the hydrogen density by a factor of $\sim 2.5$.

Larger grains lead to stronger scattering as discussed by \citet{Steinacker2010} and \citet{Ysard2018}, but the increase in grain size also increases the absorption coefficient. On the other hand, more complex or 'fluffy' grains can have a larger surface area, and have higher scattering efficiency with respect to their absorption efficiency \citep{Lefevre2016}. This can be seen with the dense dust components of the models SIGMA and DDust (Fig. \ref{fig:NIR_sim_3}), where the dense regions of the model have higher intensity and their morphology agrees better with the observations. 

Our Default model consists of only PAH, silicate, and carbon grains, however, in dense regions of the ISM, gas phase freeze-out can create mantles on the grains. The initial studies trying to explain the detected coreshine required a high fraction of dust mass in large grains \citep{Pagani2010, Steinacker2010, Andersen2013}, with maximum grain size of the order of $\sim 1 \mu$m. However, to reach such large grain size through coagulation is difficult with respect to the time-scales, cloud densities, and turbulence \citep{Steinacker2014B}. On the other hand, as discussed by \citet{Ysard2016}, a dust model including mantle formation and low-level coagulation is able to reproduce the observed cloudshine levels without the need of very large grains. Our results using the THEMIS model show that the model can fit the observed emission, the intensity of the scattered light is within $15 \, \%$ of the observed values, and the morphology of the central region of the surface brightness maps is comparable to the observed surface brightness (Fig. \ref{fig:all_spec_com}. However, the intensity of the diffuse region surrounding the cloud is a factor of 3 to 5 higher than observed. Furthermore, as discussed by \citet{Bazell1990} and \citet{Ossenkopf1991}, porosity increases the absorption efficiencies at FIR and sub-millimetre wavelengths while the compositional inhomogeneities will have an effect upon the shape and strength of broadband features. Thus, these structural differences of the grains will affect both the true temperature and color temperature derived from sub-millimetre flux ratios \citep{Kruegel1994}. 

Thus it is evident that a larger maximum grain size or the relative amount of large grains in the dust mixture is not sufficient to reproduce the observed NIR surface brightness. Evolution of dust grains, in the form of aggregates and mantle formation, appears necessary. This is supported by our core temperature estimates that tend to overestimate the core temperature by $\sim 1.5$ K. However, the THEMIS model, with three different dust populations produces a colder core, with $T_{\rm core} = 7.6$ K, similar to the estimated temperature of 8 K derived from line observations. However, even for the evolved grain models, the diffuse regions are still problematic as they are systematically brighter compared to the observations. The high surface brightness is likely related to the relative abundances of the dust populations, as for example, the THEMIS models has a relatively high abundance of the CMM population in the Veil and Striation regions, and similarly, the SIGMA model has a high abundance of the dense component in the Veil and Striations (Figs. \ref{fig:relative_abu} and \ref{fig:THEMIS_abu}). Decreasing the relative abundances of the dense dust components in these regions would decrease the surface brightness and improve the match with the observed surface-brightness morphology.

\section{Conclusions}

We have studied the cloud L1512 and modelled simultaneously the scattered NIR light at J, H, and K$_{\rm S}$ bands and the FIR emission at 250, 350, and 500 $\mu$m. We have used several dust models based on the \citet{Compiegne2011} dust and three separate dust models taking into account dust evolution. The radiation field used in the modelling is derived from the DIRBE observations. The NIR surface brightness is estimated using a density field and radiation field strength that are obtained from first fitting the FIR dust emission. The key result of our study are: 

\begin{itemize}

\item The morphologies of the observed NIR surface brightness maps are in good agreement with the column density map derived from the \textit{Herschel} observations. However, the low-column-density Veil seen above the cloud in the \textit{Herschel} observations is not visible in scattered light. 

\item In the radiative transfer modelling, we can fit the observed emission with any of the tested dust models. The average fit residual in the 350 $\mu$m band are $\pm 1.5  \, \%$ and for the 250 and 500 $\mu$m bands at most $- 15  \, \%$. Depending on the model, the estimated H$_2$ column density at the cloud centre (point 1) ranges from $4.3 \times 10^{21}$ cm$^{-2}$ to $1.6 \times 10^{22}$ cm$^{-2}$, and the relative radiation field strength from 0.3 to 0.75.

\item The core temperature of the radiative transfer models is on average $\sim 1.5$ K higher than the value $\sim 8 \, \pm \, 1$ K, suggested by $\rm N_2H^+$ line observations. With $T_{\rm core} = 7.61$ K, the THEMIS dust model gives the core temperature closest to the gas inferred value.

\item The radiative transfer models matching the dust emission predict a J band optical depth that is on average three times higher than the value measured with the background stars. The exceptions are Scaled4 and THEMIS, with $\tau_{\rm J,em} = 1.55$ and $\tau_{\rm J,em} = 1.05$, respectively, in agreement with the value $\tau_{\rm J,Card} = 1.5$ derived using the \citet{Cardelli1989} extinction curve.

\item The NICER estimates of $\tau_{\rm J,ext}$ obtained with the NIR extinction curves of the tested dust models  are mostly within $15 \, \%$ of the values obtained with the \citet{Cardelli1989} extinction curve. However, dust models containing larger grains (e.g. LG and LGM) increase $\tau_{\rm J,ext}$ by a factor of two. 
 
\item For the models based on the \citet{Compiegne2011} model, the predicted surface brightness excess $I_{\nu}^{\Delta}$ in the central region of the cloud is a factor of 2 to 4 below the observed values and the morphology of the simulated scattered light maps does not match the observations. Increasing the maximum grain size of the dust grains or extending the width of the cloud along the line-of-sight will increase the intensity of the scattered light, but not enough. Thus, dust grain evolution (e.g. aggregates) is needed.

\item Increasing the FIR emissivity by a factor of two (model Scaled2), increases the predicted NIR surface brightness by up to a factor of two. It also decreases the column density and produces more compact scattered NIR surface brightness maps. However, further increase of emissivity (model Scaled4) again decreases the NIR signal.

%\item MBB analysis of dust spectra can be problematic if the dust opacity is not a pure powerlaw. \textbf{We tested this in the case of the Default dust model, where the MBB fit gives a colour temperature of 16.5 K for an isothermal cloud with a true dust temperature of 15 K.

\item The observed H-band surface brightness and its morphology could only be matched with the Default model by making substantial modifications to the values derived from the emission fitting and observations, indicating the need of changes in the dust properties.

\item The observed thermal emission and scattered NIR surface brightness can be reasonably reproduced only by using dust models that take into account grain evolution. However, the J-band intensity and the H- and K$_{\rm S}$-band intensities of the diffuse regions are far above the observed values.  

\end{itemize}

It is easy to fit  the dust emission alone but the simultaneous fitting of emission and scattering is challenging. The dust evolution must be taken into account, to produce sufficient amounts of scattered light and with the correct morphology. The uncertainty of the sky brightness behind the studied cloud can have a significant effect on both the intensity and morphology and should be constrained with a high precision before drawing conclusions on the dust properties.

\begin{acknowledgements}
This work has made use of data from the European Space Agency (ESA) mission {\it Gaia} (\url{https://www.cosmos.esa.int/gaia}), processed by the {\it Gaia} Data Processing and Analysis Consortium (DPAC, \url{https://www.cosmos.esa.int/web/gaia/dpac/consortium}). Funding for the DPAC has been provided by national institutions, in particular the institutions participating in the {\it Gaia} Multilateral Agreement.

\end{acknowledgements}

\bibliographystyle{aa}
\bibliography{bibli}

\begin{appendix}

\section{Summary of dust models} \label{sec:SDM}

A more detailed description of all dust models used in our radiative transfer modelling is provided in this section. Unless otherwise noted, the different dust models are based on the \citet{Compiegne2011} model. We use the size distributions and optical properties as included in the DustEM\footnote{\url{https://www.ias.u-psud.fr/DUSTEM/index.html}} \citep{Compiegne2011}.

The changes that we have made to the \citet{Compiegne2011} model do not take into account any limits placed by the mineralogy or constraints placed by chemical abundances available in the ISM. The changes to for example the albedo or emissivity of the grains, are meant to be relatively small, but still large enough that differences between the models can be distinguished.

\begin{itemize}

\item Default: The \citet{Compiegne2011} model. Contains two populations of PAH grains with log-normal size distributions, a single component of small carbon grains with a log-normal distribution, and two components of large grains following power-law size distributions. The large grains consist of a population of carbon grains and a population of silicate grains. The average opacity spectral index in the range [$250,350$] $\mu$m is $\beta_{250/350} = 1.887$.

\item Albedo: The albedo of the grains at NIR wavelengths has been increased by $20  \, \%$ without changing the extinction cross sections during the scattering computations.

\item Disttest: \citet{Compiegne2011} assumed a power-law size distribution for the large silicate and carbon grains. For this model the exponent factor of the power-law $\gamma$ has been increased by $20 \, \%$ for both the silicate (original $\gamma = -3.4$) and carbon grains(original $\gamma = -2.8$) and $\beta_{250/350} = 1.888 $.

\item Gtest: The asymmetry parameter $g$ of the grains has been increased by $15 \, \%$.

\item Scaled2: The emissivity of all wavelengths longer than 60 $\mu$m has been multiplied by a factor of 2.

\item Scaled4: As Scaled2, but the emissivity has been multiplied by a factor of 4.

\item Wide: The FWHM of the Gaussian distribution describing the line-of-sight density distribution has been increased from $\sigma = 1.0$ to $\sigma = 1.73$.

\item LG: Extended the maximum grain size of the \citet{Compiegne2011} model to 5 $\mu$m. The value of $\beta_{250/350}$ is 1.903. 

\item LGM: As LG, but in addition increased the relative abundance of the large grains by a factor of 2 and decreased the relative abundance of the small carbon grains and PAH components by a factor of 2. The value of $\beta_{250/350}$ is 1.948.

\item Ddust: Combination of Default and LG dust models, the relative abundance of the latter increasing with density as

\begin{equation}\label{EQ:abundance}
S_{\rm LG} = \frac{1}{2} + \frac{1}{2} \tanh(\frac{2.0 \times \log (\rho)}{\rho_0}),
\end{equation}
where $\rho$ is the density of the modelled cloud and $\rho_0$ defines the limit where the relative abundances of the two dust components are equal. An example of the abundances is shown in Fig. \ref{fig:relative_abu}. In our model we have set $\rho_0 = 1.45$. For the average of the Default and LG models $\beta_{250/350} = 1.955$.

\begin{figure*}
%\sidecaption
\includegraphics[width=17.8cm]{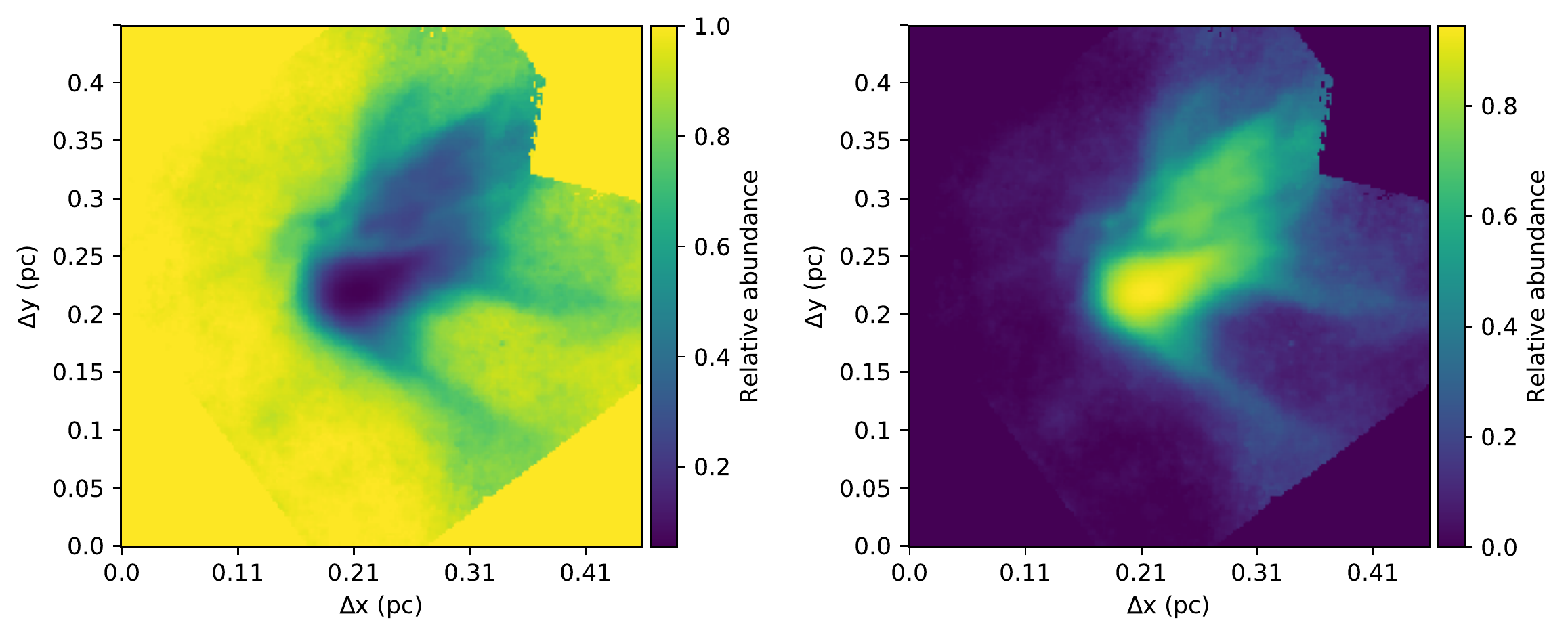}
\caption{Relative abundances of the diffuse and dense components for the DDust and SIGMA models across the innermost cells of the model cloud. Left panel: the relative abundance of the diffuse component. Right panel: relative abundance of the dense component.}
\label{fig:relative_abu}
\end{figure*}

\item SIGMA: A model with two dust components, for the diffuse component we use our Default model, and the dense component is built using SIGMA \citep[Simple Icy Grain Model for Aggregates,][]{SIGMA2020}. The dense component consists of aggregate silicate and carbon grains from the \citet{Min2016} model (note that we do not use the iron sulphide component) with added ices \citep{Pollack1994}. The final mixture consists of $56.8 \, \%$ silicates, $18.9 \, \%$ carbons, and $24.3 \, \%$ ices. The size distribution has a maximum grain size of 10.0 $\mu$m (see Fig. \ref{fig:SIGMA_size}), however for this work we have truncated the maximum grain size at 5 $\mu$m, and use a porosity factor of 0.7 for the grains. The porosity was chosen accordingly to the size distribution obtained from coagulation computation. Such a high porosity was necessary to obtain favourable conditions to stick dust grains together \citep[][Pagani et al. in prep]{Ormel2011, Hirashita2013, Hirashita2014}. In practice, there is a significant fraction of fluffy icy aggregates with a fractal degree close to 2. The dust distribution is representative of a coagulation at a constant density of $1 \times 10^6$ cm$^3$ with a constant porosity of 0.7 for 0.436 My. The simplification of dust distribution evolving at constant density and porosity will be discussed in a forthcoming paper (Pagani et al. in prep.)

The relative abundances of the dense and diffuse dust components are set according to Eq. \ref{EQ:abundance}, with  $\rho_0 = 1.45$. The threshold was so that the diffuse dust component would have a relative abundance of less than $\sim 10 \, \%$ in the core of the model. For the average of the two dust components $\beta_{250/350} = 2.024$.

\begin{figure}
%\sidecaption
\includegraphics[width=8.5cm]{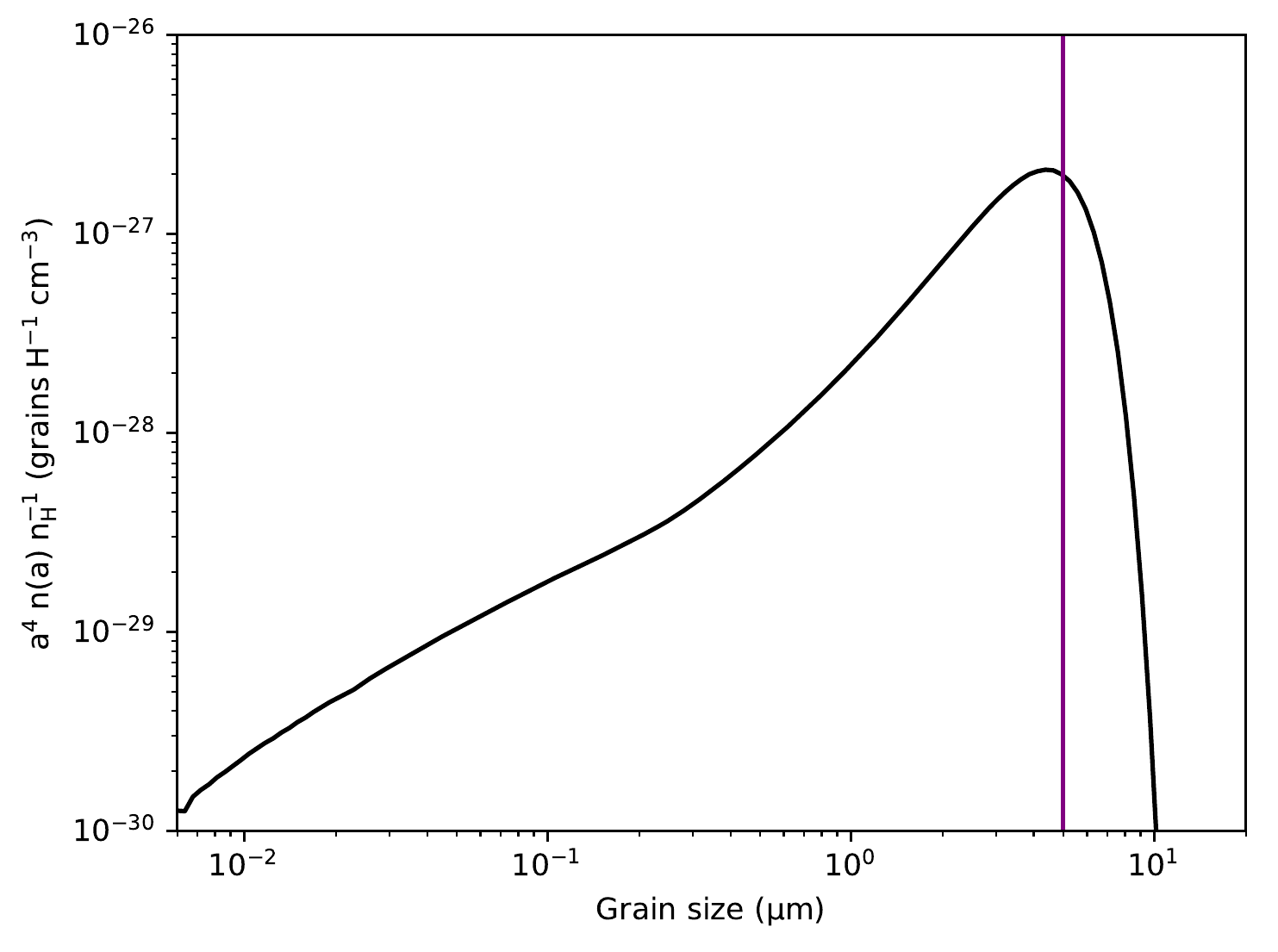}
\caption{Size distribution of of the dust grains used for the dense component of the SIGMA model. For the emission fitting and scattering computations, the size distribution was truncated at 5 $\mu$m, as indicated by the purple line. }
\label{fig:SIGMA_size}
\end{figure}

\item THEMIS: A dust model as discussed by \citet{Kohler2015} and \citet{Ysard2016}. The model consists of different dust population mixtures, that have a varying relative abundance based on the density of the model. In the following we adopt the naming convention as defined by \citet{Kohler2015}. The diffuse regions of the model consist of mostly core-mantle (CM) grains which gradually evolve to core-mantle-mantle grains (CMM) as density increases. In the densest regions of the model, we assume that the grains have further evolved and are gradually replaced by aggregate CMM grains with additional ice mantles (AMMI). The relative abundances of the CM and CMM populations are set according to Eq. \ref{EQ:abundance}, with  $\rho_0 = 1.45$. The relative abundance of the AMMI grains is then defined by taking all cells where $\rho > 7.5$, setting the relative abundance of these cells to 1.0 and smoothing the cells with a $3 \times 3$ Gaussian beam. We then reduce the relative abundance of the CMM population in these cells so that for each cell, the sum $\rm CM + CMM + AMMI = 1.0$. An example of the relative abundances of the three components across the densest part of the cloud is shown in Fig. \ref{fig:THEMIS_abu}. The size distributions of the different dust components are discussed by \citet{Kohler2015} and \citet{Ysard2016} (Figs. 1 and 2, respectively). The average $\beta_{250/350}$ over the three dust components is 2.011.

\begin{figure*}
%\sidecaption
\includegraphics[width=17.8cm]{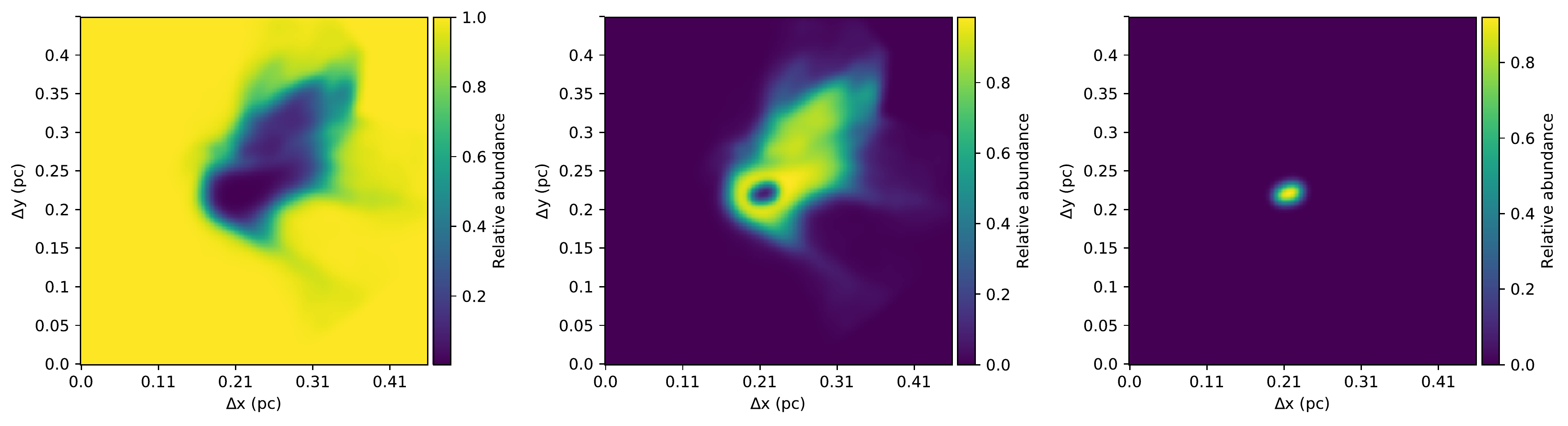}
\caption{Relative abundances of the CM, CMM, and AMMI dust populations of the THEMIS model across the innermost cells of the model cloud. Left panel: the relative abundance of the diffuse CM component. Centre: relative abundance of the CMM component. Right panel: relative abundance of the dense AMMI component.}
\label{fig:THEMIS_abu}
\end{figure*}

\end{itemize}

\section{Stochastically heated grains} \label{STOKA}

The effects of stochastic heating of small grains was not included in our modelling, because they are not directly connected to the submm emission and NIR scattering. However, in order to study how well our best-fit models agree with the observed emission in the MIR wavelengths, we computed two test cases using the models Default and THEMIS, including the stochastic heating. We did not fit the model to the observed emission, but rather used the best-fit parameters from the computations without SHGs and re-computed the emission including the SHGs.

\begin{figure*}
%\sidecaption
\includegraphics[width=17.8cm]{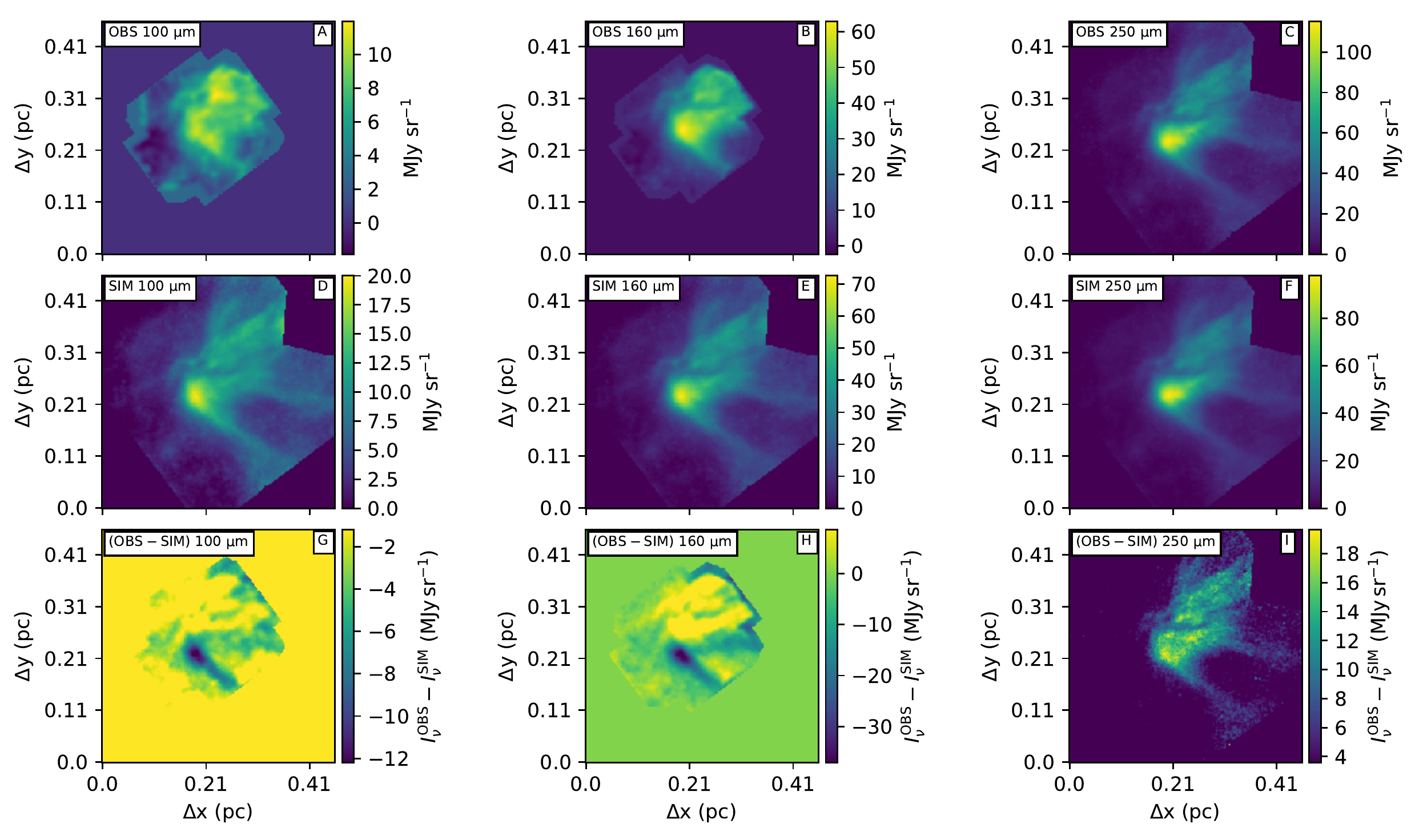}
\caption{Observed (first row) and simulated (second row) emission maps using the Default model at 100, 160, and 250 $\mu$m. The third row shows the difference between the observed and simulated maps. The simulated emission has been computed from the best fit parameters of the Default model and with taking into account stochastically heated grains.}
\label{fig:stoka_def}
\end{figure*}

\begin{figure*}
%\sidecaption
\includegraphics[width=17.8cm]{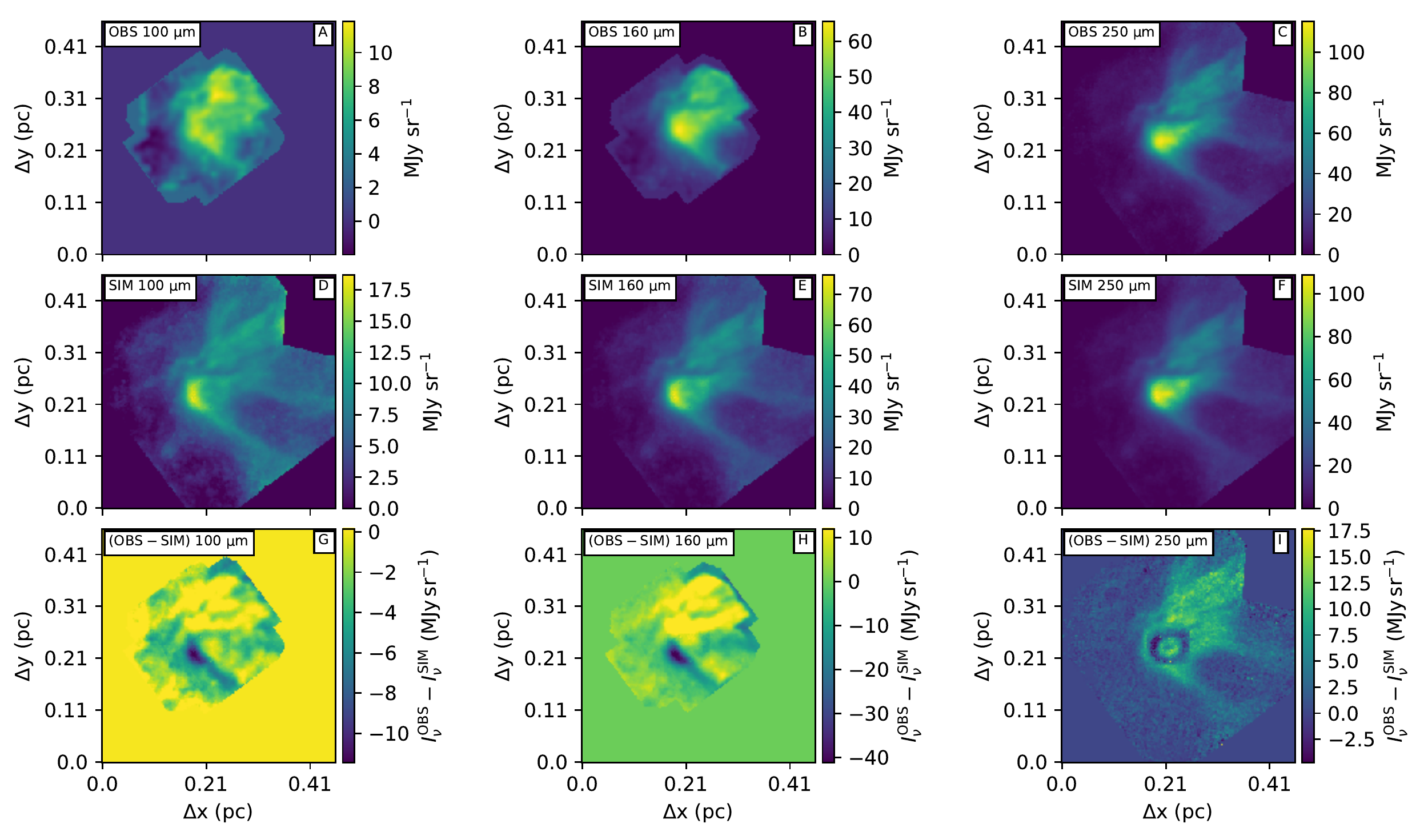}
\caption{As Fig. \ref{fig:stoka_def}, but for the THEMIS model.}
\label{fig:stoka_the}
\end{figure*}

The results of these computations are shown in Figs. \ref{fig:stoka_def} and \ref{fig:stoka_the}. For both models, the emission in the SPIRE bands has decreased, by $\sim 10 - 15 \, \%$, because some of the energy is now emitted in the MIR wavelengths. The emission is at 100 $\mu$m 30-45 $\%$ and at 160 $\mu$m ~15$\%$ above the observed values. For the Default model, the morphology of the simulated 100 and 160 $\mu$m maps agrees with the observed maps. For the THEMIS model, the bright rim in both maps is up to 40 MJy$\,$sr$^{-1}$ brighter than observed. The region where the CMM grains transition to the AMMI grains is clearly visible, especially in the 250 $\mu$m map.

\begin{figure*}
%\sidecaption
\includegraphics[width=17.8cm]{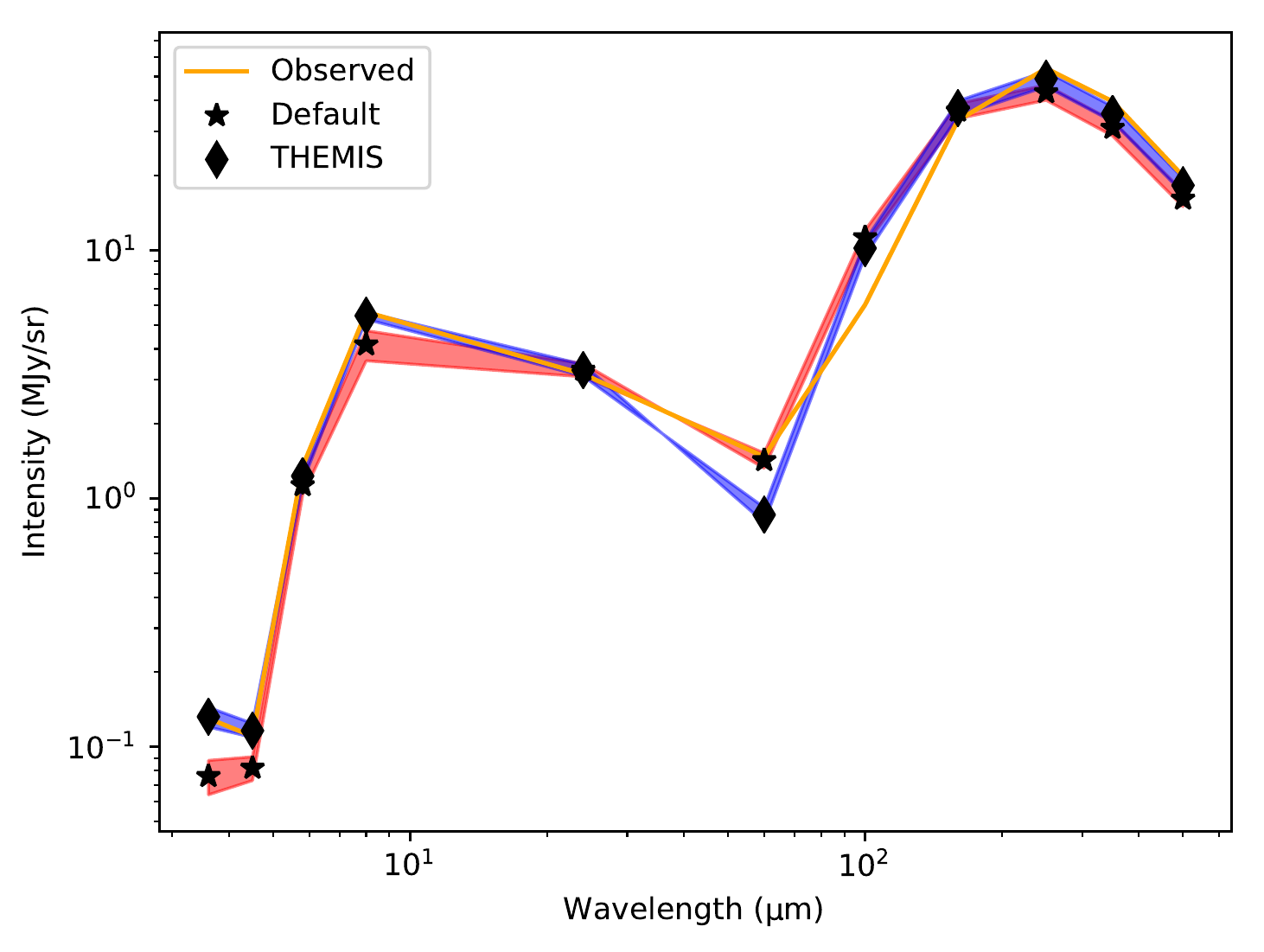}
\caption{Spectral energy distributions for the observations (orange line), the Default model (star symbols), and the THEMIS model (diamond symbols). The red and blue highlights show the assumed uncertainty in the modelled results for the Default and THEMIS models, respectively. The data points of the models are a sum of scattered surface brightness seen over the background level, estimated emission, and subtracting the background seen trough the cloud.}
\label{fig:SED}
\end{figure*}

An SED from the observations, computed as averages over a region corresponding to the red circle in Fig. \ref{fig:heating} is shown in Fig. \ref{fig:SED} (the orange line). The observations between 3.6 $\mu$m and 8 $\mu$m are from the Spitzer IRAC instrument and the 24 $\mu$m data are from the Spitzer MIPS instrument. The MIPS data are corrected with a aperture correction of 2 MJy$\,$sr$^{-1}$. For the 60 $\mu$m observations, we use the improved reprocessing of the IRAS survey (IRIS) data \citep{IRIS2005} and the 100 to 500 $\mu$m observations are from the \textit{Herschel} PACS (100 and 160 $\mu$m) and SPIRE (250, 350, and 500 $\mu$m) instruments.

An SED from the Default and THEMIS models with the SHGs included are shown in Fig. \ref{fig:SED}. The background levels of the IRAC observations have not been calibrated to an absolute level. Thus, we assume an uncertainty of $\pm \, 30 \, \%$ for the simulated surface brightness at IRAC wavelengths and because of the uncertainty of the colour corrections, we assume an uncertainty of $\pm \, 30 \, \%$ for the 24 $\mu$m data. As discussed by \citet{Launhardt2013}, the nominal calibration uncertainties for PACS and SPIRE are $\sim5 \, \%$ and $\sim7 \, \%$, respectively, thus for the simulated MIR and FIR maps we assume a flat uncertainty of 10$\, \%$. These uncertainties are highlighted with red and blue shading in Fig. \ref{fig:SED}.

It is clear that the simulations do not precisely match the observations, although the shapes of the SEDs are in agreement. The THEMIS model is closer to the observations both in the NIR and FIR, although the Default model at 60 and 160 $\mu$m is closer to the observed values.

\section{Analysis of dust models based on the default model} \label{sec:ADDmodels}

In addition to the models discussed in the main part of the text, we have tested dust models that are modifications of the Default model, increased albedo (model Albedo), increased the value of the $g$ parameter (model Gtest), and changes to the size distribution of the grains (Disttest). We have also tested a case where the LOS density of the cloud is wider compared to our assumed 'compact core' model (model Wide). A description of these models is provided in Appendix \ref{sec:SDM}, and the results are summarised in Table \ref{tab:ADDmodels}. The resulting dust emission and scattered surface brightness maps are shown in Appendix \ref{ADDFigures}. The scaling of the radiation field is similar to all models with $K_{\rm ISRF} \sim 0.45$, expect for the model Wide which has a lower scaling factor of $K_{\rm ISRF} \sim 0.37$. 

Compared to the Default model the scattered surface brightness maps (see Figs. \ref{fig:NIR_sim_0}, \ref{fig:NIR_sim_1}, and \ref{fig:NIR_sim_2}) of models Albedo, Disttest, Gtest and Wide, show only minor differences. The most notable difference is for the model Albedo, which produces $\sim 20 \, \%$ more surface brightness in all three bands compared to the Default model.

Increasing the line-of-sight extent of the cloud can produce more scattered light as the amount of illumination along the line-of-sight increases. We test a case, with the Default model, where the line-of-sight density distribution is extended, case Wide. The results indicate that the surface brightness is increased but only by $\sim 10  \, \%$ (see Fig. \ref{fig:NIR_sim_2} second row).

%\begin{table*}
\begin{sidewaystable*}
\caption{Summary of the additional radiative transfer models, including the column densities, NIR intensities, J band optical depth, 250 $\mu$m optical depth, radiation field scaling, and the core temperature.}
\centering
\begin{tabular}{c c c c c c c c c c}
\hline
\hline
& & & & & & & & &\\
Model name & Description  & $N(\rm H_2)$ \tablefootmark{\rm (1)} & J\tablefootmark{\rm (1)} & H\tablefootmark{\rm (1)} & K$\rm _S$\tablefootmark{\rm (1)} & $\tau_{\rm J, em}$\tablefootmark{\rm (1)} & $\tau_{250}$\tablefootmark{\rm (1)} & $K_{\rm ISRF}$ & $\rm T_{\rm core}$\tablefootmark{\rm (2)} \\
 &   & $(\rm cm^{-2})$ & $(\rm MJy / sr)$ & $(\rm MJy / sr)$ & $(\rm MJy / sr)$ & & ($ \times 10^{-3}$) & & $\rm (K)$ \\
\hline
& & & & & & & & &\\
OBS& Values derived from observations & $4.51 \times 10^{21}$  & 0.08 & 0.15 & 0.10 & 1.50 & 2.56 & - & 7.5 \\

& & & & & & & & &\\
\hline
& & & & & & & & &\\
& COM models & & & & & & & &\\
& & & & & & & & &\\

Default & \citet{Compiegne2011}  & $1.59 \times 10^{22}$ & 0.055 & 0.047 & 0.054 & 5.80 & 2.67 & 0.447 & 9.34\\

Albedo & Albedo of the dust grains increased by $20  \, \%$ & $1.59 \times 10^{22}$  & 0.098 & 0.074 & 0.077 & 5.77 & 2.67 & 0.447 & 9.34\\

Gtest & $g$ parameter of the dust grains increased by $15  \, \%$ & $1.58 \times 10^{22}$ & 0.052 & 0.041 & 0.051 & 5.79 & 2.66 & 0.443 & 9.34\\

Disttest & $\gamma$ of the dust size distribution increased by $20 \,  \%$  & $1.58 \times 10^{22}$ & 0.074 & 0.075 & 0.088 & 7.30 & 2.69 & 0.445 & 9.33 \\

Wide & Model cloud with a wider LOS density distribution  & $1.59 \times 10^{22}$ & 0.068 & 0.055 & 0.059 & 5.80 & 2.67 & 0.368 & 9.76\\

& & & & & & & & &\\
\hline
& & & & & & & & &\\
\end{tabular}
\tablefoot{(1) The column density, background subtracted intensity, and the optical depths of J band and 250 $\mu$m band have been computed as average values over $5 \times 5$ map pixels centred on point 1 (see left panel of Fig. \ref{fig:scat_loc_spec}). \\
(2) The value derived from observations based on the $\rm N_2H^+$ line observations by \citet{Lin2020}. The modelled values are computed as averages over $10^3$ cells centred at the core.\\}

\label{tab:ADDmodels}
%\end{table*}
\end{sidewaystable*}

\section{NIR and MIR observations}

The MIR Spitzer observations were acquired from the Spitzer Heritage Archive. The cloud L1512 has been covered by three programs, Program id. 94 (PI Charles Lawrence) and Program id. 90109 (PI Roberta Paladini), both using the InfraRed Array Camera (IRAC) (see Fig. \ref{fig:MIR_obs}) and in Program id. 53 (PI George Rieke) with the Multiband Imaging Photometer for Spitzer (MIPS). However, the program 90109 was carried out during the warm mission, thus only the 3.6 and 4.5 $\mu$m channels were available. The IRAC observations are discussed in more detail by \citet{Stutz2009} and \citet{Steinacker2015}, for the cold and warm missions, respectively. The MIPS observations are described by \citep{Rieke2004} and \citet{Stutz2007}.

The Spitzer 3.6 $\mu$m and 4.5 $\mu$m maps show extended surface brightness towards the central region of the cloud, but in both 5.8 $\mu$m and 8.0 $\mu$m maps the region is seen in absorption (see Fig. \ref{fig:MIR_obs}). The surface brightness in the 3.6 and 4.5 $\mu$m maps can be caused by thermal emission by small grains, but the surface brightness is only seen from the dense central regions. If the surface brightness is caused by thermal emission because of the high optical depth for the radiation heating the dust grains, one would expect it to be bright in the more extended region, not in the cloud centre. On the other hand, as discussed by \citet{Steinacker2010}, in the current PAH emission models, the emission should increase towards the longer wavelengths and the Spitzer 4.5 and 5.8 $\mu$m bands, but in the Spitzer images the opposite is seen as the cloud in seen in absorption towards the longer wavelengths. Thus, we can conclude that the extended surface brightness seen in the 3.6 $\mu$m band, and at shorter wavelengths, is scattered light.

\begin{figure*}
%\sidecaption
\includegraphics[width=17.8cm]{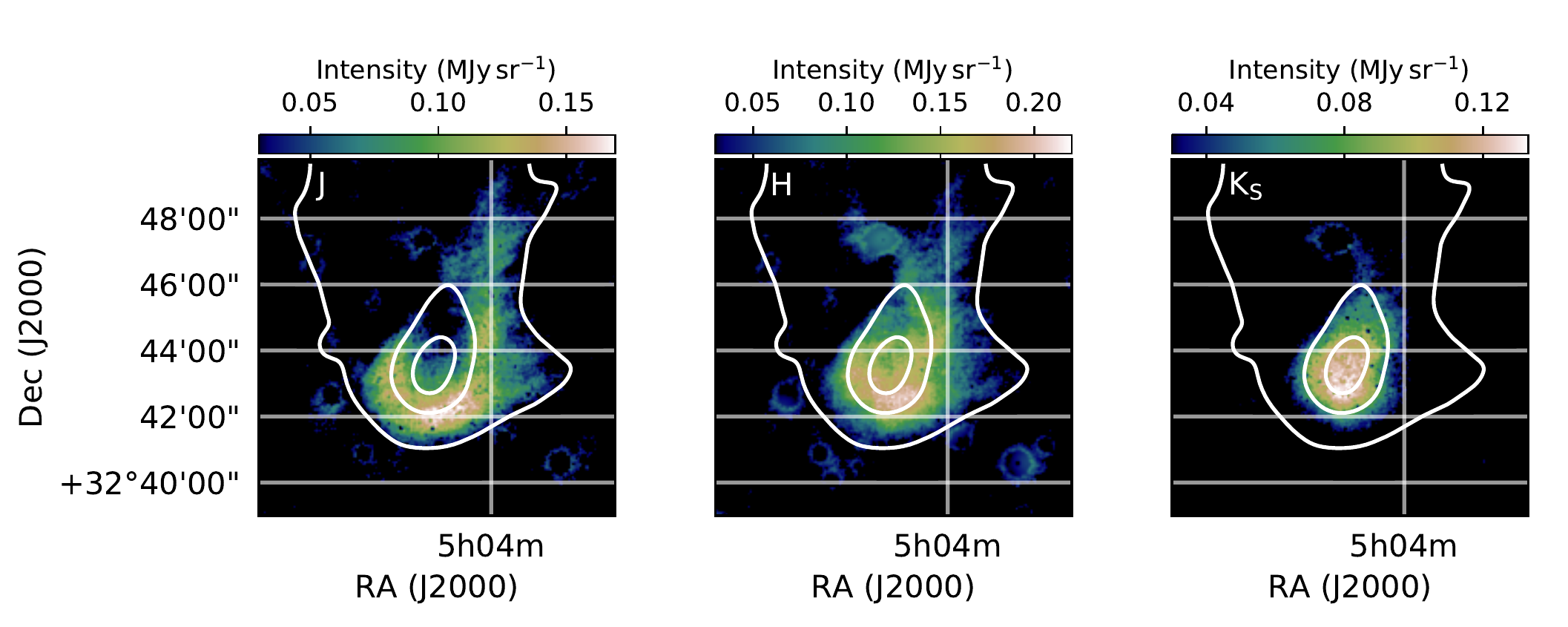}
\caption{Colour maps show the NIR surface brightness at J band (left frame), H band (centre), and K$\rm _S$ band (right). The surface brightness maps have been smoothed over 6 $\times$ 6 map pixels. The white contours show the $N (\rm H_2)$ column density derived from the \textit{Herschel} observations. The contour levels are 15$ \, \%$, 45$ \, \%$, and 75$ \, \%$ of the peak column density of $1.12 \times 10^{22}$ cm$^{-2}$.}
\label{fig:NIR_obs}
\end{figure*}

\begin{figure*}
%\sidecaption
\includegraphics[width=17.8cm]{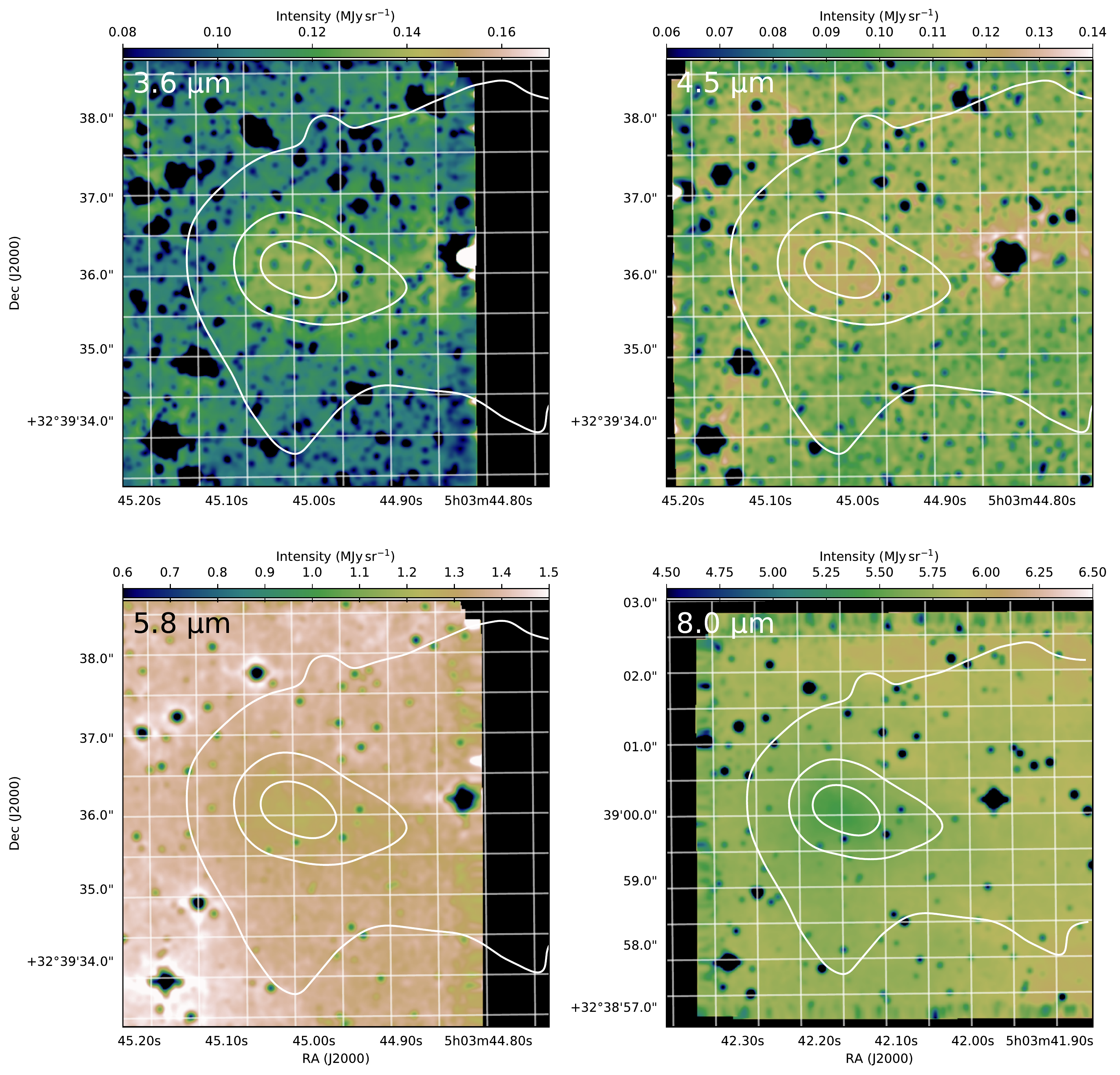}
\caption{Spitzer observations covering the 3.6, 4.5, 5.8, and 8.0 $\mu$m bands. The surface brightness maps have been smoothed over 6 $\times$ 6 map pixels. The white contours show the $N (\rm H_2)$ column density derived from the \textit{Herschel} observations. The contour levels are 15$ \, \%$, 45$ \, \%$, and 75$ \, \%$ of the peak column density of $1.12 \times 10^{22}$ cm$^{-2}$.}
\label{fig:MIR_obs}
\end{figure*}

Shown in Figs. \ref{fig:NIR_obs} and \ref{fig:MIR_obs} are the observations from the WIRCam instrument and Spitzer space telescope. To study the diffuse signal, we have used the Source-Extractor \citep[][SExtractor]{Bertin1996} and Point Spread Function Extractor \citep[][PSFEx]{Bertin2011} to remove point sources. The extraction was carried out in three steps, in the first step, we use SExtractor to detect only the brightest point sources in each image. The detected bright sources are then analysed by PSFEx to construct a PSF for each detected source. The PSFs are then used in a second run with SExtractor to detect all point sources in the images. This second run produces an image of  objects which we subtract from the original observations resulting in an image in which stars appear as smooth holes. The NIR images are further smoothed over $6 \times 6$ map pixels to better show the morphology of the scattered light.

\section{Additional figures} \label{ADDFigures}

\begin{figure*}
%\sidecaption
\includegraphics[width=17.8cm]{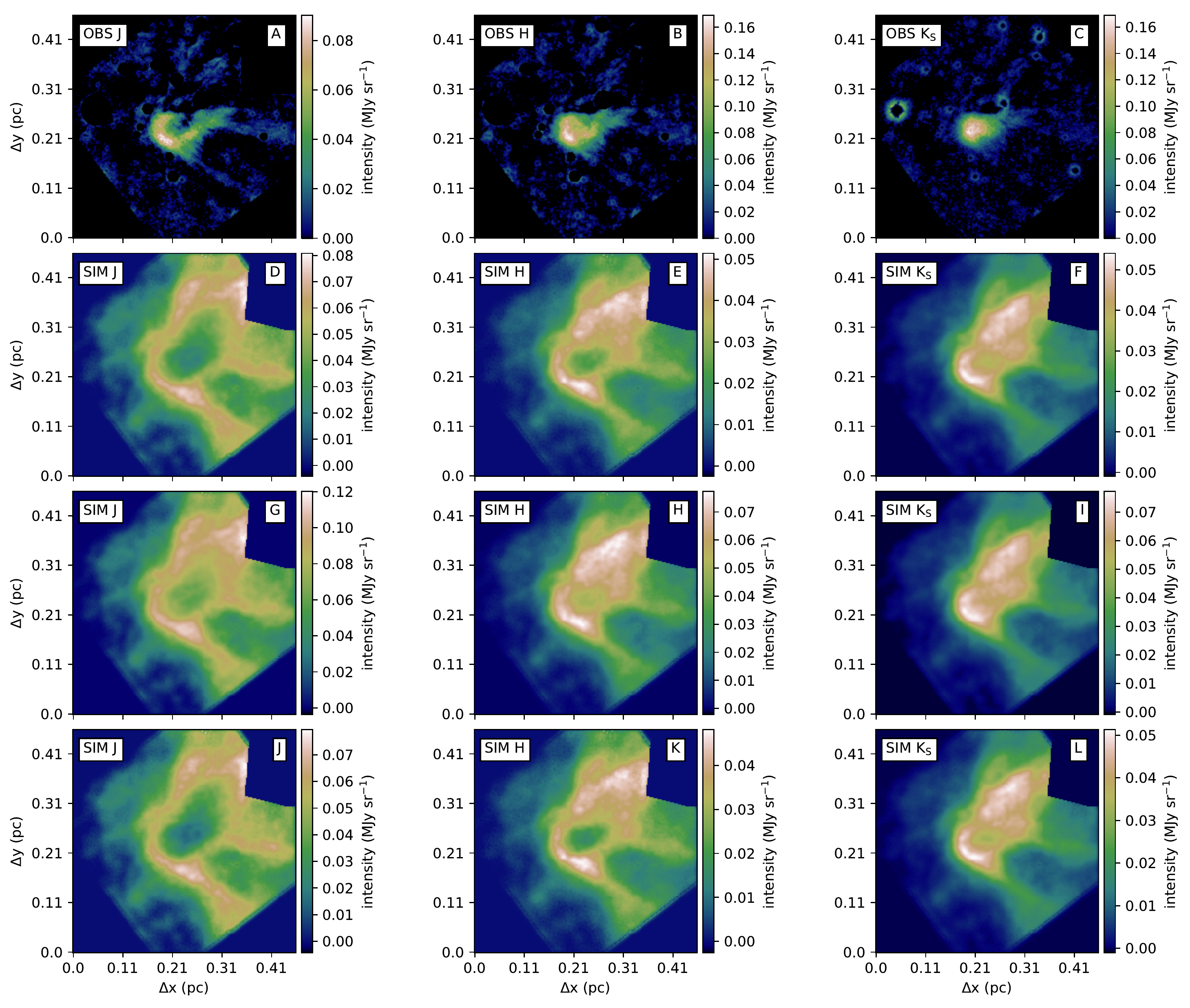}
\caption{Observed NIR surface brightness (first row) compared to model predictions with the Default (second row), Albedo (third row), and Disttest (fourth row) models.}
\label{fig:NIR_sim_0}
\end{figure*}

\begin{figure*}
%\sidecaption
\includegraphics[width=17.8cm]{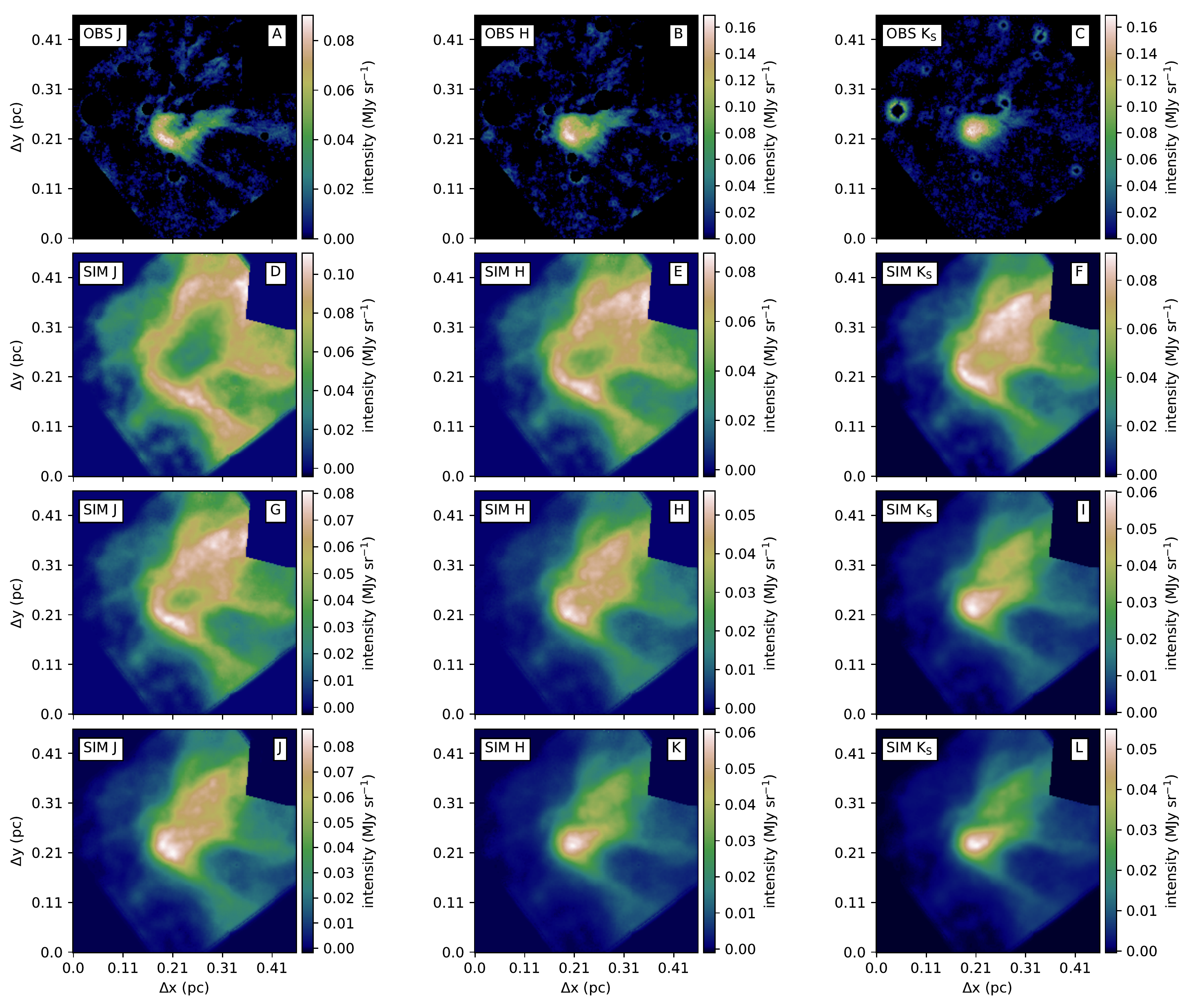}
\caption{As Fig. \ref{fig:NIR_sim_0}, but for models Gtest (second row), Scaled2 (third row), and scaled4 (Fourth row).}
\label{fig:NIR_sim_1}
\end{figure*}

\begin{figure*}
%\sidecaption
\includegraphics[width=17.8cm]{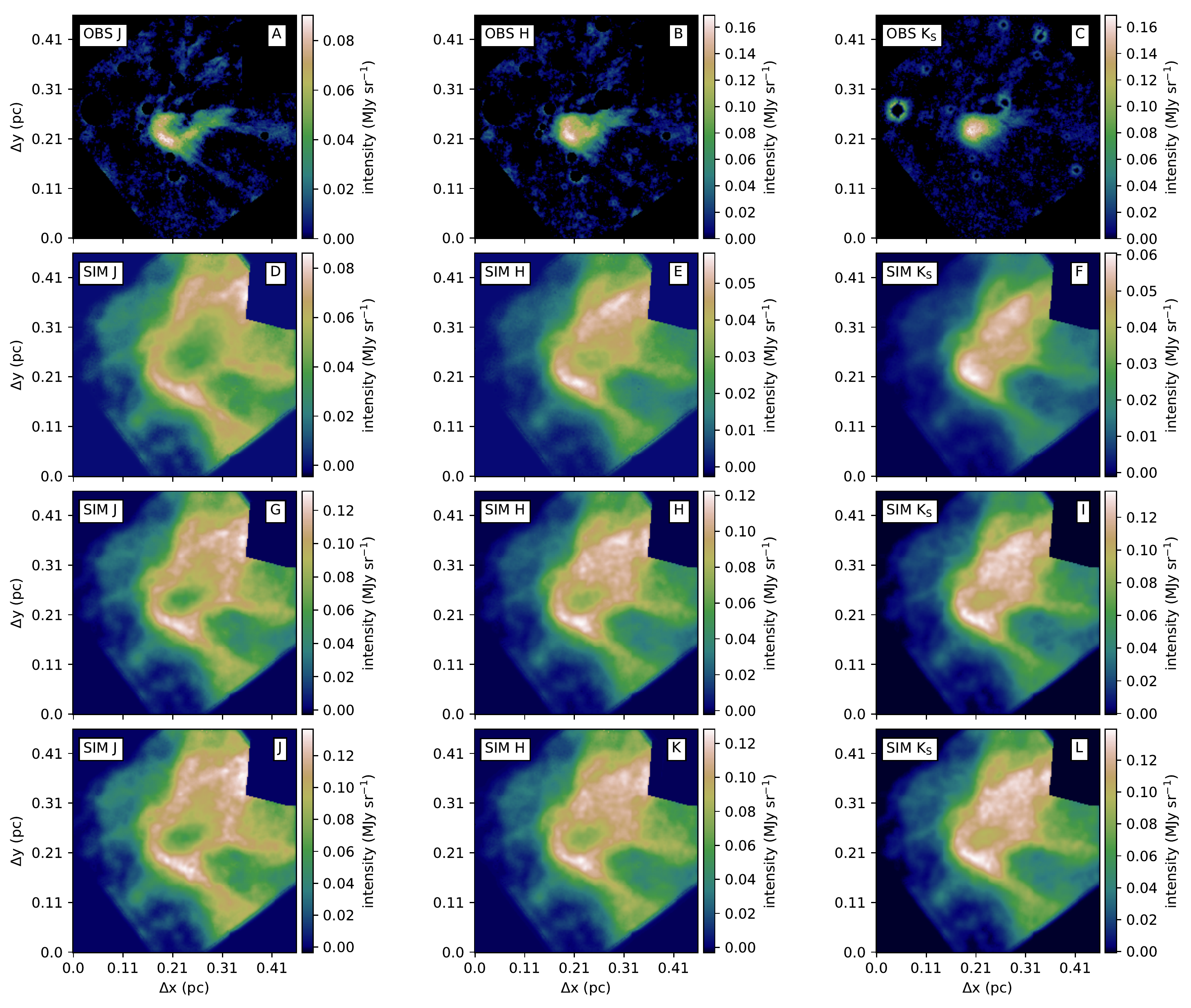}
\caption{As Fig. \ref{fig:NIR_sim_0}, but for models Wide (second row), LG (third row), and LGM (Fourth row).}
\label{fig:NIR_sim_2}
\end{figure*}

\begin{figure*}
%\sidecaption
\includegraphics[width=17.8cm]{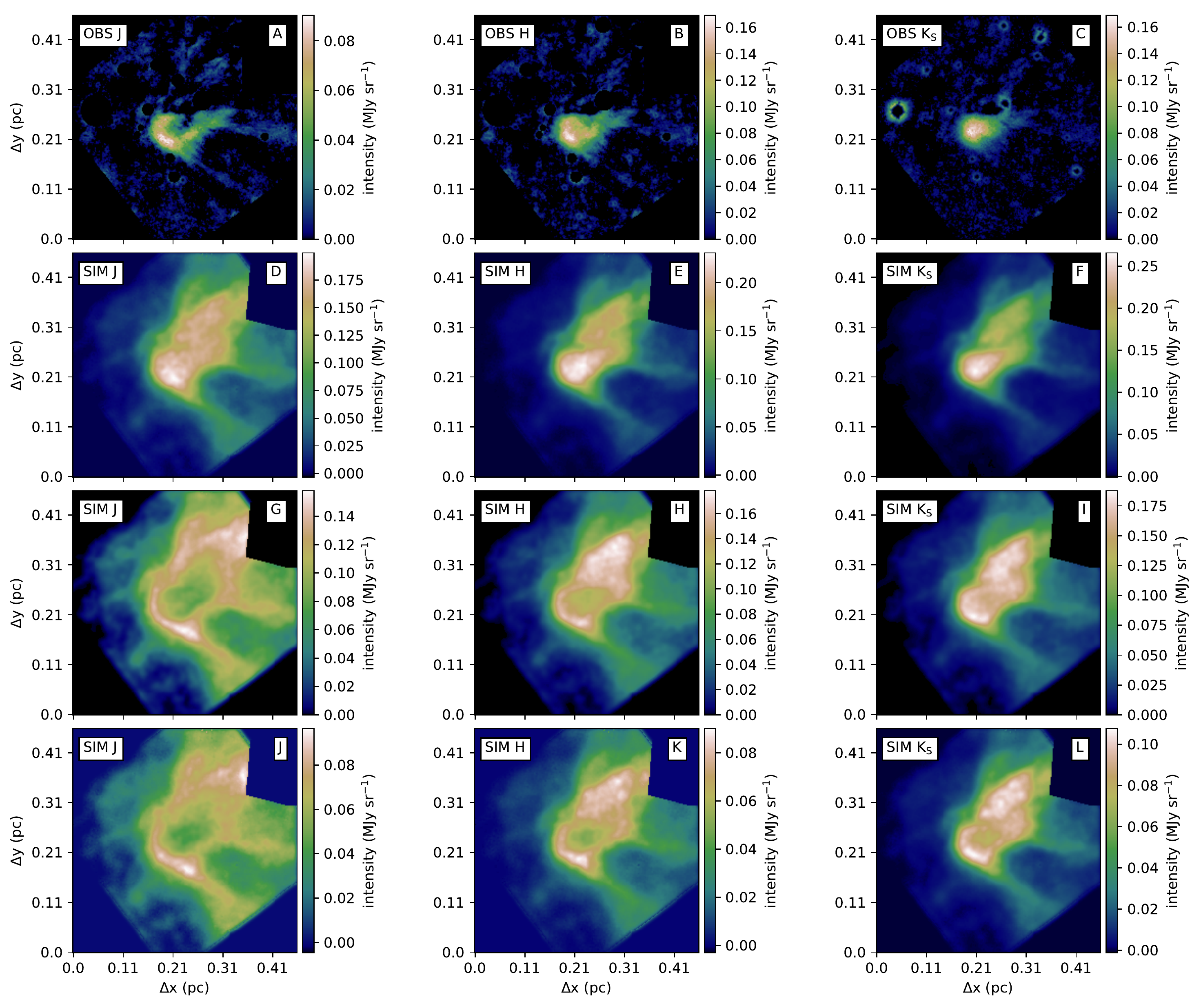}
\caption{As Fig. \ref{fig:NIR_sim_0}, but for models SIGMA (second row), THEMIS (third row), and DDust (Fourth row).}
\label{fig:NIR_sim_3}
\end{figure*}

\begin{figure*}
%\sidecaption
\includegraphics[width=17.8cm]{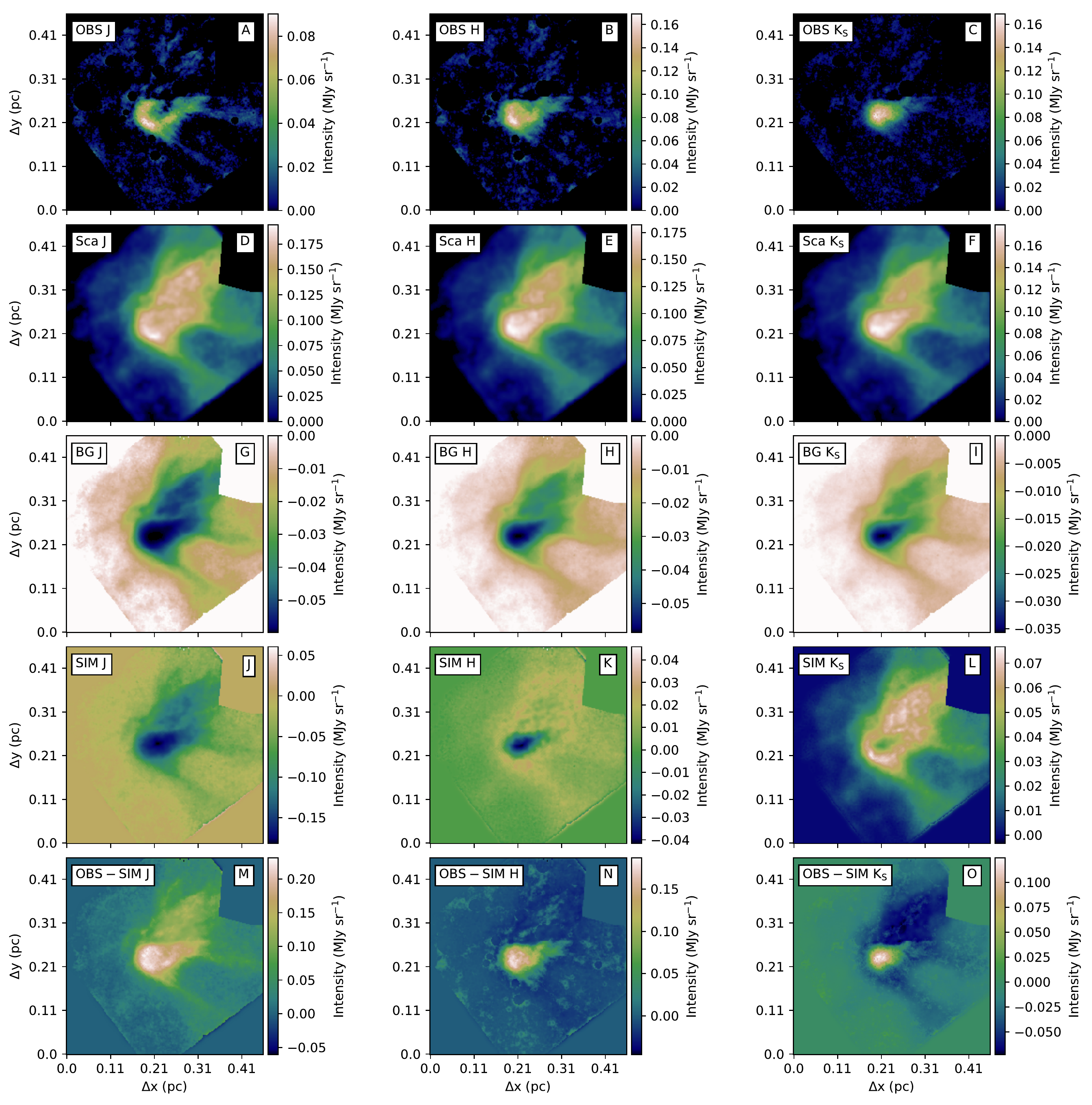}
\caption{Different components in the surface brightness maps for the Default model. Shown on the first row are the background subtracted observed surface brightnesses and the second row shows the simulated scattered light without the background subtraction. Shown on the third row is the component of the background that is seen trough the cloud (attenuated by a factor of $e^{-\tau}$) and the fourth row shows the background subtracted simulated surface brightness. The last row shows the difference between the observed and simulated background subtracted surface brightness.}
\label{fig:NIR_comparison}
\end{figure*}

The Figs. \ref{fig:NIR_sim_0} to \ref{fig:NIR_sim_3} show all of our surface brightness maps for the different dust models. Shown in Fig. \ref{fig:NIR_comparison} are the individual components of the simulations. 

\begin{figure*}
%\sidecaption
\includegraphics[width=17.8cm]{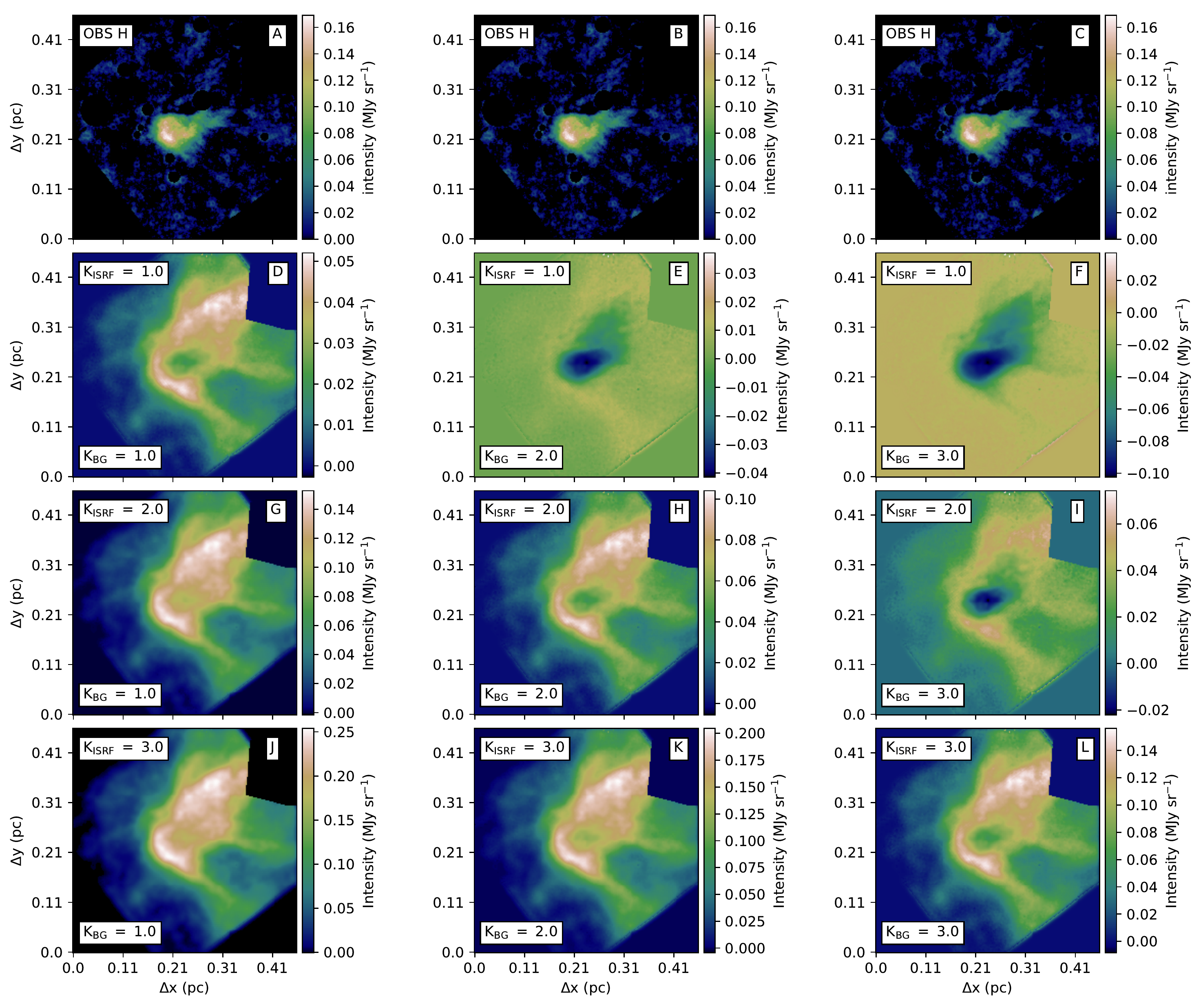}
\caption{Simulated H band surface brightness maps for the Default model with different assumptions on the strength of the radiation field and on the sky brightness behind the cloud. Shown on the first row are the observed surface brightness maps. Shown on the rows 2 to 4 are the simulated H band maps, for which the strength of the radiation field has been scaled with a factor between 1 and 3, and the intensity of the background has been scaled between factors 1 and 0.3. The density of the cloud is the same as in the Default model.}
\label{fig:ISRF_test1}
\end{figure*}

\begin{figure*}
%\sidecaption
\includegraphics[width=17.8cm]{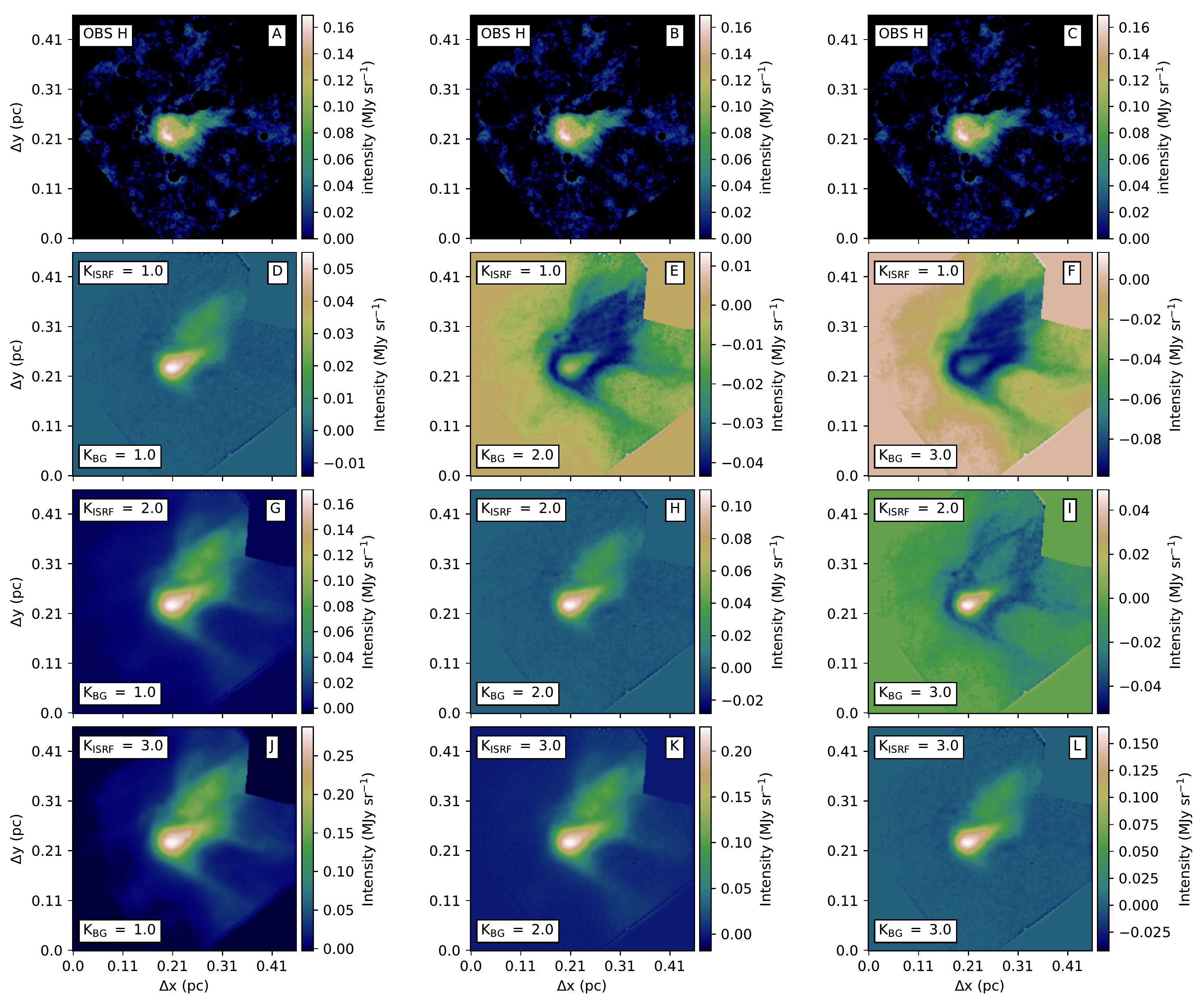}
\caption{As Fig. \ref{fig:ISRF_test1}, but the cloud model has a $70 \, \%$ lower column density compared to the Default model.}
\label{fig:ISRF_test3}
\end{figure*}

Figs. \ref{fig:ISRF_test1} to \ref{fig:ISRF_test3} are the simulated surface brightness maps of the H band using the Default dust model. The rows and columns of the figures corresponding to different assumptions on the strength of the radiation field and the sky brightness behind the cloud. The model cloud column density is scaled by 1, 0.6, and 0.3 for the three figures, respectively.

\begin{figure*}
%\sidecaption
\includegraphics[width=17.8cm]{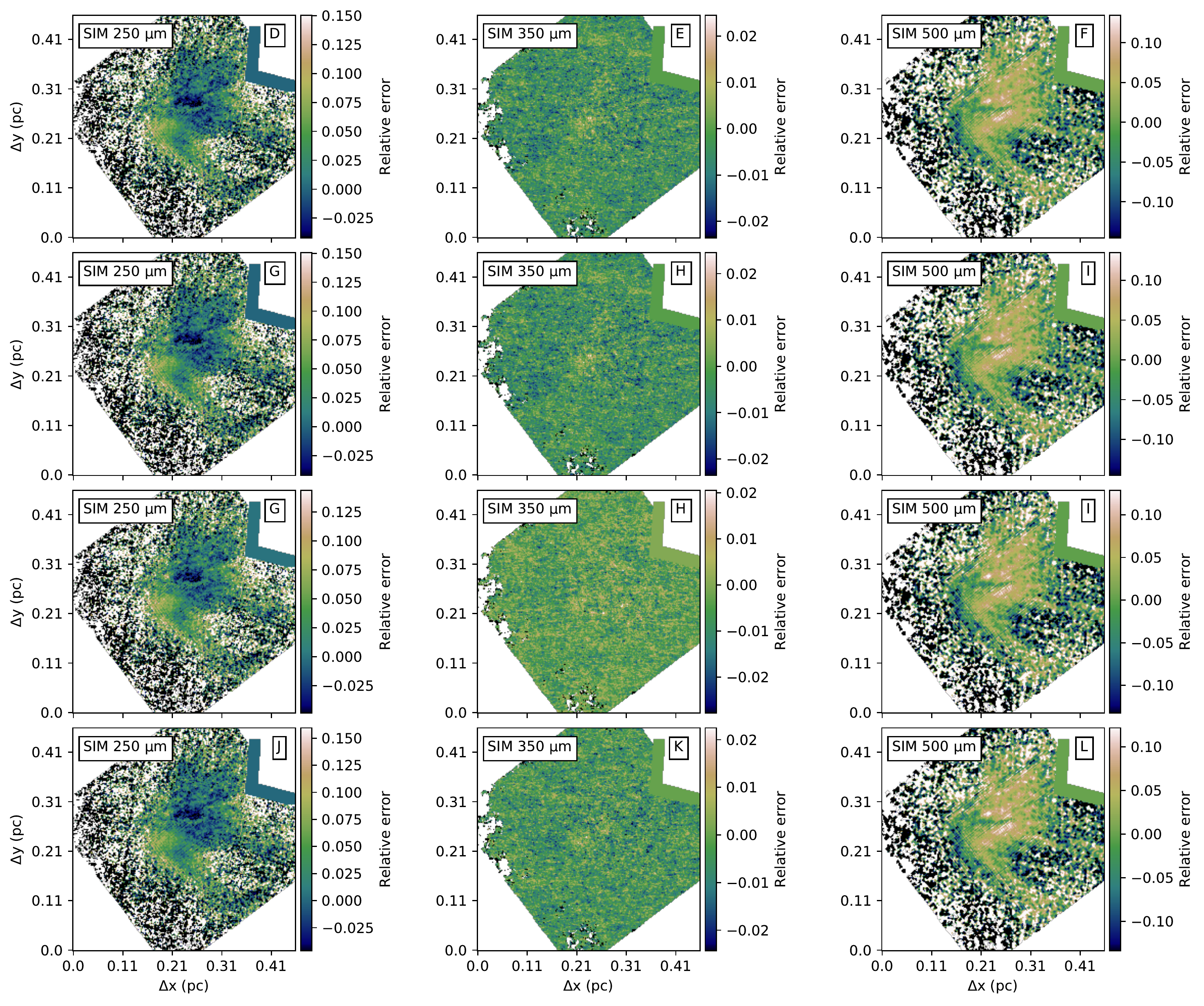}
\caption{Relative errors between the observed emission and our model predictions for 250, 350, and 500 $\mu$m bands. Each row corresponds to a different dust model. The rows from top to bottom correspond to models Default, Albedo, Disttest, and Gtest. }
\label{fig:EM_1}
\end{figure*}

\begin{figure*}
%\sidecaption
\includegraphics[width=17.8cm]{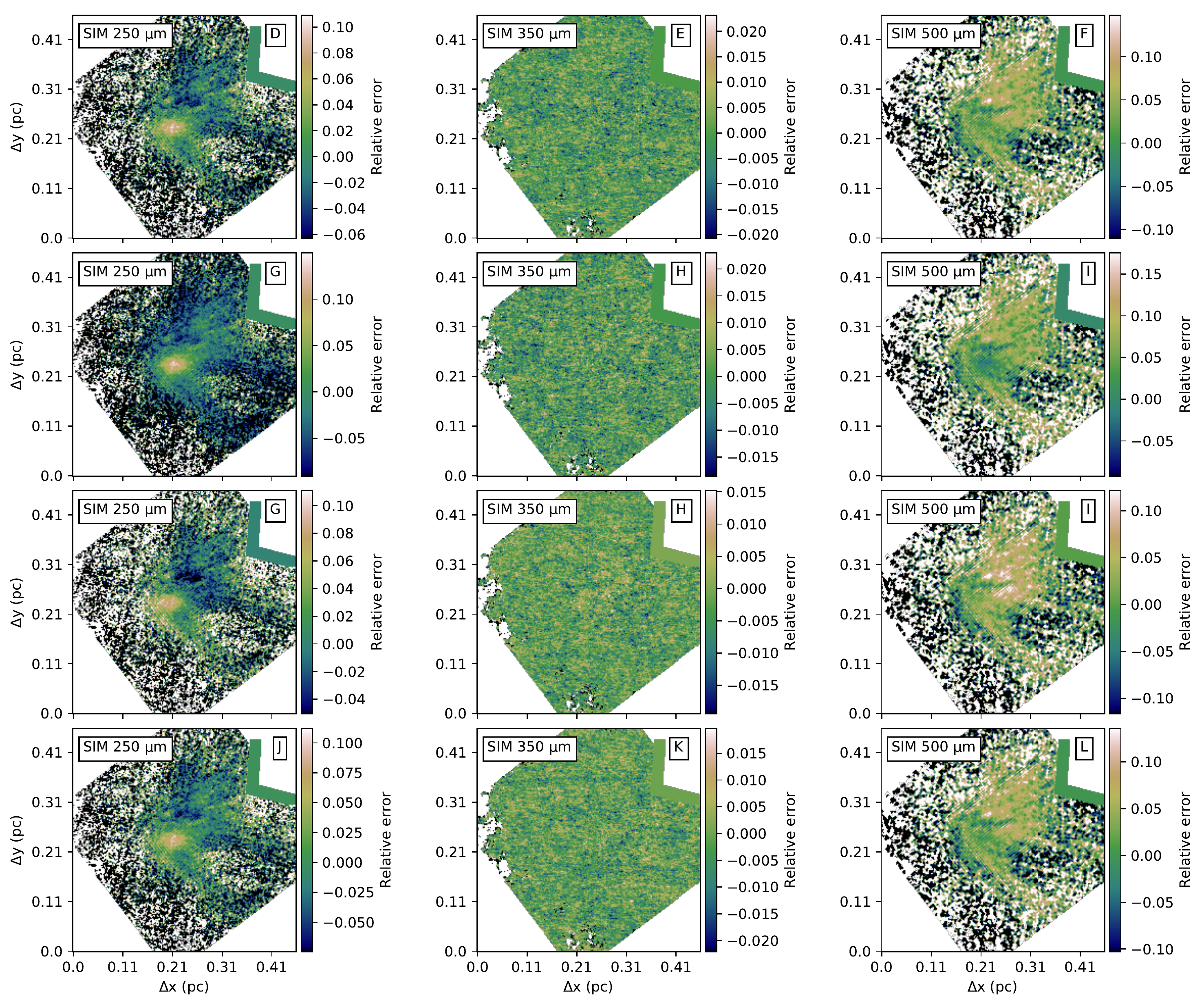}
\caption{As Fig. \ref{fig:EM_1}, but for models Scaled2, Scaled4, Wide, and LG.}
\label{fig:EM_2}
\end{figure*}

\begin{figure*}
%\sidecaption
\includegraphics[width=17.8cm]{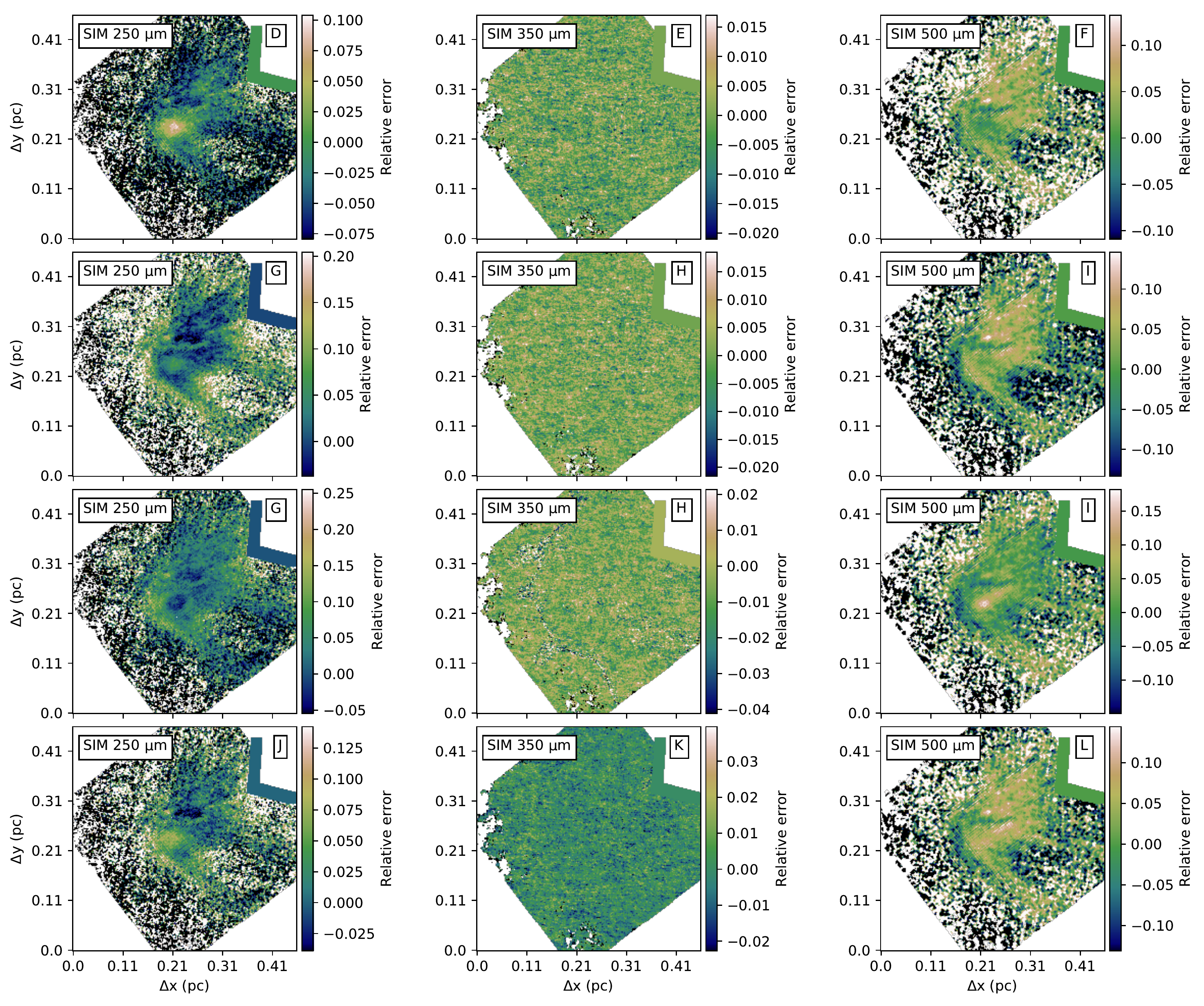}
\caption{As Fig. \ref{fig:EM_1}, but for models LGM, SIGMA, THEMIS, and DDust.}
\label{fig:EM_3}
\end{figure*}

Shown in Figs. \ref{fig:EM_1} to \ref{fig:EM_3} are the residuals between our models and the observed emission (observed value minus the model prediction) for all dust models. Each row in the figures corresponds to a single model and the column show the reiduals in the 250, 350, and 500 $\mu$m bands.

\begin{figure*}
%\sidecaption
\includegraphics[width=17.8cm]{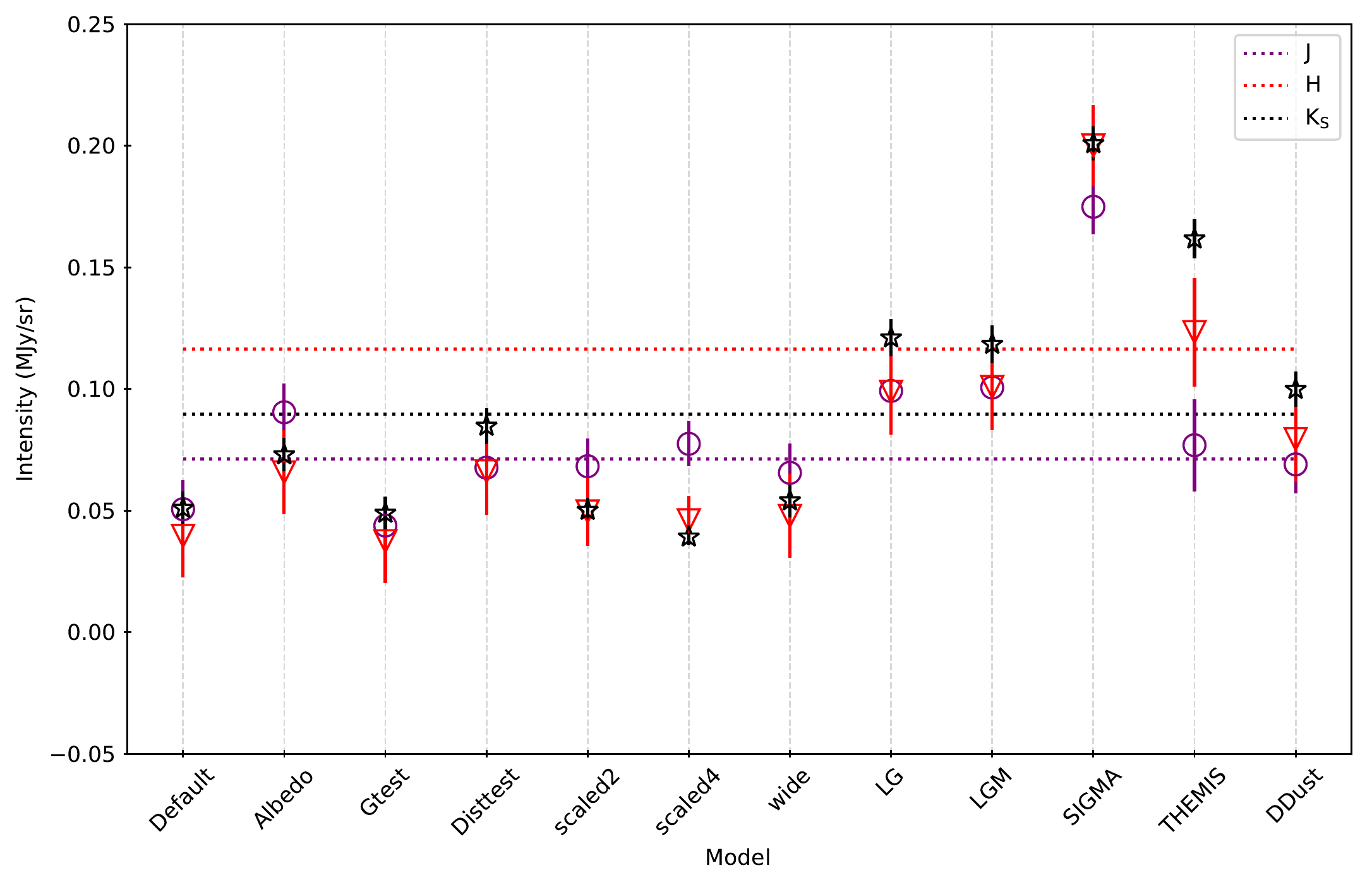}
\caption{J, H, and K$_{\rm S}$ band intensities from observations (horizontal lines) and different models (symbols) for the map position 2. The colours correspond to the J (purple), H (red), and K$_{\rm S}$ (black) bands. All intensity values have been background subtracted. We assume a $20 \, \%$ uncertainty in background sky estimates for the J and K$_{\rm S}$ bands and an uncertainty of $30 \, \%$ for the H band.}
\label{fig:all_spec_P2}
\end{figure*}

\begin{figure*}
\sidecaption
\includegraphics[width=11.5cm]{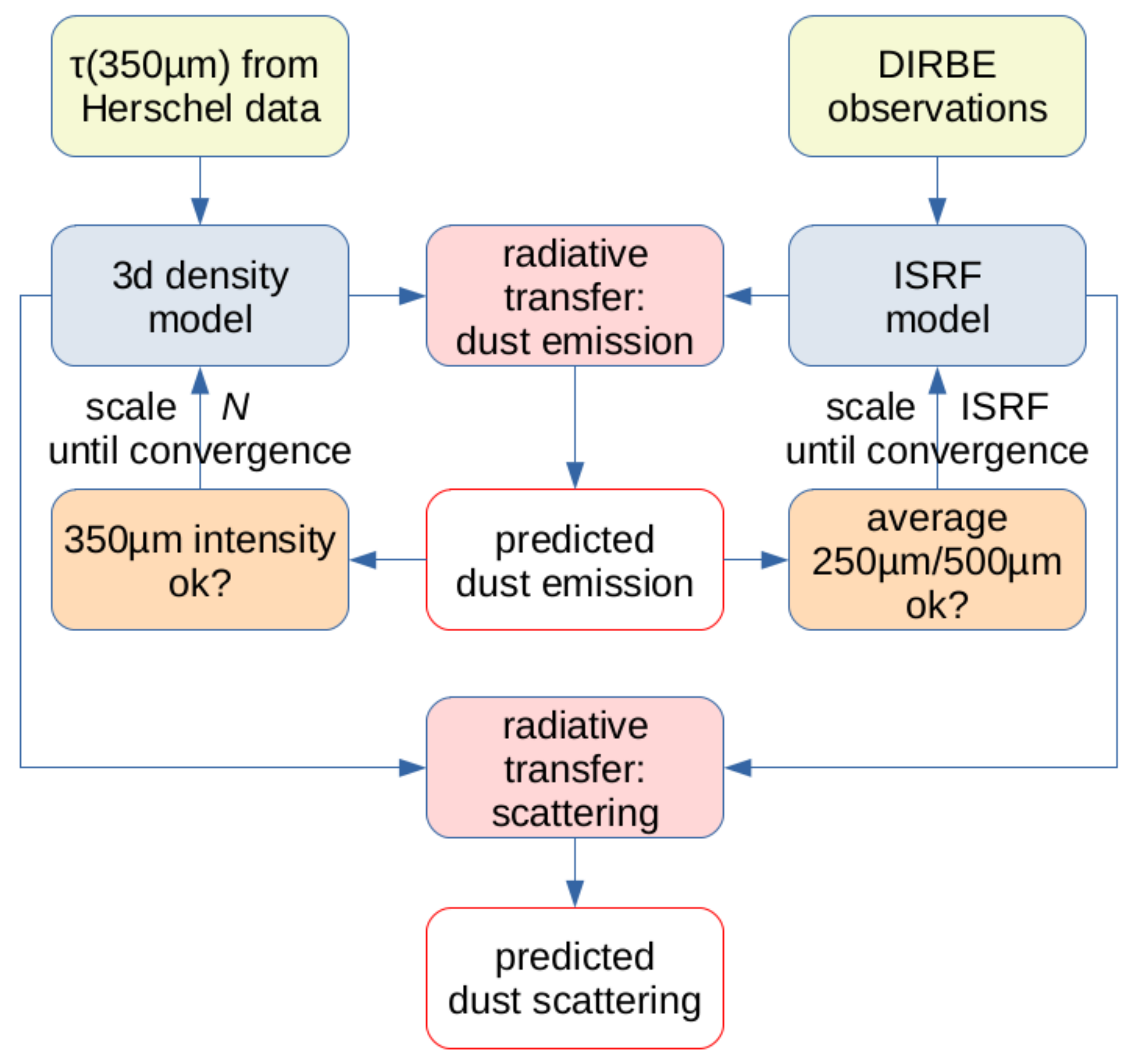}
\caption{Schematic overview of our modelling process to fit the dust emission and to compute an estimate for the scattered surface brightness.}
\label{fig:work_flow}
\end{figure*}

Additional NIR spectra extracted from point 2, see Fig. \ref{fig:scat_loc_spec}, is shown in Fig. \ref{fig:all_spec_P2} and a schematic overview of our modelling process is shown in Fig. \ref{fig:work_flow}.

\end{appendix}
\end{document}